\documentclass[fleqn,usenatbib]{mnras}

\usepackage{newtxtext,newtxmath}
\usepackage[T1]{fontenc}
\DeclareRobustCommand{\VAN}[3]{#2}
\let\VANthebibliography\thebibliography
\def\thebibliography{\DeclareRobustCommand{\VAN}[3]{##3}\VANthebibliography}

\usepackage{graphicx}	% Including figure files
\usepackage{amsmath}	% Advanced maths commands
\usepackage{orcidlink}
\usepackage{stfloats}
\usepackage{float}

\usepackage{stfloats}

\title{Eclipse Mapping with Ariel: Future Prospects for a Population-Level Mapping Survey}

\author[D. Valentine et al.]{Daniel Valentine,$^{\orcidlink{0000-0002-2643-6836} ,1,2}$\thanks{E-mail: daniel.valentine@bristol.ac.uk}
Hannah R. Wakeford,$^{\orcidlink{0000-0003-4328-3867},1}$
Mark Hammond,$^{\orcidlink{0000-0002-6893-522X},3}$
Ryan C. Challener,$^{\orcidlink{0000-0002-8211-6538},4}$
\newauthor
Billy Edwards,$^{\orcidlink{0000-0002-5494-3237},5}$
Theresa L\"uftinger,$^{2}$
Maximillian N. G\"unther$^{\orcidlink{0000-0002-3164-9086},2}$
\\
$^{1}$School of Physics, University of Bristol, HH Wills Physics Laboratory, Tyndall Avenue, Bristol BS8 1TL, UK\\
$^{2}$European Space Agency (ESA), European Space Research and Technology Centre (ESTEC), Keplerlaan 1, 2201 AZ Noordwijk, The Netherlands\\
$^{3}$Department of Physics, University of Oxford, Oxford OX1 3PU, UK\\
$^{4}$Department of Astronomy, Cornell University, 122 Sciences Drive, Ithaca, NY 14853, USA\\
$^{5}$SRON Netherlands Institute for Space Research, Niels Bohrweg 4, 2333 CA Leiden, The Netherlands\\
}

\date{Accepted 2025 September 26. Received 2025 September 05; in original form 2025 May 27}

\pubyear{\the\year{2025}}

\begin{document}
\label{firstpage}
\pagerange{\pageref{firstpage}--\pageref{lastpage}}
\maketitle

% Abstract of the paper
\begin{abstract}
Eclipse mapping is a powerful tool for measuring 3D profiles of exoplanet atmospheres. To date, only JWST has been capable of widely applying this technique, but as a general observatory, it is too time-limited to conduct population-level mapping studies. Ariel, on the other hand, is a dedicated exoplanet mission set to observe 1000 transiting exoplanets, making it a natural candidate for this. To assess Ariel's mapping potential, we quantitatively benchmark its abilities against those of JWST using a simulation-and-retrieval framework with existing JWST eclipse maps as test cases. We find that for high-ranking targets, Ariel will be able to derive qualitatively similar maps to JWST using the same amount of observations; for mid-ranking targets, Ariel will be able to compete using as few as 3$\times$ as many observations; and for lower-ranking targets, the use of phase curves overcomes the need for an impractical number of repeated eclipse observations. We find that while Ariel is unlikely to have extensive latitudinal mapping abilities, it will have wide-ranging longitudinal abilities, from which the first-order atmospheric dynamics can be constrained. Using an analytically-derived metric, we determine the best eclipse mapping targets for Ariel, finding that it will be able to map nearly 100 targets using full phase curves in only quarter of its lifetime. This would be the largest mapping survey to date, and have enormous ramifications for our demographic understanding of exoplanet atmospheric dynamics. Finally, we rank all the best mapping targets for JWST and Ariel in order to encourage future eclipse mapping studies.
% 250/250 words.

\end{abstract}

% Select between one and six entries from the list of approved keywords.
% Don't make up new ones.
\begin{keywords}
methods: observational -- exoplanets -- planets and satellites: atmospheres
\end{keywords}

%%%%%%%%%%%%%%%%%%%%%%%%%%%%%%%%%%%%%%%%%%%%%%%%%%

%%%%%%%%%%%%%%%%% BODY OF PAPER %%%%%%%%%%%%%%%%%%

\section{Introduction}
\label{sec:intro}
Exoplanet atmospheres are complex, heterogeneous environments, but past studies have tended to treat them as one-dimensional (1D) structures, primarily due to limitations associated with the precision, temporal coverage, and/or cadence of observations. The unprecedented quality of JWST data overcomes these limitations and now allows us to characterise the multidimensional profiles of exoplanet atmospheres \citep[e.g.,][]{wasp18eclipsemap, wasp43miri_eclipsemap, murphy2024, espinoza2024, wasp43nirspec_eclipsemap, wasp17eclipsemap, lally2025}. However, as an oversubscribed general observatory, the opportunities to do this with JWST are limited.

The multidimensional profiles of exoplanet atmospheres have previously only been characterisable through the inversion of phase curves into {flux} maps \citep{cowan_agol_2008}. We refer to this technique as ``phase mapping'', in which the entire planetary orbit is observed, allowing the flux of the visible hemisphere at every orbital phase to be measured and used to construct a global {flux} map. However, only differing longitudes are revealed by the planetary rotation, whilst the observed latitudes are constant. Additionally, this technique relies on whole-hemisphere measurements, which therefore only yield hemispherically-averaged representations of the true spatial {flux} profile at each phase of rotation.
Hence, only large-scale longitudinal information can be constrained by these so-called ``phase maps''. The use of phase curves also naturally limits the technique and subsequent inferences to planets with short enough periods to justify full-orbit observations; see \citet{dang2025} for a comprehensive summary of Spitzer 4.5 $\upmu$m phase curves.

Eclipse mapping takes this technique one step further in order to unlock the full three-dimensional (3D, longitude-latitude-pressure) profiles of exoplanet atmospheres \citep{rauscher2007}. During eclipse ingress, the dayside of the planet is gradually occulted by the star, leading to gradual flux decreases in the light curve until the planet is fully eclipsed. The inverse is also true during egress as the planet is revealed from behind the star. With high-cadence data, these partial eclipse phases are well sampled, wherein the star is essentially observed to be obscuring/revealing successive ``slices'' of the dayside profile. The change in flux between successive measurements can therefore be attributed to these slices, enabling slice profiles of the dayside atmosphere to be derived from both ingress and egress. Because the eclipse geometry of ingress is approximately inverse to that of egress for planets that orbit off the stellar equator, these slice profiles are oppositely oriented. Overlaying them therefore results in a grid which slices over both the longitudes and latitudes of the dayside hemisphere, within which the flux of each cell is measurable, thereby producing a two-dimensional (2D) map of the dayside thermal profile \citep{rauscher2018}. Performing this mapping technique as a function of pressure with spectroscopic observations further extends the profile to 3D. By exploiting the geometry of eclipse with high-cadence observations, eclipse mapping can therefore measure the latitudinal profiles of exoplanet atmospheres, and overall constrain smaller-scale structure than is possible with phase mapping. However, eclipse mapping requires higher quality data than phase mapping, and can only map the dayside. For a comparison of phase versus eclipse mapping, see Figure 1 of \citet{dewit2012}.

Spherical harmonics are traditionally used in the modelling framework of eclipse mapping because they can reproduce any pattern on a sphere, with the degree, $l_{\max}$, of the model denoting the complexity of the map. The first exoplanet to be eclipse mapped was HD 189733b, using 8 Spitzer/IRAC eclipses and a partial phase curve \citep{majeau2012, dewit2012}. However, even with this large dataset, the map could only constrain the first-order harmonics, corresponding to large-scale longitudinal patterns, and contained no robust latitudinal information, which is equivalent to what one would obtain with phase mapping alone. The unprecedented quality of JWST data now allows higher-quality maps to be derived for more targets using fewer data.

A handful of exoplanets have now been eclipse mapped using JWST. The first of these was WASP-18b, mapped using a single NIRISS/SOSS eclipse as part of the ERS program \citep{wasp18eclipsemap}. This map used up to fifth-order harmonics and contained evidence of magnetic field interactions with the atmosphere. Following this, WASP-43b was mapped using a MIRI/LRS phase curve \citep{wasp43miri_eclipsemap}. This was the first eclipse map wherein the latitudinal profile was robustly constrained, revealing indications of a latitudinal hotspot offset. These indications were confirmed in an additional map of WASP-43b constructed from a NIRSpec/G395H phase curve, which provided a 4$\sigma$ detection of this latitudinal asymmetry in the dayside atmosphere \citep{wasp43nirspec_eclipsemap}. WASP-17b was mapped from a single MIRI/LRS eclipse observation \citep{wasp17eclipsemap}, revealing a substantial day-night temperature contrast with a marginal eastward hotspot offset, indicative of inefficient heat recirculation for this hot Jupiter. At the cooler end of the hot Jupiter regime, HD 189733b was mapped once again with the inclusion of two MIRI/LRS eclipse observations \citep{lally2025}, revealing a much larger hotspot offset and smaller day-night temperature gradient than WASP-17b, indicative of efficient heat recirculation. WASP-69b was also preliminarily mapped from a MIRI/LRS eclipse, but degeneracies with the ephemeris led to an unconstrained hotspot location \citep{schlawin2024}. These examples demonstrate the power of eclipse mapping in extracting dynamical information, with a number of other programs having been accepted to carry out such studies with JWST (e.g., KELT-8b with MIRI/LRS, GO 5687, PI: Valentine; KELT-20b with NIRSpec/G395H, GO 6978, PI: Wardenier; TOI-2490b with NIRSpec/G395H, GO 7686, PI: Mullens).

In order to interpret the atmospheric dynamics of an exoplanet as inferred by the observables of an eclipse map, one must compare to general circulation models (GCMs), tuning different parameters until the same thermal structure is derived as that observed in the eclipse map. However, many of these parameters can be degenerate with one another; for example, both enhanced metallicity and atmospheric drag can seek to reduce the east-west hotspot offset, whilst rotation rate and irradiation temperature both dominate the heat recirculation regime at first order \citep{kataria2016, showman2020}. {Additionally}, such comparisons {carry a} model dependence because the computational complexity of GCMs necessitate a number of a-priori assumptions and approximations to be made, the different choices of which can yield significantly different end results even when the input parameters are the same \citep{wasp43miri_eclipsemap}. Degeneracies and model-dependencies such as these can be alleviated with a large and diverse enough sample to isolate control parameters and qualitatively test model assumptions across a wide parameter space \citep{lewis_and_hammond, roth2024}. Hence, a population-level eclipse mapping study is needed in order to truly utilise the wealth of atmospheric dynamic information that exoplanet eclipse maps contain. Whilst JWST has proved fruitful for pioneering the eclipse mapping technique, it is not conducive for such a time-costly study; a dedicated exoplanet mission is required.

An observatory must be capable of high-cadence, high-flux precision, and precise pointing observations in order to facilitate eclipse mapping. In order to perform 3D mapping, the observatory must also have spectroscopic capabilities. The natural candidate for such requirements is the Ariel mission \citep{ariel_citation}. Set to launch in late 2029 as the next ESA medium-class (M4) science mission, Ariel will perform a census of the atmospheric chemistry and thermodynamics of hundreds of transiting exoplanets over a four-year primary mission lifetime. Designed for precise spectroscopic measurements on an L2 orbit, Ariel meets the above requirements for eclipse mapping, with continuous spectroscopic coverage from $1.1-7.8 \ \mu$m and photometric coverage down to 0.5 $\mu$m, sampling the majority of the flux output for irradiated exoplanets. Crucially, as a dedicated exoplanet mission, Ariel has the observing time that JWST does not to conduct a population-level mapping study.

Spitzer proved the baseline requirements for eclipse mapping from its ability to map the highest-ranking eclipse mapping target \citep{boone2023}, HD 189733b, at the most fundamental level \citep{majeau2012, dewit2012}. The capabilities of the Ariel mission for such observations will supersede that of Spitzer, indicating that Ariel will at least have some degree of eclipse mapping abilities, the extent of which therefore needs to be determined.

The eclipse mapping capabilities of an observatory can be determined quantitatively via a simulation-and-retrieval framework, or qualitatively via analytic methods \citep[e.g., using the eclipse mapping metric (EMM) proposed by][]{boone2023}. We apply both methods here to robustly assess Ariel's eclipse mapping capabilities. Using the aforementioned JWST eclipse maps as test cases, we post-process them into simulated Ariel light curves for mock eclipse mapping retrievals in order directly benchmark Ariel's eclipse mapping abilities against those of JWST. We then use the methods of \citet{boone2023} to rank all the best eclipse mapping targets and provide recommendations for the {objectives, targets, and design} of {a} population-level mapping survey with Ariel.

The structure of this paper is as follows. In Section \ref{sec:methods}, we outline our simulation-and-retrieval framework for our test case analysis, and verify that it produces consistent results with the input JWST maps. In Section \ref{sec:ariel_sims}, we use this framework to conduct our Ariel test case analysis, comparing the results to JWST in order to both assess Ariel's mapping abilities, and directly benchmark the mapping abilities of the two observatories. In Section \ref{sec:emm_ranking}, we analytically rank the best mapping targets for both Ariel and JWST, and use the results of this and our test cases to devise a recommended target list and observational strategy for a population-level mapping survey with Ariel. Finally, in Section \ref{sec:conclusions}, we give our conclusions.

\section{Simulation-and-Retrieval Framework}
\label{sec:methods}

In this section, we detail our simulation-and-retrieval framework. In order to directly benchmark the eclipse mapping capabilities of Ariel against those of JWST, we take every published JWST eclipse map to date\footnote{{The maps can be obtained through the Zenodo links contained within each reference.} We exclude the WASP-69b eclipse map \citep{schlawin2024} due to the uncertainties reported by the authors.} and use them as test cases. These are listed in Table \ref{tab:test_cases}, and span a range of JWST instruments, observational set-ups, and mapping-model complexities, which allow us to robustly assess Ariel's mapping abilities across the observational parameter space. We note that we also include a {map generated from the 160mbar level of a \texttt{THOR} GCM} of HD 209458b \citep[L4 control case in][]{hd209gcm}, {post-processed as it would be observed by a MIRI/LRS observation,} since this planet has not yet been observed in eclipse with JWST. We choose to include it here because as one of the canonical hot Jupiters, it is a natural high-tier target for Ariel, therefore simulations of its eclipse mapping potential are imperative. 

\begin{figure}
    \centering
    \includegraphics[width=1\linewidth]{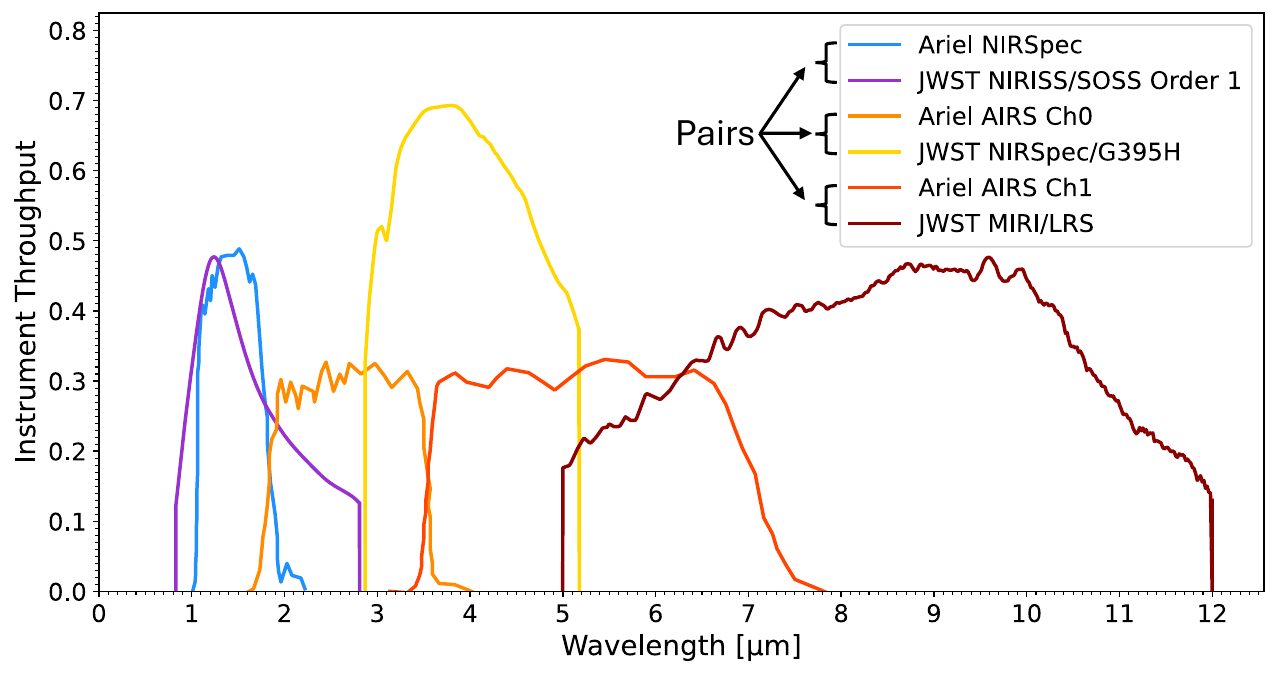}
    \caption{JWST versus Ariel spectrograph bandwidths and throughputs, {obtained from \citet{exotic-ld} and \citet{arielrad}, respectively}.}
    \label{fig:jwst_vs_ariel_wvs}
\end{figure}

We then determine the data quantity that Ariel would need to derive a qualitatively equivalent map for each test case, and compare the quantitative constraints. {To do this, we take the JWST eclipse map, post-process it into a simulated Ariel light curve and retrieve on it, continuously stacking observations if necessary until the input mapping signal is recovered.} Whilst Ariel simultaneously observes using all of its spectrographs and photometric filters for every observation, {which we highlight is highly advantageous for 3D mapping}, we choose to conduct our simulations in the closest Ariel spectrograph to the input JWST spectrograph for fairness of comparison and ease of analysis. For JWST NIRISS/SOSS Order 1 ($0.85-2.83 \ \upmu$m), this is Ariel NIRSpec ($1.10-1.95 \ \upmu$m); for JWST NIRSpec/G395H ($2.84-5.14 \ \upmu$m), it is Ariel AIRS Ch0 ($1.95-3.90 \ \mu$m); and for JWST MIRI/LRS ($5.0-12.0 \ \upmu$m), it is Ariel AIRS Ch1 ($3.9-7.8 \ \upmu$m). A comparison between the bandwidths {and throughputs} of these instruments is shown in Figure \ref{fig:jwst_vs_ariel_wvs}. {We work with broadband-integrated (i.e., ``white-light'') light curves for the entirety of this analysis.}

\begin{table*}
	\centering
	\caption{Test cases. These are the planets which have been reliably eclipse mapped with JWST to date, alongside a GCM output of HD 209458b. We use these to benchmark the eclipse mapping capabilities of the Ariel mission versus those of JWST.}
	\label{tab:test_cases}
	\begin{tabular}{lccr}
		\hline
		Planet & JWST instrument & Observations & Reference\\
		\hline
		HD 189733b & MIRI/LRS & 2 Eclipses & \citet{lally2025}\\
		HD 209458b & MIRI/LRS* & 1 Eclipse* & \citet{hd209gcm}\\
		WASP-18b & NIRISS/SOSS & 1 Eclipse & \citet{wasp18eclipsemap}\\
        WASP-17b & MIRI/LRS & 1 Eclipse & \citet{wasp17eclipsemap}\\
        WASP-43b & MIRI/LRS & 1 Phase Curve & \citet{wasp43miri_eclipsemap}\\
         & NIRSpec/G395H & 1 Phase Curve & \citet{wasp43nirspec_eclipsemap}\\
         \hline
         \multicolumn{4}{p{10.1cm}}{\footnotesize{*HD 209458b has no JWST eclipse observations, but as one of the canonical hot Jupiters, we include a map {post-processed from the 160 mbar level of a GCM of the planet, corresponding to pressures probed in the mid-IR, and integrated over the MIRI/LRS bandpass.}}}
	\end{tabular}
\end{table*}

\subsection{Simulation Framework}
\label{sec:sim_framework}
We use the mapping software \texttt{starry} \citep{starry} to post-process these input maps into simulated light curves. To construct a light curve in \texttt{starry}, one sets up a star-planet system with user-defined orbital parameters; we adopt the same orbital parameters as those used in the original work for each test case (see Table \ref{tab:test_cases}). For HD 209458b, we adopt system parameters from the JWST NIRCam analysis presented in \citet{hd209_system_params}. In order to inject the mapping signal into the simulated light curve, we take the input eclipse map from each original reference in units of $F_{\mathrm{p}}/F_{\mathrm{*}}$
in \texttt{.npy} format and load it directly into {the \texttt{starry} \textit{planet} object}. This effectively constructs a constant output star, a planet with the same {spatial flux} profile as the input map, and simulates the orbit {and rotation} of the planet and the resultant eclipse that would be observed {by the original instrument}, generating the desired simulated light curve.

Because the input maps are in units of $F_{\mathrm{p}}/F_{*}$ as observed by the original JWST spectrograph, when simulating the light curves as they would be observed by the equivalent Ariel spectrograph, we analytically rescale the relative flux values by multiplying through the flux ratios (Ariel/JWST) of the instruments.
{We analytically calculate the relative flux for each instrument using
\begin{equation}
    \frac{F_p}{F_*}=\frac{I_p}{I_*}
    (\frac{R_p}{R_*})^2,
\label{eq:flux}
\end{equation}
\noindent where $F$ is flux, $I$ is intensity, and $R$ is radius, with the subscripts $p$ and $*$ denoting the planet and star, respectively. The intensity is calculated with a blackbody assumption using the Planck function as
\begin{equation}
    {I_\alpha=\frac{2hc^2}{\lambda_{\mathrm{thr}}^5 e^{(\frac{hc}{\lambda k_B T_\alpha}-1)}}},
\label{eq:intensity}
\end{equation}
\noindent where the wavelength-dependence, $\lambda_{\mathrm{thr}}$, carries our scaling between JWST and Ariel, and is weighted by the relevant instrument throughput. 
We use the equilibrium temperature of each test case for the planetary temperature, $T_{p}$, and the stellar effective temperature for $T_*$, again adopted from the relevant citation in Table \ref{tab:test_cases}.}
To additionally account for differences between the measured and analytically-predicted JWST eclipse depth, we further take their ratio and multiply this through as a correction factor when scaling the Ariel flux.

For fairness of comparison, we adopt the same observing time and baseline distribution of the original JWST observations for our Ariel simulations. For HD 209458b, we assume a JWST comparison observation made up of one MIRI/LRS eclipse observation using a standard 1:1 baseline, with an additional hour pre-eclipse to account for the MIRI/LRS persistence ramp \citep{dyrek2024}, and half an hour post-eclipse to emulate a scheduling window buffer.

To set the cadences and flux precisions of the Ariel simulated light curves, we use outputs from \texttt{ArielRad} \citep{arielrad}, which is an end-to-end radiometric tool designed to simulate realistic Ariel observations.
\texttt{ArielRad} provides the saturation time within every wavelength channel of a given spectrograph for a given target. We adopt the minimum saturation time across the channels of a given spectrograph for a given target as the cadence of our simulated observations in order to ensure that there is no saturation across the entire wavelength range. We then calculate the native flux error from \texttt{ArielRad} using this cadence in order to derive the precision of our simulated light curve (i.e., the flux error on each data point).

However, due to Ariel's smaller collecting area compared to JWST \citep[1 m versus 6.5 m diameter mirror,][]{mirror_reference}, the saturation times can run into the tens of minutes, which leads to poor spatial sampling of the dayside. \citet{nullspace} numerically determined that a 16/16 sampling rate of ingress/egress (32 in total) is sufficient to scan over all spatial signals up to 4th spherical harmonic degree. To facilitate eclipse mapping with Ariel, we therefore introduce a decision loop into our calculations that if the saturation time is short enough to provide $\geq$32 samplings of ingress/egress, then we adopt it as the cadence (true for our bright HD test cases). If the saturation time is longer than this (true for our dimmer WASP test cases), then we adopt a cadence such that the criterion of 32 samplings of ingress/egress is achieved, and rescale our flux precision accordingly by substituting this value back into \texttt{ArielRad}.

When simulating stacking additional Ariel observations, we rescale the flux precision as $\sqrt{N}$ the number of observations following a first-order photon noise approximation. In doing so, we adopt a noise floor of 10 ppm, as set by \citet{arielrad}, to prevent the flux precision from scaling down indefinitely.

\subsection{Mapping Framework}
To extract the mapping signal from the simulated light curves and construct eclipse maps from them, we use the eclipse mapping software \texttt{ThERESA} \citep{theresa}. Here, we briefly summarise the methodology of \texttt{ThERESA}, and refer the reader to \citet{theresa} for further details.

To construct an eclipse map, one fits to signals in the light curve during ingress/egress which correspond to deviations from the occultation of a uniformly-emitting planet, and converts them to the brightness pattern to which they correspond. To do this, one therefore requires a set of basis light curves which are translatable to basis maps. A natural formulation for this is spherical harmonics, as was first adopted for exoplanet eclipse mapping \citep{majeau2012, dewit2012}. However, pure spherical harmonic mapping suffers from degeneracies in that while the set of basis maps themselves are orthogonal and thus independent of one another, their corresponding basis light curves are not. {This means that the same feature in the light curve can be produced by an ensemble of different basis map configurations, leading to signal degeneracy complications which inhibit definitive mapping solutions.}

{The problem here can be attributed to the fact that the orthogonality of the modelling basis is located on the output side (i.e., the maps) rather than the observable side (i.e., the light curve).} \citet{rauscher2018} therefore proposed the method of ``eigenmapping'' in order to overcome such signal degeneracies {by inverting the order of the orthogonality}. By performing principal component analysis (PCA) on the spherical harmonic basis, this method transforms it to a new basis in which the basis light curves (``eigencurves'') {are now the ones that are orthogonal to one another}. This facilitates unique mapping solutions to each light curve signal {using the newly created ``eigenmaps'', overcoming the signal degeneracy complication. This is} the modelling framework that \texttt{ThERESA} adopts.

\texttt{ThERESA} optimises for spherical harmonic degree, $l_{\mathrm{max}}$ and number of components, $N$, selecting the appropriate model complexity based on the information content of the data via the Bayesian Information Criterion \citep[BIC,][]{Raftery1995BIC}.
We use the notation LXNX to refer to our eigenmapping models, which denotes the maximum spherical harmonic degree used in the eigenbasis input (LX), and the number of eigenmap components used to construct the overall eclipse map (NX).

\texttt{ThERESA} ranks the eigenmaps in terms of their observability, fitting the most observable signals first. The NX of a model therefore denotes its complexity. In general, the first (N1) eigenmap corresponds to a day-night temperature contrast; the second (N2) eigenmap to an east-west hotspot offset; the third (N3) to north-south brightness patterns; and higher-order (N4+) maps to smaller-scale patterns from less dominant atmospheric dynamics, such as the magnetic drag effects in the WASP-18b eclipse map \citep{wasp18eclipsemap}.  Hence, only models with N3+ can contain any latitudinal structure.

{Since the modelling framework is built on spherical harmonics, the derived mapping model will naturally extend over the entire sphere of the planet. However, \texttt{ThERESA} uses a visibility function to calculate which longitudes of the planet were scanned over during the observation, which may extend beyond the dayside due to the rotation of the planet during the observation. It is only in these observed regions that the model is in fact data-driven, and so we only plot our maps and infer constraints over these regions.}

The maps are finally converted from relative flux units to brightness temperature, {weighted by the instrument throughput and} calculated per pixel of the user-set resolution of the eclipse map (we elect to use $48_{\mathrm{lon}}\times24_{\mathrm{lat}}$ pixels for runtime efficiency). We model the stellar flux using Phoenix models \citep{phoenix} with stellar parameters adopted from the original reference (see Table \ref{tab:test_cases}) for each test case (using \citet{hd209_system_params} again for HD 209458b). Phoenix models were used in the original analysis for all of the test cases other than WASP-18b, where the star was modelled as a blackbody; for ease of comparison with the original work, we therefore model it the same way here. {To prevent non-physical solutions, \texttt{ThERESA} enforces a physicality constraint that the flux in every cell must be greater than zero. This constraint is not enforced on the regions of the planet not scanned over during the observation.}

\begin{table*}
	\centering
	\caption{Comparison of the literature parameters of the test cases versus those we recover from our simulation-and-retrieval framework. {We quote the median dayside temperature, in addition to the temperature range across the observed regions of the planet in square brackets.} We list the longitudinal hotspot offset for each test case, with the latitudinal hotspot offset below if one was measured (only true in the case of WASP-43b).}
	\label{tab:jwst_sims}
	\begin{tabular}{lccccccc} % four columns, alignment for each
		\hline
        \hline
         & & \multicolumn{3}{c}{Literature} & \multicolumn{3}{c}{Simulations} \\
        \hline
		Planet & {Observation} & Model &  Hotspot [deg] & {$T_{\mathrm{Bright}}$} [K] & Model &  Hotspot [deg] & {$T_{\mathrm{Bright}}$} [K]\\
		\hline
        \hline
        
		HD 189733b & {2 MIRI/LRS} & L5N2 & $33.0{^{+2.3}_{-2.5}}$* & {1210} & L5N2 & {$33.7\pm3.1$} & {1200$\pm$10}\\

         & {Eclipses} & & & {$[1000-1330]$} & & & {$[1010-1320]$}\\

        \hline
        
		HD 209458b & {1 MIRI/LRS} & / & 48.8 & {$1480$} & L5N2 & $54.3{^{+6.2}_{-6.1}}$ & {1430$\pm$20}\\

        & {Eclipse (GCM)} & & & {$[1150-1600]$} & & & {$[1240-1570]$}\\

        \hline
        
		WASP-18b & {1 NIRISS/SOSS} & L5N5 & $\sim$-40$-$40** & {$2900$} & L5N5 & $\sim$-40$-$40* & {$2860\pm40$}\\

        & {Eclipse} & & & {$[1250-3100]$} & & & {$[1250-3140]$}\\

        \hline
        
        WASP-17b & {1 MIRI/LRS} & L2N2 & $18.7^{+11.1}_{-3.8}$ & {$1600$} & L2N2 & $18.1^{+8.5}_{-1.1}$ & {$ 1600\pm60$}\\

        & {Eclipse} & & & {$[600-2170]$} & & & {$[600-2180]$} \\

        \hline
        
        WASP-43b & {1 MIRI/LRS} & L2N6*** & $7.8\pm0.4$ & {$1530$} & L2N6 & $6.6^{+0.4}_{-0.5}$ & {$1520\pm50$}\vspace{2pt}\\
        
        (Hammond+24) & {Phase Curve} & & $-10.7^{+4.1}_{-4.7}$ & {$[750-1790]$} & & $-14.3^{+9.8}_{-8.5}$ & {$[730-1760]$}\\

        \hline
        
        WASP-43b & {1 NIRSpec/G395H} & L3N6 & $6.9\pm0.5$ & {$1640$} & L3N6 & $7.2\pm0.3$ & {$1630\pm 25$}\\
        (Challener+24) & {Phase Curve} & & $-13.4^{+3.2}_{-1.7}$ & {$[710-1975]$} & & $-12.6^{+2.4}_{-1.7}$ & {$[720-1950]$}\\
        
		\hline
        \hline
        \multicolumn{8}{p{0.875\linewidth}}{\footnotesize{*We quote the median value derived from the canonical analysis, which used both the JWST and Spitzer data, but the uncertainties from only fitting the JWST data for fairer comparison to our simulation.}}\\
        \multicolumn{8}{p{0.875\linewidth}}{\footnotesize{**No significant hotspot offset was measured for WASP-18b, but rather a brightness plateau around the substellar point - we therefore instead note the ranges of this plateau here.}}\\
        \multicolumn{8}{p{0.875\linewidth}}{\footnotesize{***The canonical analysis of this data used spherical harmonic mapping, but a secondary eigenmapping analysis with \texttt{ThERESA} was also conducted; that model is stated here for comparison, but the other values and input map are from the primary analysis.}}
	\end{tabular}
\end{table*}

\subsection{Verification of the JWST Inputs}
\label{sec:jwst_sims}

To first verify that our simulation-and-retrieval framework yields both quantitatively- and qualitatively-equivalent maps as were injected, we first post-process these input maps into simulated versions of the JWST light curves originally used to derive them, and test that we recover the same map. For HD 209458b, we simulate the light curve flux precision and cadence using \texttt{PandExo} \citep{pandexo}, and adopt the same baseline as described in Section \ref{sec:sim_framework}.

Since any attempt to simulate Ariel systematics would be an approximation prior to commissioning tests conducted in space following launch, we elect to conduct our simulations with no injected systematic signal. These JWST simulations therefore give us a fairer set of maps against which to eventually compare our Ariel simulations, since they remove any systematic biases \citep[][]{schlawin2023} between the original work and our framework, in addition to any potential modelling biases.

We tabulate the results of our JWST simulations, including the mapping models, temperatures, and hotspot offsets, against those of the literature values in Table \ref{tab:jwst_sims}. No literature mapping model is listed for HD 209458b as this is a GCM output, not a data-derived JWST eclipse map (see Section \ref{sec:methods} for details). We note that the temperature ranges quoted in the square brackets are only defined over the longitudes that were visible during the observation; whilst our models naturally extend over the entire planet, they are not data-validated over these unobserved longitudes, and hence we do not place constraints on nor plot them. {The median temperatures quoted are only defined across the dayside longitudes (i.e., between $\pm90^{\circ}$), since these are the regions most formally constrained by eclipse maps. This also enables more intuitive interpretation for cases where a significant portion of the nightside is observed. These median dayside temperatures are weighted by the cosine squared of the latitudes to account for the decreasing emitting area and reduced viewing geometry.}

We find that for each of our test cases, we recover the same mapping model from our simulations as was derived in the original analyses, indicating that the mapping signal is being correctly injected. Below, we discuss the consistency of our simulation-retrieved hotspot offsets and temperatures with the literature values.

\subsubsection{Hotspots}
In general, our simulation-recovered hotspots are all within 1$\sigma$ of the literature values, with similar magnitude uncertainties consistent with random scatter.
The only notable discrepancy is in the case of the WASP-43b MIRI/LRS map \citep{wasp43miri_eclipsemap}, for which the longitudinal hotspot offset is only consistent to within 2$\sigma$, and the uncertainty on the latitudinal hotspot offset is $\sim$2$\times$ larger in our simulation.

Both of these can be attributed to the differing modelling framework we employ compared to the original analysis; this is the only test case where spherical harmonic mapping rather than eigenmapping was used to derive the input map. A secondary eigenmapping analysis with \texttt{ThERESA} was carried out in that work, and our longitudinal hotspot offset is in fact 1$\sigma$ consistent with that map. Figure 5 of \citet{wasp43miri_eclipsemap} also shows that the eigenmapping analysis derived a less-constrained latitudinal profile, consistent with what we recover here. Since we will also use \texttt{ThERESA} to model the Ariel simulations, this modelling bias is therefore removed by adopting this simulated JWST result as our comparison case. We also note that the discrepancies between the models are largely exacerbated by the small uncertainties on the hotspot offset in this case, which is driven by the inclusion of the full phase curve signal in the map; morphologically, the maps are essentially identical, and yielded identical inferences on the atmospheric dynamics \citep{wasp43miri_eclipsemap}.

For WASP-17b, we conversely recover slightly smaller uncertainties on the hotspot offset from our simulation than was originally derived. This can be attributed to the lack of systematics in the simulated light curve; the original MIRI/LRS eclipse observation of WASP-17b \citep{wasp17eclipsemap} showed a large magnitude persistence ramp of a morphology similar to that of the phase signal associated with the hotspot offset, leading to degeneracies between these two parameters. This effect was more significant than has been observed for most other JWST MIRI/LRS observations, which tend to show a faster decaying and oppositely oriented ramp in the white light curve \citep[e.g.,][]{grant2023, bell2024}, leading to a larger degeneracy for the WASP-17b eclipse map. The systematic-free nature of our simulation breaks this degeneracy and results in a better constrained hotspot offset. 

Both of these examples support our decision to remove both modelling- and systematic-dependencies from our test cases in order for fairer eventual comparison to our Ariel simulations. Hence, we conclude that our recovered hotspot offsets are consistent with the literature values, indicating that the mapping signal is being correctly injected into the light curves.

\subsubsection{Temperatures}
{In order to compare to ``literature'' temperature values for HD 209458b, the map of which is post-processed from a GCM, we use the \texttt{{utils.fmap\_to\_map()}} functionality of \texttt{ThERESA} to convert the GCM flux map to brightness temperature, using a Phoenix model \citep{phoenix} with stellar parameters from \citet{hd209_system_params} to model the stellar flux. For our data-derived test cases, we compare directly to the literature values.}

As in the case of the hotspot offsets, we find that our simulation-recovered temperatures are also largely consistent with the literature values (see Table \ref{tab:jwst_sims}). {We do not quote uncertainties on the literature values here they are not quoted in all cases, and not comparative for the GCM-derived map of HD 209458b.} The only notable discrepancy between the literature-and-simulation temperatures is in the case of HD 209458b, for which the limb temperatures are recovered to be {$\sim$100$\ \mathrm{K}$} hotter than the input map, {whilst the median dayside temperature is recovered slightly too cold.} 

{There are complexities between directly comparing a GCM map and a data-derived eclipse map, which can appear to be different but in reality be consistent; for further details, see \citet{nullspace}. However, we were able to show that these differences can} be attributed to this map having the largest hotspot offset with the most extended profile. Hence, in the most reliable central-longitudes region of the map, from which the day-night contrast is largely derived, the temperature gradient is observed to be quite small, and the sharper drop-off to the limbs is missed. We tested this theory by extending the observation to a full phase curve in order to more accurately measure the limb temperatures, and found that the day-night gradient was more successfully recovered in this case, {in addition to producing a more consistent median dayside temperature}. For a map derived from an eclipse-only observation, these discrepant limb temperatures are of least concern because they are largely unconstrained due to being only momentarily observed and not optimally viewed. {However, we recommend that for a data-derived eclipse map of HD 209458 with JWST, a long baseline would be advisory in order to account for this large hotspot offset.}

In all other cases, the recovered temperatures from these JWST simulations are consistent with the input maps within the uncertainties.
Coupled with the consistent hotspot offsets, this validates that the mapping signal is being correctly injected. We therefore now proceed to construct our Ariel simulations.

\section{Ariel Test Case Simulations}
\label{sec:ariel_sims}

\begin{table*}
	\centering
	\caption{JWST vs Ariel test case results. We outline the results of our Ariel simulations, in particular the recovered models, hotspot offsets and temperatures, as a function of the number and type of observations required in order to derive them. For ease of comparison, we also re-tabulate our simulated JWST results from Table \ref{tab:jwst_sims}. In the case of WASP-43b, we first list the longitudinal and then the latitudinal hotspot offset below. Temperatures are quoted over the regions of the planet visible during the observation in square brackets, alongside the median dayside temperature and uncertainty.}
	\label{tab:jwst_vs_ariel}
	\begin{tabular}{lcccccccc}
        \hline
        \hline
    	 & \multicolumn{4}{c}{JWST} & \multicolumn{4}{c}{Ariel} \\
        \hline
		Planet & Observation & Model & Hotspot [deg] & {$T_{\mathrm{Bright}}$} [K] & Observation & Model & Hotspot [deg] & {$T_{\mathrm{Bright}}$} [K]\\
		\hline
        \hline
		HD 189733b & 2 MIRI/LRS & L5N2 & $33.7\pm3.1$ & $1200\pm10$ & 2 AIRS Ch1 & L5N2 &{$31.5^{+4.2}_{-6.3}$} & {$1210\pm20$}\\

         & Eclipses & & & $[1010-1320]$ & Eclipses & & & $[1050-1290]$\\
        
        \hline
        
		HD 209458b & 1 MIRI/LRS & L5N2 & $54.3^{+6.2}_{-6.1}$ & 1430$\pm$20 & 1 AIRS Ch1 & L5N2 & {$58.2^{+20.8}_{-20.6}$} & {$1400\pm 50$}\\

         & Eclipse (GCM) & & & $[1240-1570]$ & Eclipse & & & $[1280-1510]$\\
        
        \hline
        
		WASP-18b & 1 NIRISS/SOSS & L5N5 & $\sim$-40$-$40 & $2860\pm40$ & 3 Ariel NIRSpec & L5N5 & $\sim$-40$-$40 & $2890\pm40$\\

         & Eclipse & & & $[1250-3140]$ & Eclipses & & & $[1270-3130]$\\

        \hline
        
        & & & & & 5 AIRS Ch1 & L2N2 & $33.7^{+38.7}_{-21.9}$ & $1675\pm180$\\
        & & & & & Eclipses & & & $[1250-2000]$\vspace{3pt}\\

        WASP-17b & 1 MIRI/LRS & L2N2 & $18.1^{+8.5}_{-1.1}$ & $1600\pm60$ & 20 AIRS Ch1 & L2N2 & $24.6^{+12.3}_{-7.8}$ & $1650 \pm 100$\\
        & Eclipse & & & $[600-2180]$ & Eclipses & & & $[1050-2050]$\vspace{3pt}\\

        & & & & & 1 AIRS Ch1 & L2N4 & $18.7^{+1.2}_{-3.0}$ & $1620\pm30$\\
        & & & & & Phase Curve & & & $[700-2100]$\\

        \hline
        
        & & & & & 1 AIRS Ch1 & L2N4 & $10.9\pm0.8$ & {$1530\pm50$} \\
        
        & & & & & Phase Curve & & / & $[800-1700]$\vspace{3pt}\\
        
        WASP-43b & 1 MIRI/LRS & L2N6 & $6.6^{+0.4}_{-0.5}$ & $1520\pm50$ & 18 AIRS Ch1 & L2N6 & $6.9^{+0.5}_{-0.7}$ & $1510\pm50$\vspace{1pt}\\        
        
        (Hammond+24) & Phase Curve & & $-14.3^{+9.8}_{-8.5}$ & $[730-1760]$ & Phase Curves & & $-11.8^{+8.2}_{-9.6}$ & $[800-1720]$\vspace{3pt}\\

         & & & & & 18 AIRS Ch1 & L2N2 & $9.6\pm1.0$ & $1520\pm 15$\\

        & & & & & Eclipses & & / & $[950-1720 ]$\\
        
        \hline
        
        & & & & & 1 AIRS Ch0 & L3N4 & $11.3^{+0.7}_{-0.6}$ & $1640\pm25$ \\
         
        & & & & & Phase Curve & & / & $[750-1950]$\vspace{3pt}\\
         
        WASP-43b & 1 NIRSpec/G395H & L3N6 & $7.2\pm0.3$ & $1630\pm25$ & 18 AIRS Ch0 &  L3N6 & $6.5^{+0.5}_{-0.6}$ & $1650\pm25$\vspace{1pt}\\
         
        (Challener+24) & Phase Curve & & $-12.6^{+2.4}_{-1.7}$ & $[720-1950]$ & Phase Curves & & $-14.1^{+2.1}_{-2.2}$ & $[730-1950]$ \vspace{3pt}\\

         & & & & & 18 AIRS Ch0 & L3N2 & $10.9\pm1.0$ & $1640\pm 10$\\

        & & & & & Eclipses & & / & $[950-1950]$\\
        
		\hline
        \hline
	\end{tabular}
\end{table*}

Here, we inject our test case eclipse maps into simulated Ariel light curves.
We then retrieve on these simulated light curves using \texttt{ThERESA}, seeking to recover
the same eclipse mapping model, LXNX, that we recover from our JWST simulations (Table \ref{tab:jwst_sims}). If the number of observations is insufficient to recover the input map, then we simulate stacking additional Ariel observations by scaling the flux precision by $\sqrt{N}$ the number of observations (up to a noise floor of 10 ppm) until we recover the input model. 

Our primary goal is to test the data quantity required by Ariel in order to derive the same qualitative mapping model as JWST, and then compare the quantitative inferences that are derivable from those maps between the observatories.
In the following subsections, we outline the results of our test cases, and the primary conclusion that we draw from each. We tabulate the comparisons between JWST and Ariel in Table \ref{tab:jwst_vs_ariel}.

\subsection{HD 189733b \& HD 209458b}

HD 189733b and HD 209458b are the two canonical hot Jupiters which paved the way for many advancements in the field as a result of their system brightness and therefore high data quality \citep[e.g.,][]{charbonneau2000, brown2001, bouchy2005, deming2006, deming2013, knutson2007, knutson2012, showman2008, sing2011, majeau2012, dewit2012, zellem2014, flowers2019, kilpatrick2020, inglis2024, hd209_system_params}. As such, much of our theoretical understanding and modelling frameworks are based upon these targets. They are two of the highest-ranking targets in terms of their eclipse mapping metric \citep[EMM, ][]{boone2023}. Hence, they are the natural first test cases for a study such as this.

\subsubsection{HD 189733b}

HD 189733b is the planet with the overall highest EMM \citep[][]{boone2023}, with its best value for MIRI/LRS. This is therefore our fundamental test case; an observatory with baseline eclipse mapping capabilities should be capable of eclipse mapping it in the mid-infrared, as Spitzer was capable of doing \citep{majeau2012, dewit2012}. The input JWST MIRI/LRS map \citep{lally2025} is a two-component (N2) model, broadly corresponding to a day-night temperature contrast and longitudinal hotspot offset, with no constrained latitudinal structure. We take this input map and post-process it into an Ariel AIRS Ch1 light curve, our MIRI/LRS equivalent, {using the methods outlined in Section \ref{sec:sim_framework}}.

Since our JWST reference case was produced using two JWST MIRI/LRS eclipses, we compare to what we would recover using two Ariel AIRS Ch1 eclipses by scaling our flux uncertainties down by $\sqrt{2}$. We list our recovered model, hotspot location, and temperatures in Table \ref{tab:jwst_vs_ariel}. We find that with the same amount of data as JWST, Ariel is able to recover the input mapping model, proving that it has, at minimum, baseline eclipse mapping capabilities. We also note that, the same as JWST, Ariel can also map HD 189733b using only a single eclipse observation, but we discuss the results of our two-eclipse example here for fairness of comparison to the results of \citet{lally2025}.

\begin{figure*}
    \centering
    \includegraphics[width=0.3275\linewidth]{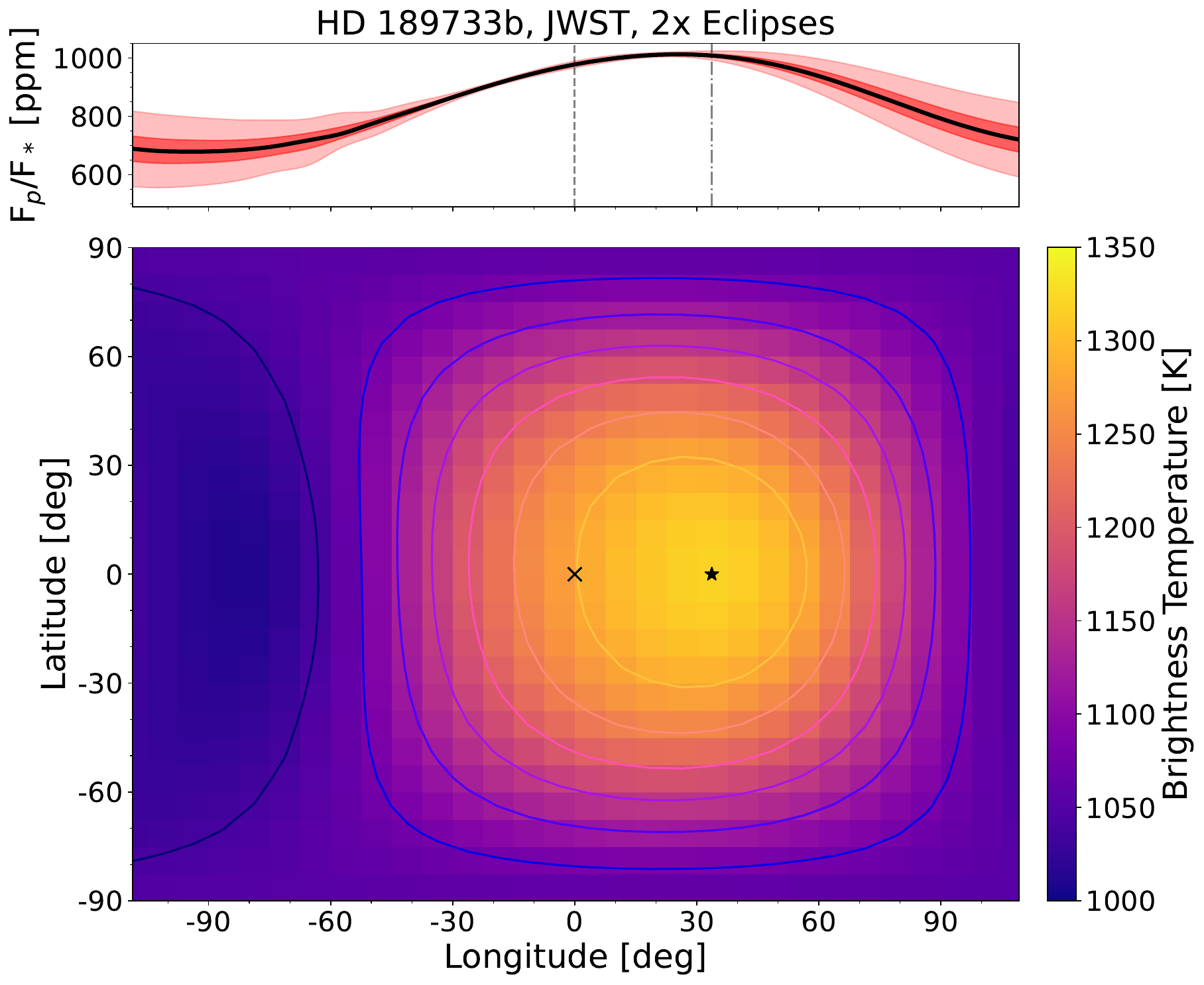}
    \includegraphics[width=0.3275\linewidth]{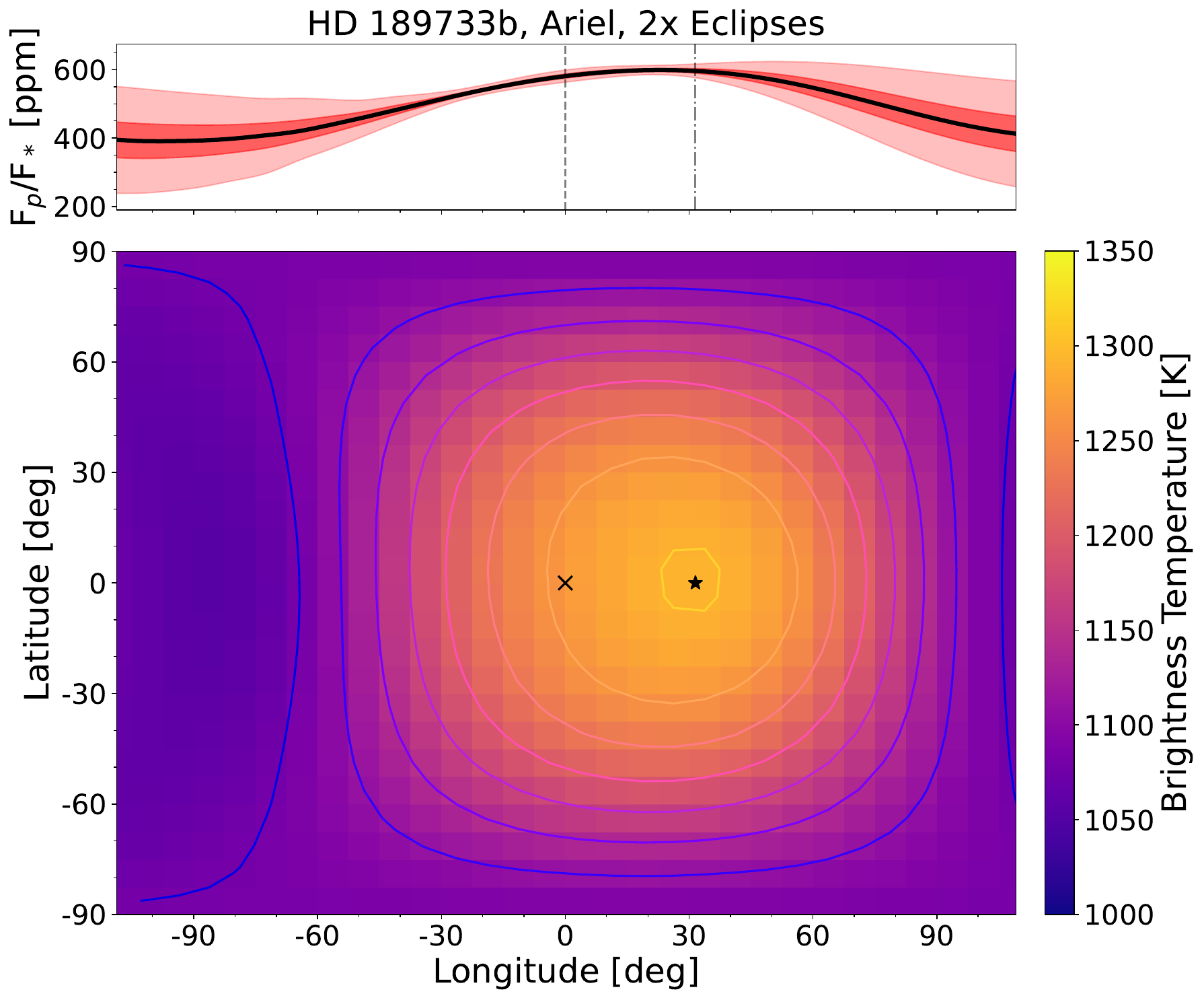}
    \includegraphics[width=0.335\linewidth]{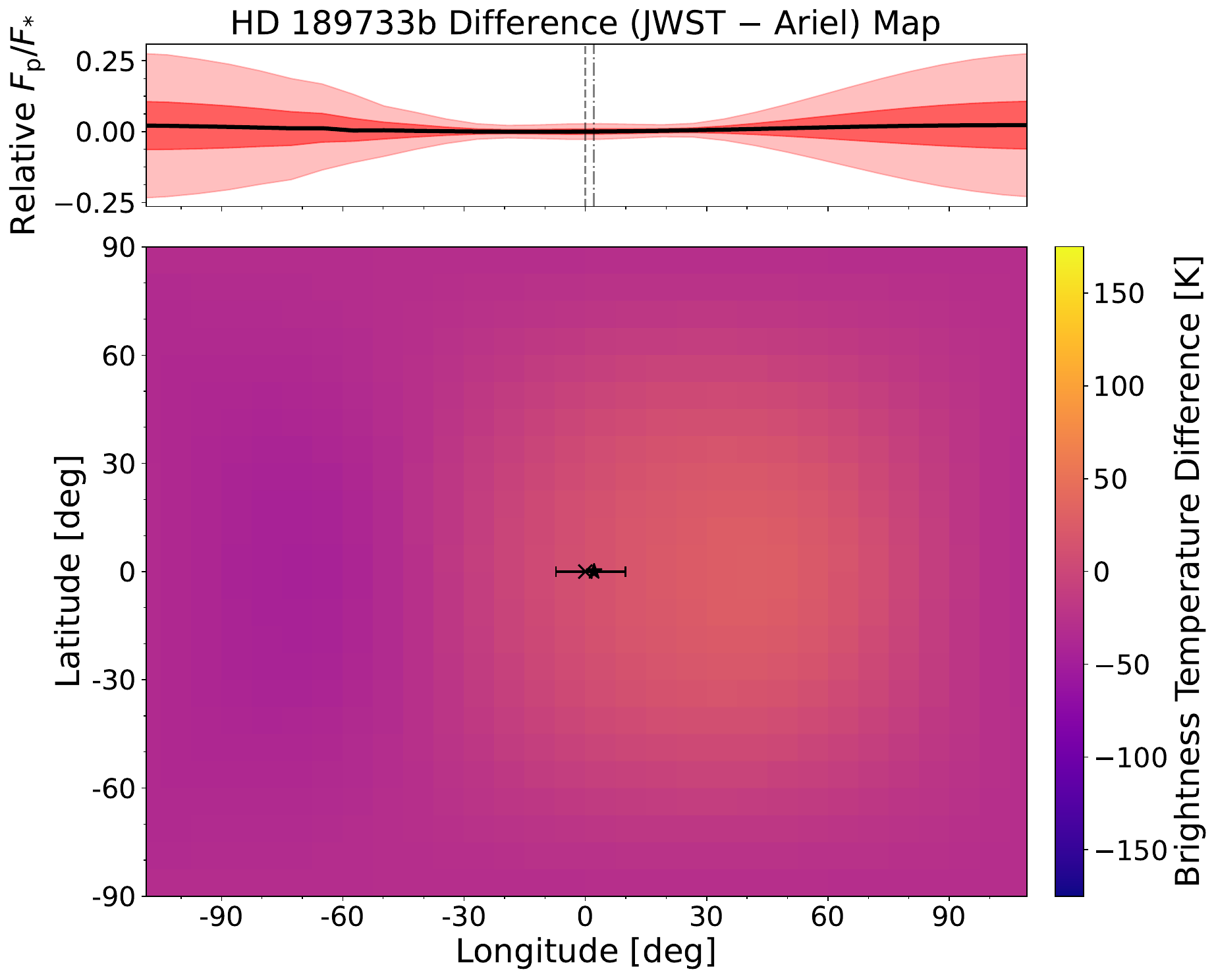}
    \caption{HD 189733b eclipse maps, {with flux profiles shown above. The cross and dashed line mark the location of the substellar point, whilst the star and dash-dotted line mark the location of the hotspot}. \textit{Left:} Simulated version of the JWST MIRI/LRS map, derived from two eclipses \citep{lally2025}. \textit{Middle:} Equivalent Ariel AIRS Ch1 map, also derived from two eclipse observations. \textit{Right:} Difference map (JWST$-$Ariel). The flat structure shows that we recover consistent thermal structures between the instruments, although with a slightly shallower day-night gradient for the latter as a consequence of its smaller bandwidth (see text). The overlap of the differenced hotspot with the substellar point shows that we accurately and precisely recover this key parameter.}
    \label{fig:hd189_maps}
\end{figure*}

Figure \ref{fig:hd189_maps} shows our simulation-recovered JWST map on the left, Ariel map in the middle, and difference map (JWST$-$Ariel) on the right, {with flux profiles plotted above to give a sense of the temperature constraints as a function of longitude.}
The difference plot reveals any structural differences between the JWST and Ariel maps.
For identical maps, the difference map {and flux profile} would be flat, and the differenced hotspot would be located at the substellar point. The uncertainties quoted on the differenced hotspot are the sum of the JWST and Ariel uncertainties, {whilst for the differenced flux profile, we sum the uncertainties in quadrature.}

{We plot the maps in absolute temperature units since we rescale the Ariel values from the JWST ones using a blackbody assumption, therefore we expect to recover comparable temperatures. The measured flux, on the other hand, will differ due to the different wavelengths of the spectrographs, therefore we plot the differenced flux profile in relative units.}

\begin{figure}
    \centering
    \includegraphics[width=1\linewidth]{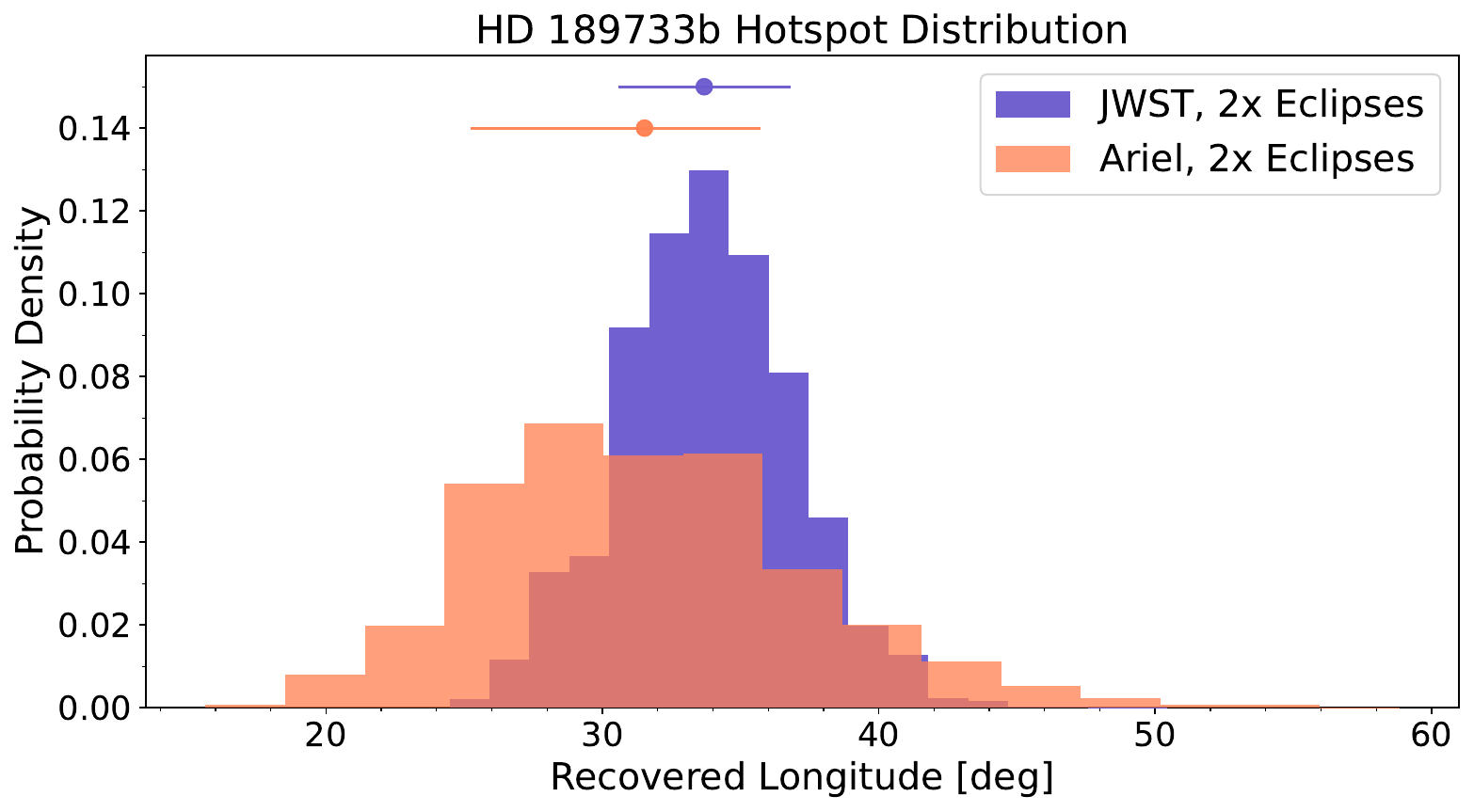}
    \caption{HD 189733b hotspot location posteriors. In blue is what we recover from the JWST MIRI/LRS map, and in orange what we recover from the Ariel AIRS Ch1 map. Using the same amount of data, Ariel is able to recover a highly consistent hotspot offset with uncertainties only $\sim$2$\times$ as large as JWST for this bright target.}
    \label{fig:hd189_hotspot_hist}
\end{figure}

These plots show that overall, we recover a highly consistent {thermal structure} between JWST and Ariel for this test case. 
% {as evidenced by the similar contour profiles of the maps, lack of prominent structure in the difference map, and flatness of the differenced flux profile.
{However, the AIRS Ch1 spans a slightly narrower temperature range than the MIRI/LRS map, leading to a slightly shallower day-night temperature gradient. This is consistent with theoretical expectations: AIRS Ch1 spans a narrower bandwidth than MIRI/LRS, and therefore samples less of the planet's spectral energy distribution.}

In Figure \ref{fig:hd189_hotspot_hist}, we show the posteriors of our recovered longitudinal hotspot offsets between the observatories. The overlap between the posteriors, and the fact that the difference map in Figure \ref{fig:hd189_maps} shows a differenced hotspot offset consistent with zero, demonstrates that we accurately recover the injected hotspot offset with Ariel, and at only $\sim${2}$\times$ lower precision than JWST.

{The flux profiles above the maps are highly consistent in terms of both morphology and precision, with Ariel tightly constraining the flux and thus temperature distribution of the central longitudes in a manner comparable to JWST.
In total, Ariel achieves a median temperature precision of 20 K across the dayside, compared to the 10 K precision achieved with JWST.}

{The precision achieved on both the recovered hotspot and dayside temperature is encouraging, and can likely be attributed to the brightness of the HD 189733 system. This results in fast saturation times for JWST due to its large collecting area, necessitating a $<$1 second cadence. Conversely, Ariel's smaller collecting area allows it to expose for approximately twice as long, resulting in a comparable flux precision to JWST, but still with a short enough cadence to finely sample the dayside flux profile. This enables Ariel to rival JWST in terms of not only qualitative, but also quantitative mapping ability for brighter targets, using fewer observations than one would expect from a simple scaling of their collecting areas. This test case therefore gives promising indications of Ariel's potential to eclipse map bright targets at high precision.}

\subsubsection{HD 209458b}

HD 209458b is the planet with the second highest EMM for MIRI/LRS \citep{boone2023}. However, no eclipse observations have yet been conducted with JWST. We therefore use a GCM output of the planet \citep{hd209gcm} to inform our test case here (see Section \ref{sec:methods} for details). Similar to our previous test case, our JWST reference map here is a two-component (N2) model with no constrained latitudinal structure. To test what we would recover with Ariel, we post-process this GCM output into an equivalent Ariel AIRS Ch1 light curve {using the methods outlined in Section \ref{sec:sim_framework}}.

\begin{figure*}
    \centering
    \includegraphics[width=0.3275\linewidth]{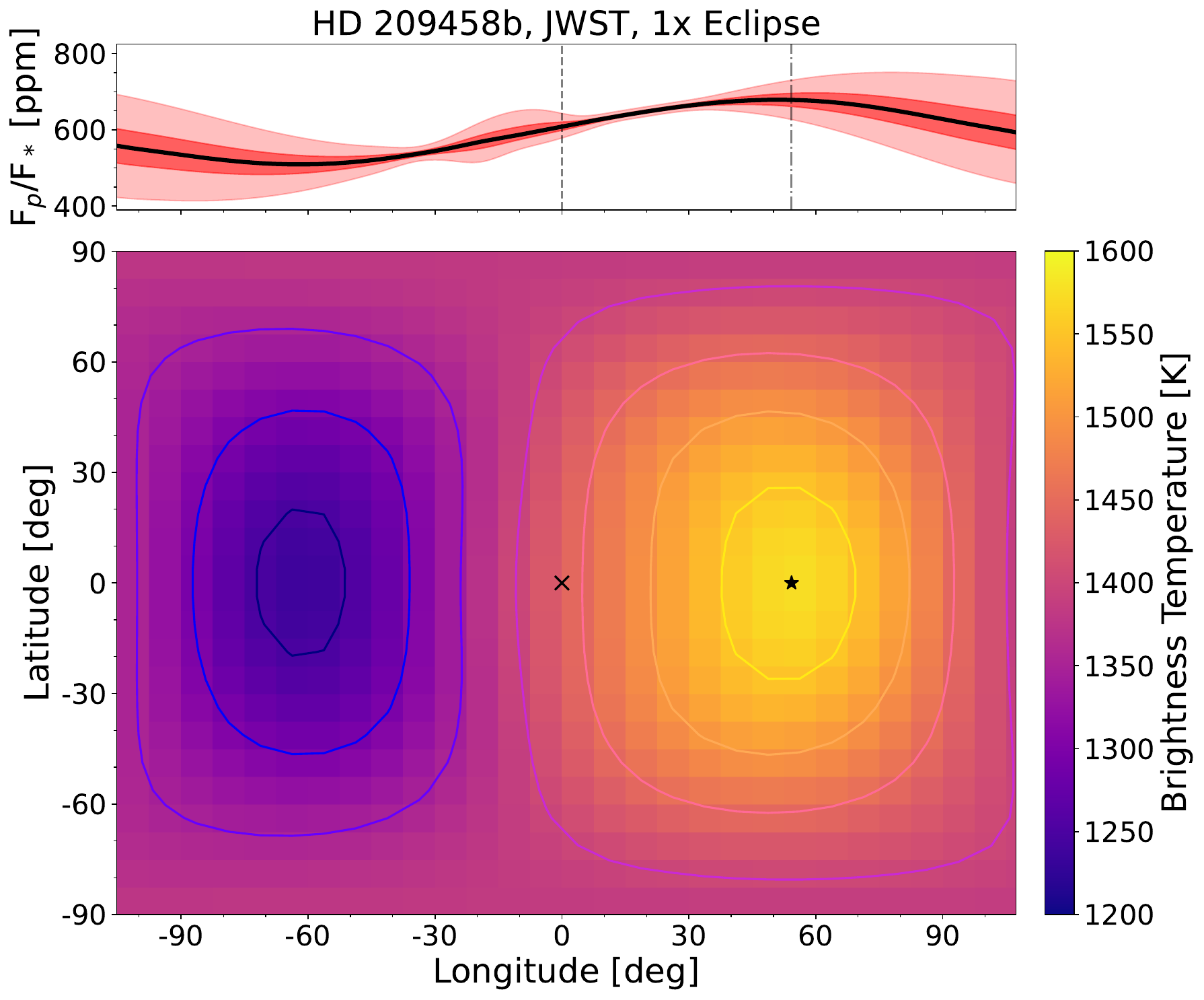}
    \includegraphics[width=0.3275\linewidth]{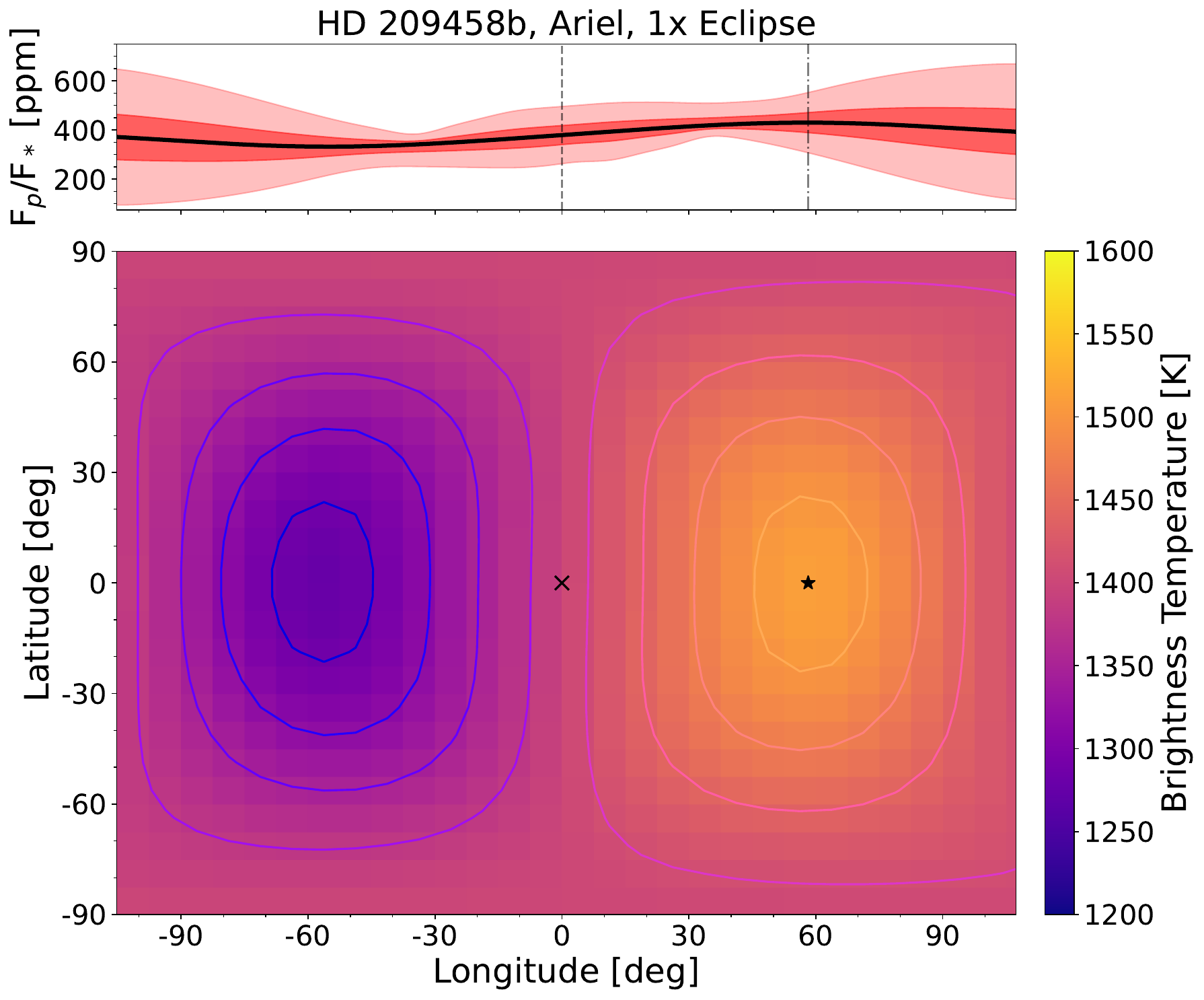}
    \includegraphics[width=0.335\linewidth]{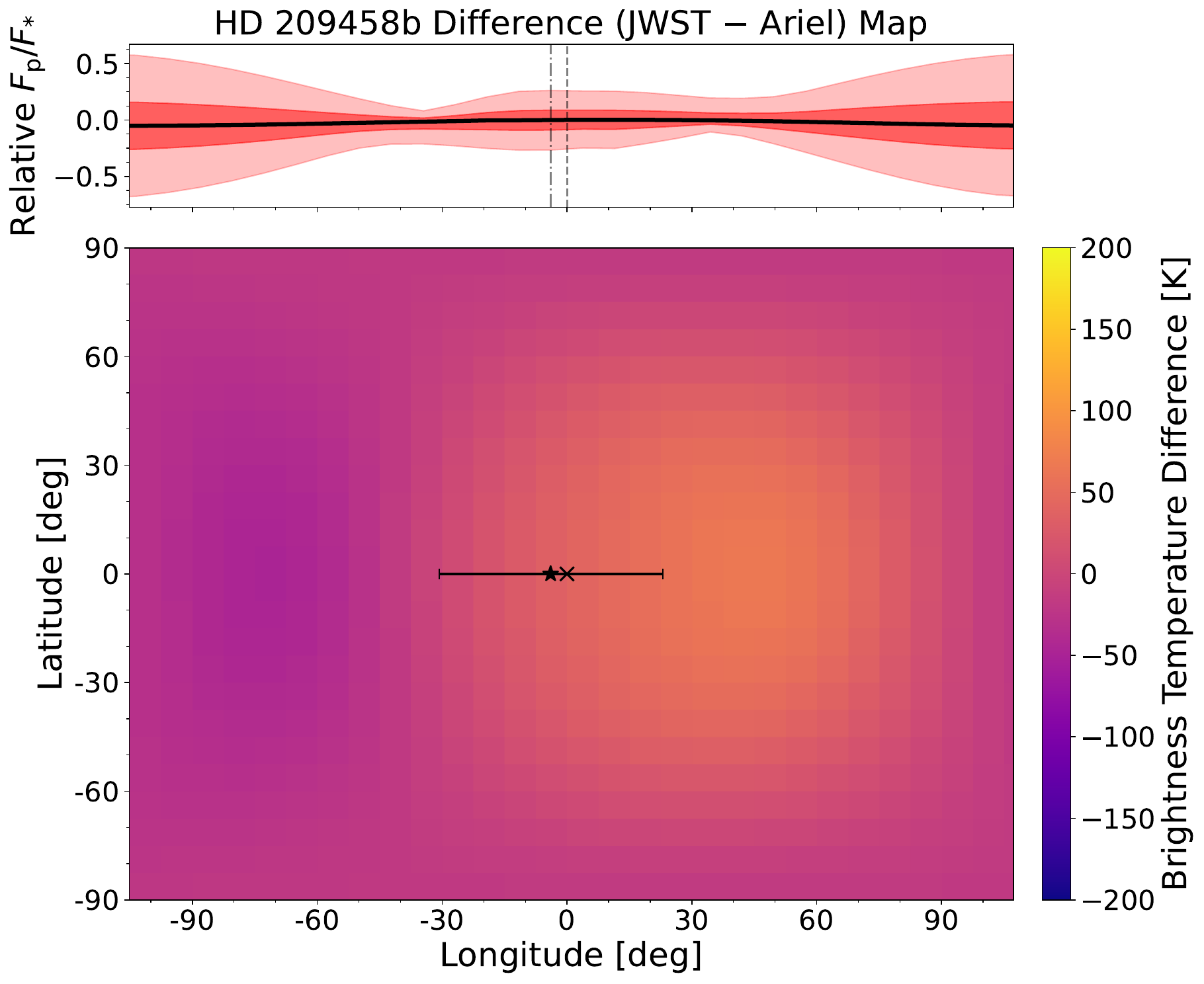}
    \caption{HD 209458b eclipse maps, {with flux profiles shown above. The cross and dashed line mark the location of the substellar point; the star and dash-dotted line mark the location of the hotspot}. \textit{Left:} Simulated JWST MIRI/LRS map, derived from a single eclipse observation post-processed from a GCM of the planet \citep[][see Section \ref{sec:methods} for details]{hd209gcm}. \textit{Middle:} Equivalent Ariel AIRS Ch1 map, also derived from a single eclipse observation. \textit{Right:} Difference map (JWST$-$Ariel). The mostly flat structure shows that we accurately recover the JWST input map with Ariel, but with an expected shallower day-night gradient due to the smaller Ariel instrument bandwidth. The overlap of the differenced hotspot with the substellar point shows that we recover this key parameter to better than 1$\sigma$.}
    \label{fig:hd209_maps}
\end{figure*}

\begin{figure}
    \centering
    \includegraphics[width=1\linewidth]{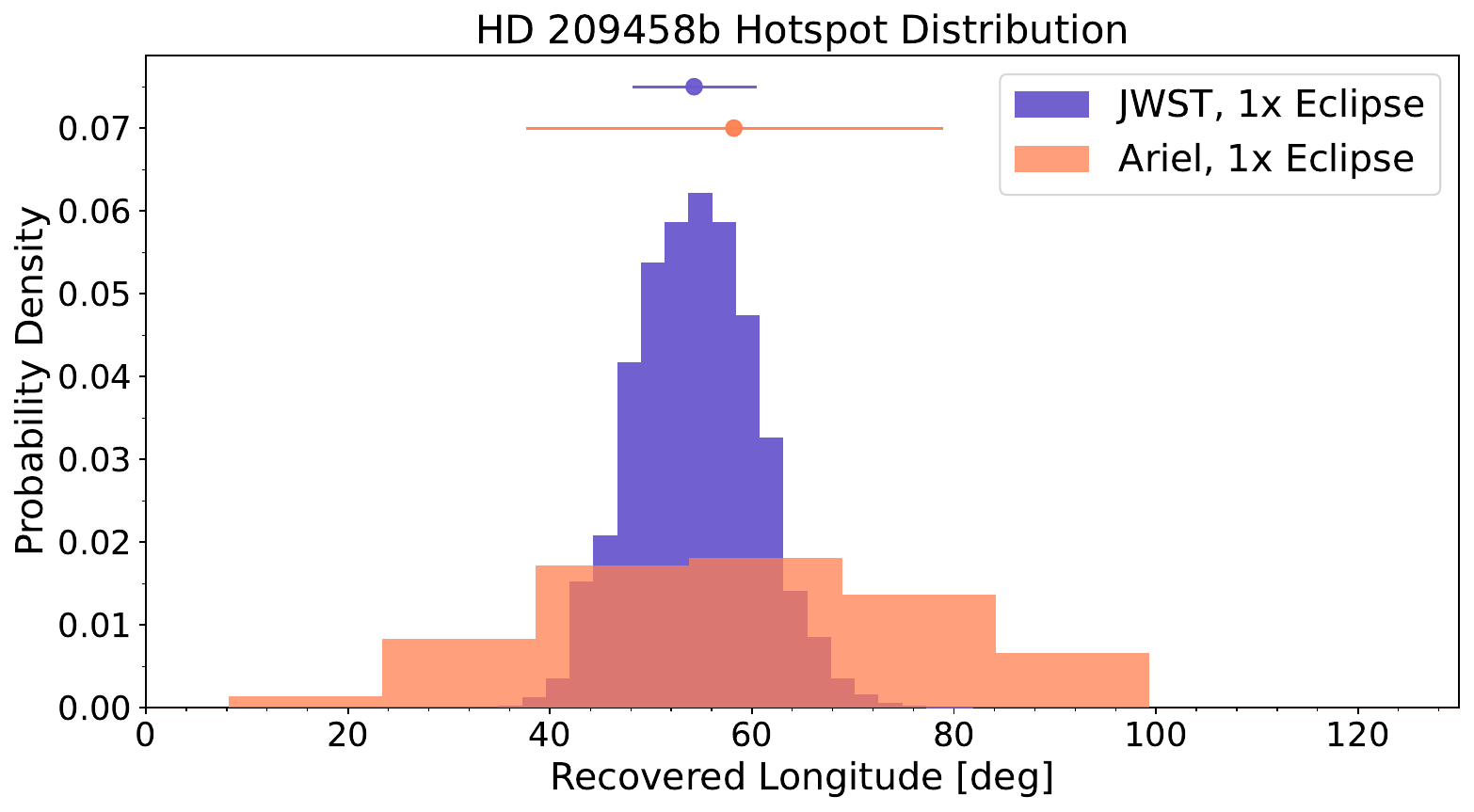}
    \caption{HD 209458b hotspot location posteriors. In blue is what we recover from the JWST MIRI/LRS map, and in orange what we recover from the Ariel AIRS Ch1 map. Using the same amount of data, and only a single eclipse observation at that, Ariel is able to recover a consistent hotspot offset as JWST with uncertainties only $\sim$3$\times$ as large for this bright target.}
    \label{fig:hd209_hotspot_hist}
\end{figure}

With this test case, we now benchmark Ariel's eclipe mapping abilities versus those of JWST using only a single eclipse observation. We plot the JWST, Ariel, and difference maps for HD 209458b in Figure \ref{fig:hd209_maps}, the longitudinal hotspot posteriors in Figure \ref{fig:hd209_hotspot_hist}, and tabulate these values in Table \ref{tab:jwst_vs_ariel}. We find that even this minimal amount of data is sufficient to recover the input mapping model. Ariel's ability to map high-ranking JWST mapping targets using only single eclipse observations, compared to the eight required by Spitzer to map even the highest-ranking target \citep{majeau2012, dewit2012}, is encouraging for its future mapping prospects.

We once again derive a highly consistent thermal structure, as evidenced by the similar contour profiles and mostly flat difference map in Figure \ref{fig:hd209_maps}. {The residual structure in the latter is of the same origin as was found for the previous test case: a comparable spatial temperature profile, but over a narrower range for AIRS Ch1 compared to MIRI/LRS due to the narrower bandpass. This lends further credence to this effect being systemic rather than systematic. The median dayside temperature is constrained at 50 K by Ariel in this case, compared to 20 K for JWST.}
The hotspot posterior plot shows that Ariel accurately recovers the hotspot offset at 1$\sigma$ agreement with the JWST value, and at only $\sim$3$\times$ lower precision.

{
The quality of these constraints can again be attributed to the brightness of the system inhibiting JWST's integrations times compared to Ariel's, allowing the latter to increase its flux precision whilst still maintaining good spatial scanning.
The relative precision on mapping parameters between JWST and Ariel is slightly lower here than in the previous test case because the HD 209458 system is relatively dimmer, meaning that even with this longer cadence, Ariel is not able to rival JWST's flux precision to the same degree. However, this $\sim$3$\times$ scaling still exceeds first-order expectations for how one might expect the much smaller Ariel telescope to compare against JWST. This test case therefore further supports our prior conclusion that Ariel shows promising potential to map bright targets.}

\subsubsection{{3D Mapping of HD 189733b and HD 209548b}}

Since Ariel simultaneously observes using all of its spectrographs, {giving it inherent 3D mapping abilities}, we also test its ability to map the atmospheres of these high-ranking mapping targets using its AIRS Ch0 and NIRSpec instruments. {We rescale the input flux maps to the values that would be observed in the bandwidth and throughput of these instruments using the methods of Section \ref{sec:sim_framework}, and retrieve on them with \texttt{ThERESA}. We find that we again recover the same input mapping model, with consistent thermal structure and 1$\sigma$ consistent hotspot offsets as we find with our AIRS Ch1 examples.}

Out of our JWST spectrographs of interest for eclipse mapping \citep[see Table \ref{tab:test_cases} and][]{boone2023}, these two canonical hot Jupiters can only be mapped in the mid-IR, using MIRI/LRS, as they are too bright and saturate in NIRISS/SOSS and NIRSpec/G395H\footnote{We note that HD 189733b only partially saturates in NIRSpec/G395H, but this is still using only two groups per integration.}. Ariel's lower collecting area means that it can observe these targets in the near-IR without saturating, allowing it map them across its entire wavelength range from $0.5-7.8 \ \upmu$m.
Hence, for these two canonical hot Jupiters, Ariel is the only way to access and map these pressures, which is imperative if we are to build a full 3D picture of their atmospheres. This further adds to the need for another eclipse mapping observatory in order to provide synergy to what is achievable with JWST, and the advantages of using Ariel to do so.

Hence, the primary conclusion that we reach from these test cases is that Ariel has, at minimum, baseline-level eclipse mapping capabilities. Its ability to directly compete with JWST using the same amount of data for these highest-ranking targets is a testament to the fact that JWST light curves are of a quality that far supersedes the threshold for eclipse mapping, at least for longitudinal signals of bright targets. In such cases, Ariel therefore does not need to directly match JWST light curve quality in order to observe the same mapping signals. Hence, for bright, high-ranking mapping targets, Ariel light curves will have the precision necessary to measure eclipse maps in the same observational time as JWST.

\subsection{WASP-18b}
\begin{figure*}
    \centering
    \includegraphics[width=0.3275\linewidth]{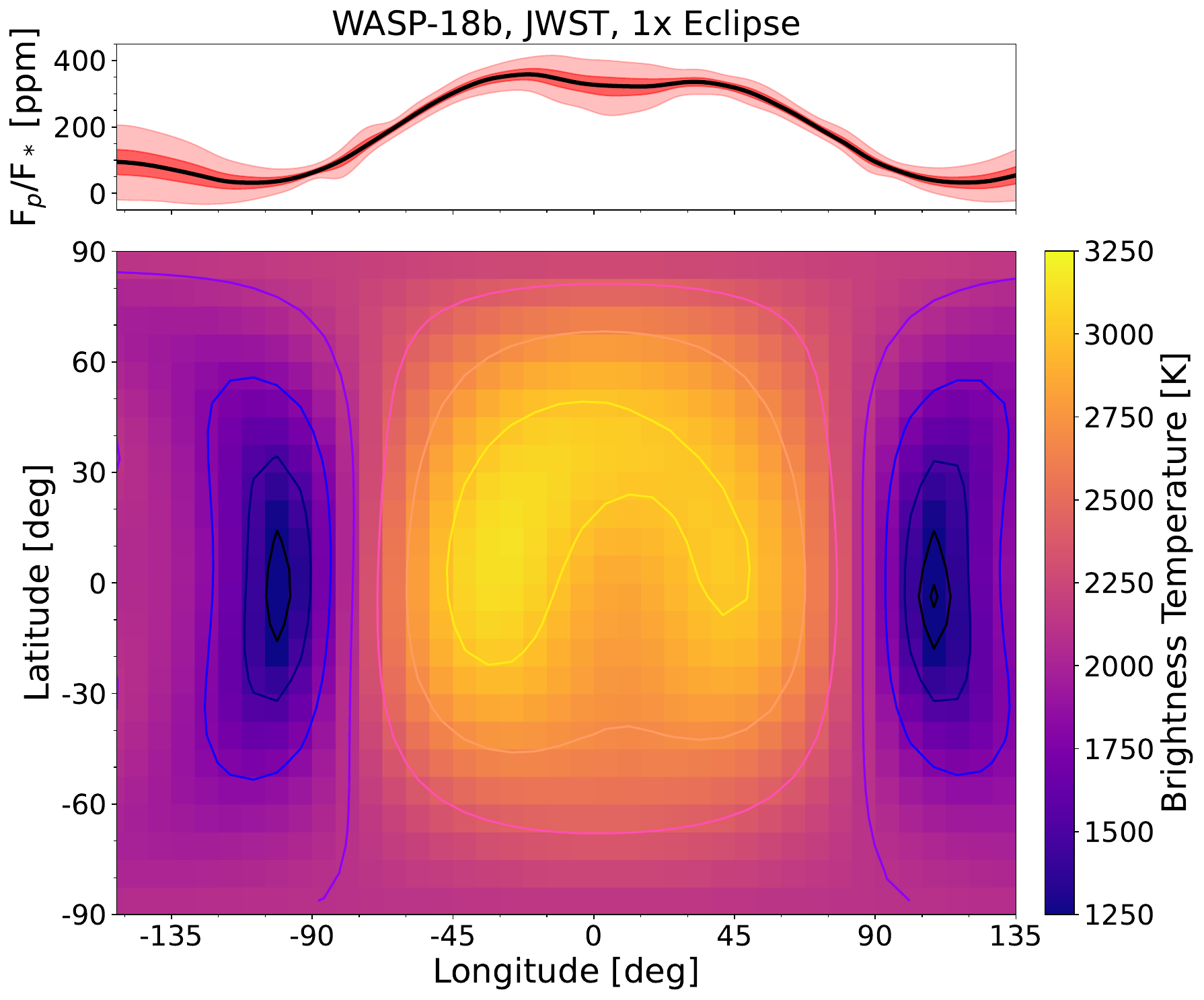}
    \includegraphics[width=0.3275\linewidth]{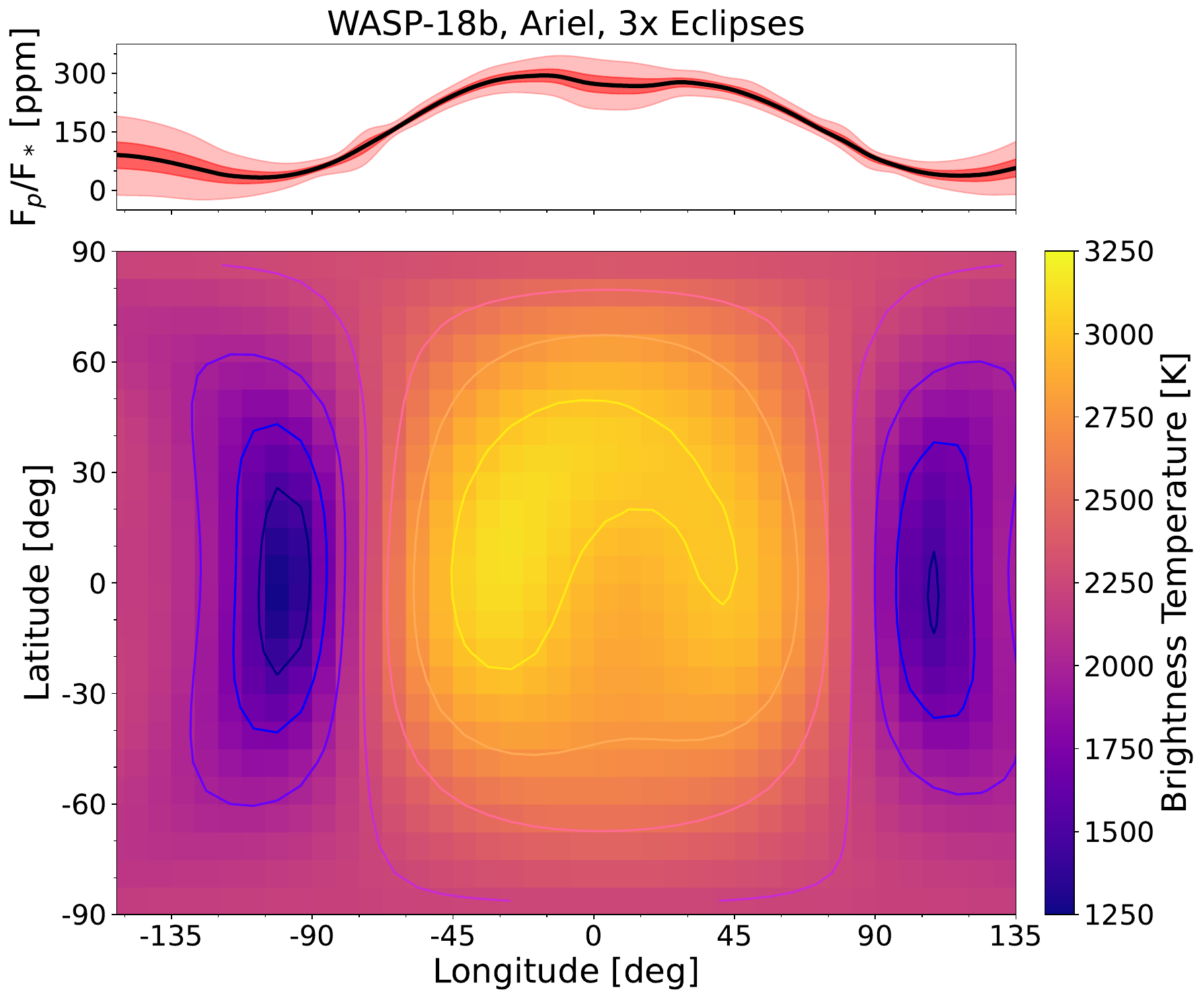}
    \includegraphics[width=0.335\linewidth]{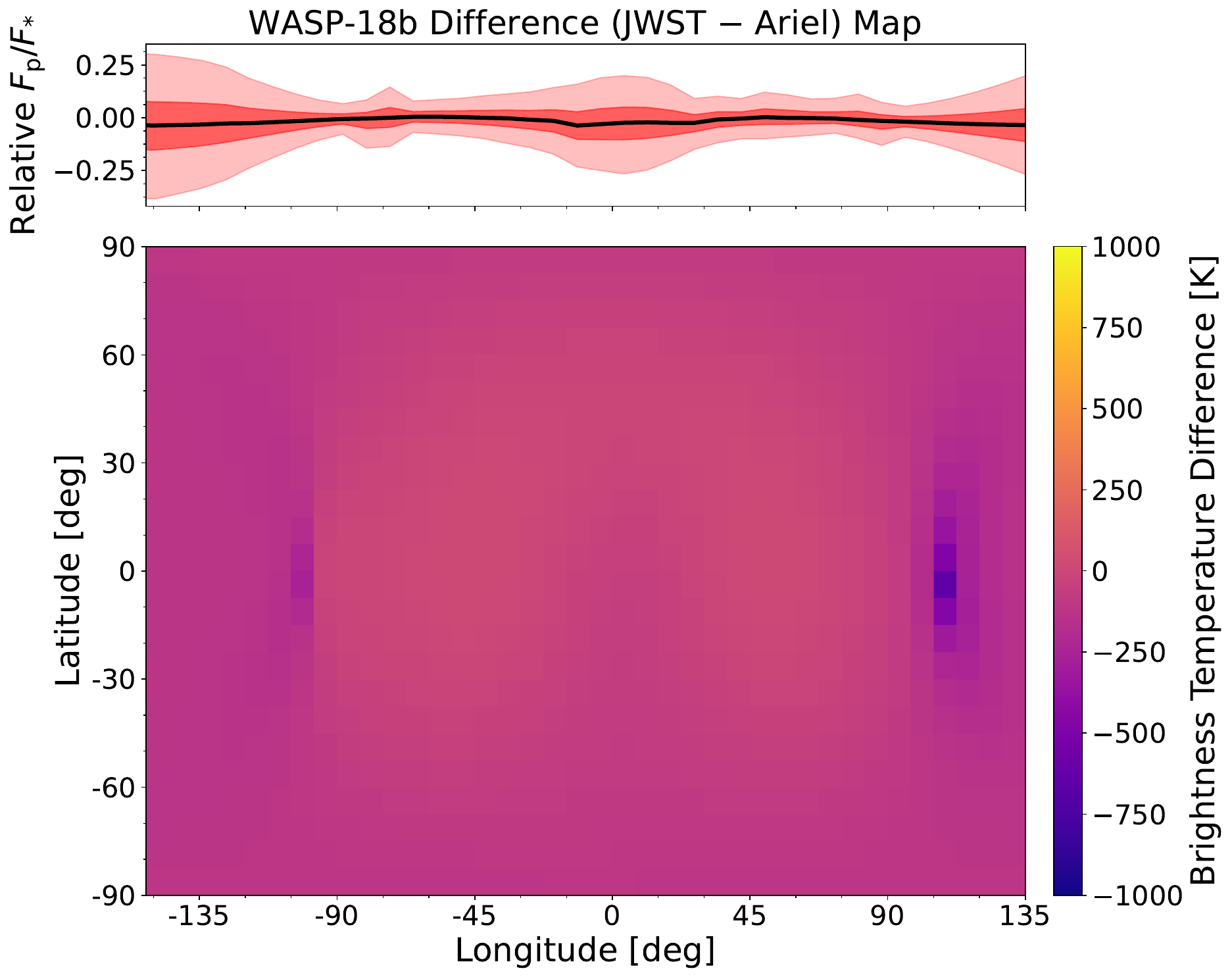}
    \caption{WASP-18b eclipse maps, {with flux profiles shown above.} \textit{Left:} Simulated version of the JWST NIRISS/SOSS map, derived from one eclipse observation \citep{wasp18eclipsemap}. \textit{Middle:} Equivalent Ariel NIRSpec map, derived from three eclipse observations. \textit{Right:} Difference map (JWST$-$Ariel). The flat structure of the latter shows that with Ariel, we recover the thermal structure of the input map with high accuracy. The artefacts beyond the limbs are of least concern since, as {the flux profiles show}, they contribute negligible signal to the overall map.}
    \label{fig:w18_maps}
\end{figure*}

\citet{boone2023} rank the top 15-to-16 eclipse mapping targets for JWST. WASP-18b is the target that achieves the lowest numeric score, placing 16th in the NIRISS/SOSS ranking. We showed in the previous subsection that our ``best of the best'' test cases are mappable using as little as single eclipse observations with Ariel. We now assess how well Ariel performs for our ``worst of the best'' test case. We take the input WASP-18b map, derived from a single JWST NIRISS/SOSS eclipse \citep{wasp18eclipsemap}, and post-process it into an equivalent Ariel NIRSpec light curve {using the methods outlined in Section \ref{sec:sim_framework}}.

The input map in this case is a higher-order L5N5 model\footnote{We note that an L2N5 model was only slightly less preferred and formally statistically indistinguishable \citep{wasp18eclipsemap}.}, which, as evident in the left panel of Figure \ref{fig:w18_maps},
contains a degree of complex, small-scale structure that we seek to replicate with Ariel. There is no hotspot offset to compare in this case because WASP-18b, as an ultra-hot Jupiter, has a hotspot localised to the substellar point, a consequence of the much shorter radiative timescales preventing efficient advection of heat, and/or magnetic field interactions with the ionised flow imparting a strong drag \citep{wasp18eclipsemap}. However, the input eclipse map does have a distinctive overturning-morphology around the substellar region, with an approximately double-peaked brightness plateau extending between $\sim\pm40^{\circ}$ longitude.

Given the lower EMM of this target, we find that in this case, Ariel requires more observations compared to JWST in order to recover the same mapping model. {This is further exacerbated by the fact that the WASP-18 system is dimmer than our previous test cases, which results in long native saturation times for Ariel, and therefore a cadence that is too long for adequate spatial scanning of the dayside atmosphere. We therefore rescaled the cadence per the methods of Section \ref{sec:sim_framework} in order to maximise the spatial information content of the light curve, which necessarily leads to a reduced flux precision.}

Despite this, we find that as few as three eclipse observations are sufficient in this case to accurately recover this high-order L5N5 model, as evidenced by the consistent contour profiles between the maps, and flat difference map in Figure \ref{fig:w18_maps}. The most notable differences are slight artefacts beyond the limbs, which {the flux profiles show} contribute negligible signal to the overall map and therefore are the most unconstrained.

{The Ariel NIRSpec and JWST NIRISS/SOSS instruments have the most comparable wavelength coverage of our designated Ariel-JWST pairs. Consequently, we recover a highly consistent thermal structure between the maps, with near-identical temperature ranges and thermal gradients. In terms of precision, we find that with 3$\times$ as many observations, Ariel is able to constrain the median dayside temperature to the same precision as JWST, at 40 K in both cases (see Table \ref{tab:jwst_vs_ariel}). The 2D profile is similarly well constrained, as evidenced by the similar magnitude flux profile uncertainties.}

Whilst three eclipse observations may be difficult to request on the oversubscribed JWST, it is more than feasible for a dedicated exoplanet mission like Ariel, which can dedicate up to twenty observations for high-ranking targets in its census \citep{ariel_target_list}. This
lends further credence to our prior conclusion: that JWST light curves far supersede the minimal quality required for eclipse mapping, therefore directly matching its data quality is not required in order to replicate its mapping capabilities. {This validates Ariel as an observatory capable of deriving both qualitatively and quantitatively similar multidimensional mapping results as JWST, even for dimmer targets.}

\subsubsection{{Time Sampling Test}}

{Here, we test whether our simplifying assumption of fitting a single noise-scaled light curve yields the same results as jointly fitting separate light curves with slightly different time sampling, as would be obtained in reality. The time sampling of an observation is critical for eclipse mapping since it sets the spatial scanning of the atmosphere. Using our optimised cadence calculation detailed in Section \ref{sec:sim_framework}, we find that for WASP-18b, cadences below 47 s maximise the spatial information content of the light curve. We therefore test shifting the additional observations by 10 s in either direction in order to provide a non-integer time sampling offset between the observations. From jointly fitting these light curves, we find that we are able to derive the same qualitative map with identical quantitative constraints as following $\sqrt{N}$ statistics.}

{We attribute this to our use of the optimised cadence when designing the Ariel light curves, meaning that no matter the placement of the data points, they each sample the spatial profile of the atmosphere at the maximum achievable resolution. This demonstrates that if the cadence of observations is optimised, then specific phase placements of the data points is not necessary to combine datasets and increase the signal-to-noise (SNR). By designing the observations in this way, Ariel will therefore have the ability to combine datasets in order to map lower-ranking mapping targets.}\\

The fact that Ariel can compete with JWST for both the ``best of the best'' and the ``worst of the best'' targets in the ranking of \citet{boone2023}, using a reasonable amount of data, indicates that Ariel will have the capability of mapping all of these highest-ranking targets, of which there are 22 unique ones. This gives promising indications of Ariel's potential to expand the parameter space of eclipse mappable datasets and facilitate population-level mapping studies at a faster rate than what is feasible with JWST.

\subsection{WASP-17b}
\label{sec:ariel_sims_w17}
\begin{figure*}
    \centering
    \includegraphics[width=0.3275\linewidth]{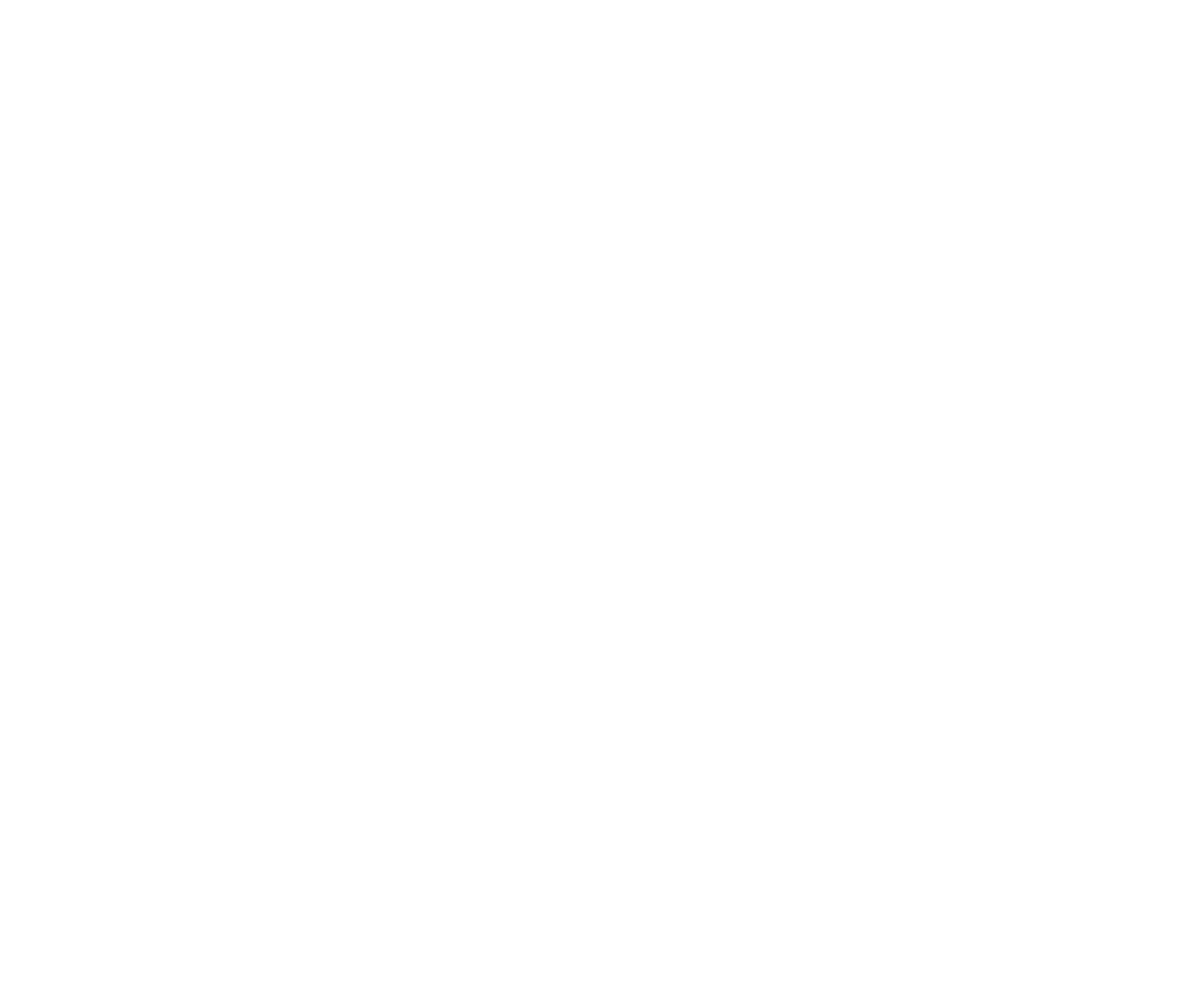}
    \includegraphics[width=0.3275\linewidth]{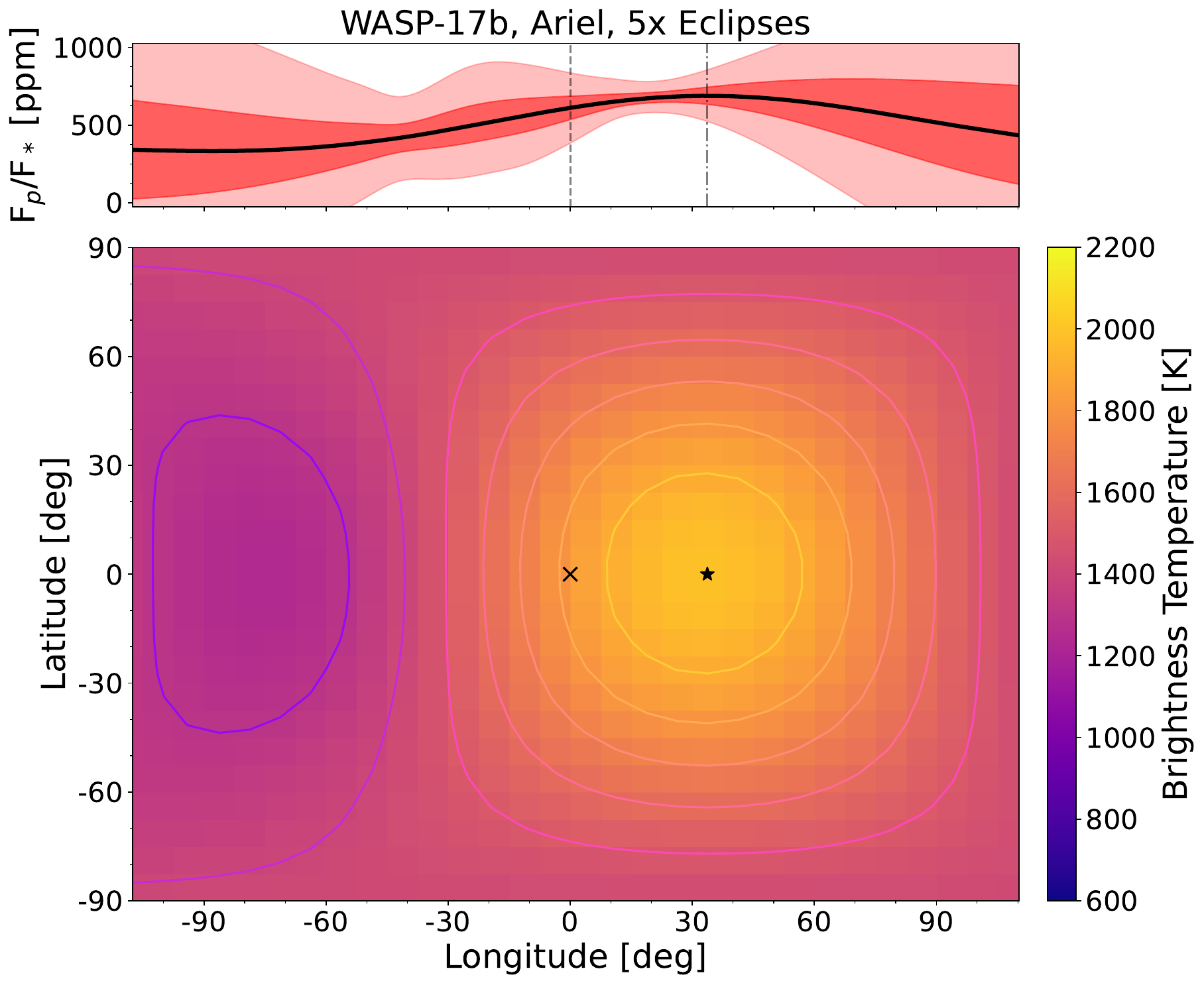}
    \includegraphics[width=0.335\linewidth]{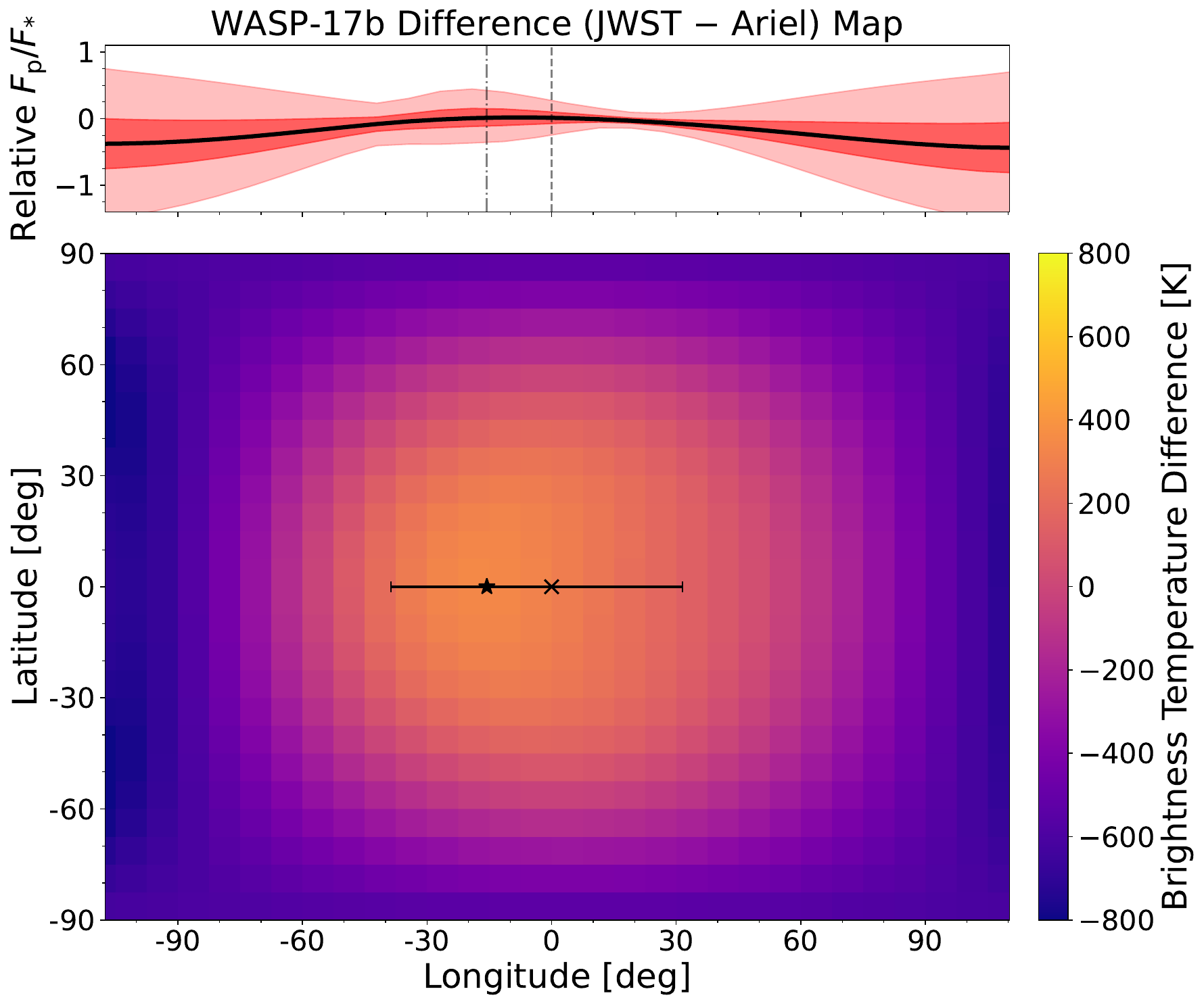}
    \includegraphics[width=0.3275\linewidth]{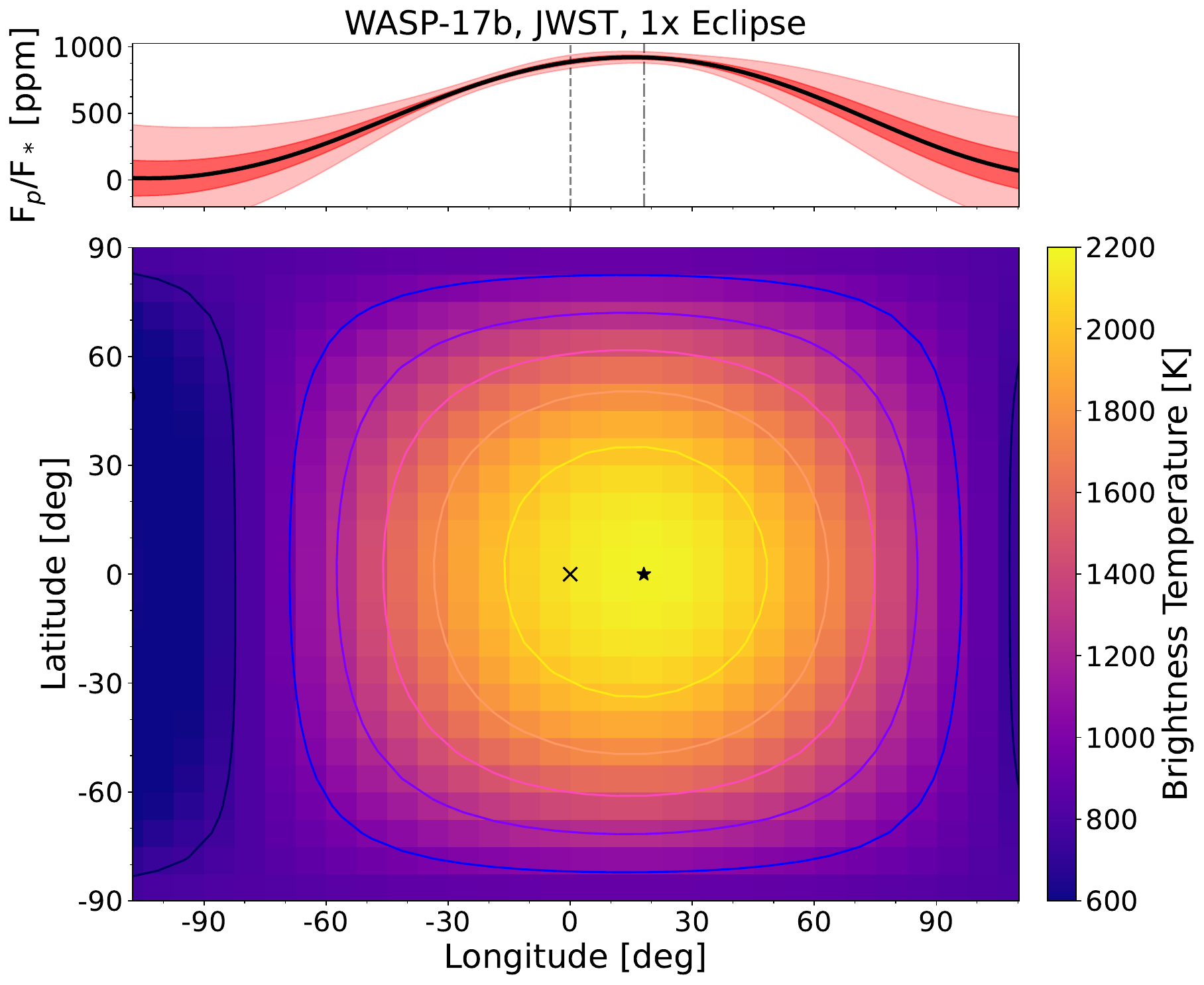}
    \includegraphics[width=0.3275\linewidth]{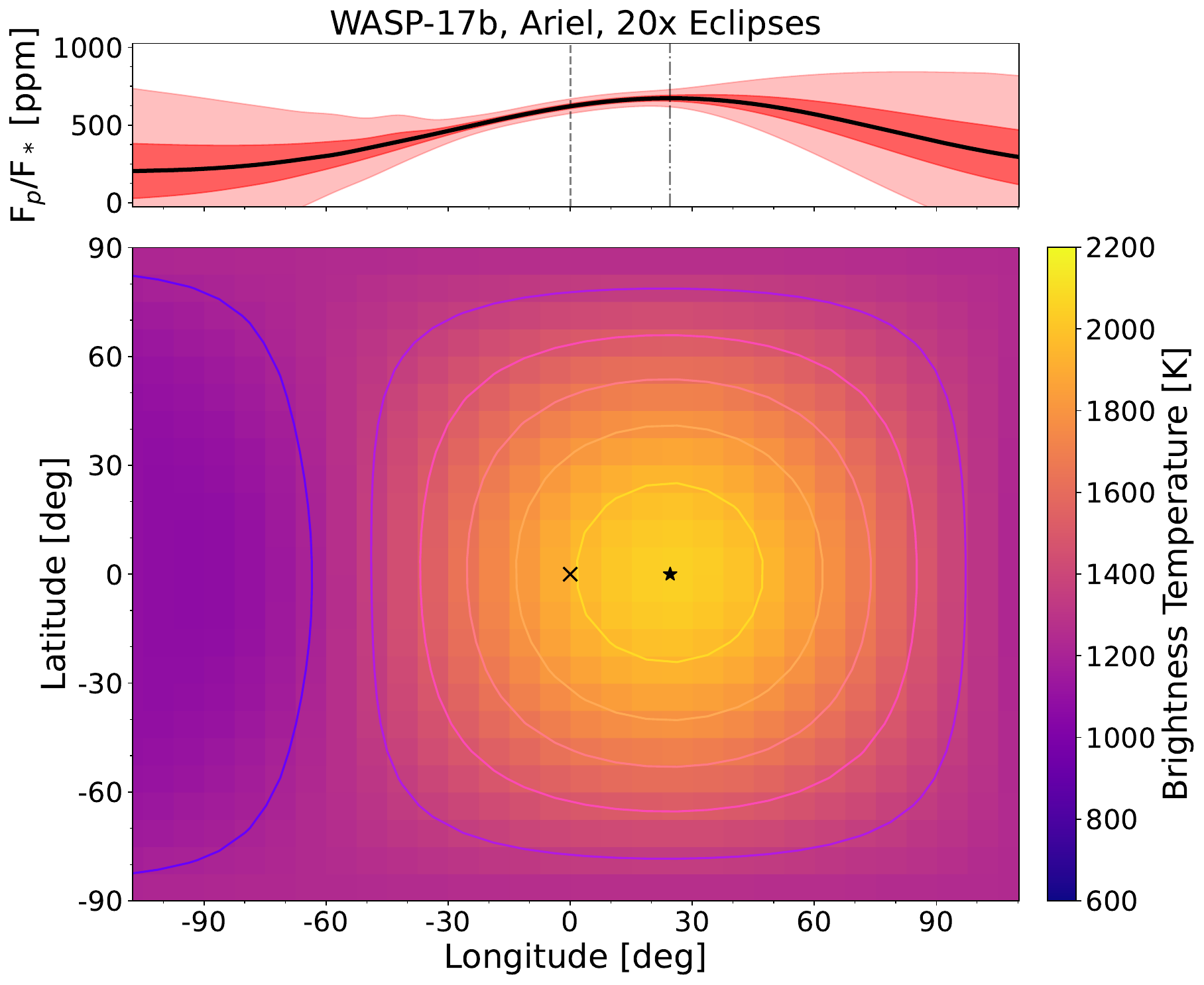}
    \includegraphics[width=0.335\linewidth]{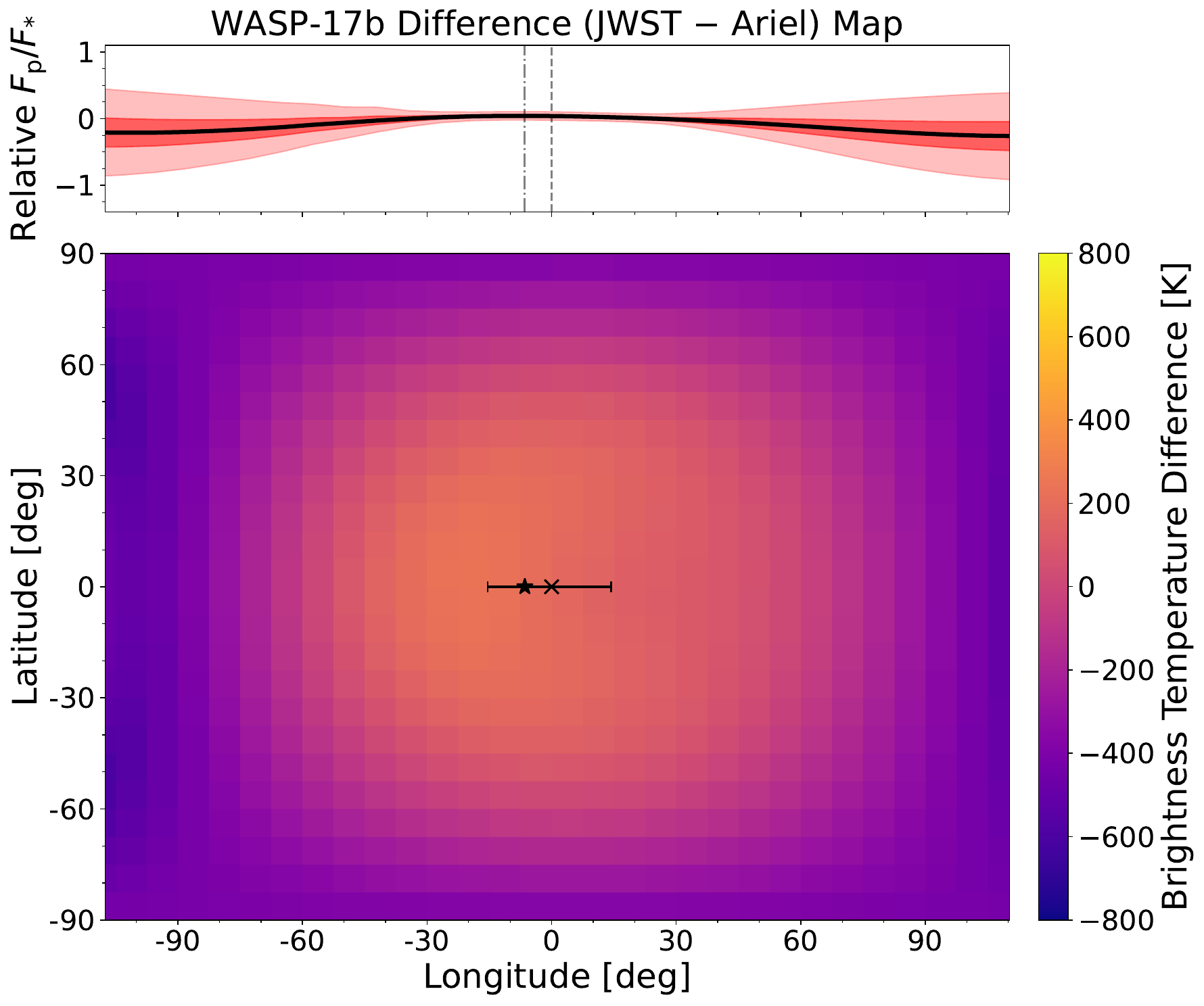}
    \includegraphics[width=0.3275\linewidth]{maps_w_flux/w17/filler.pdf}
    \includegraphics[width=0.3275\linewidth]{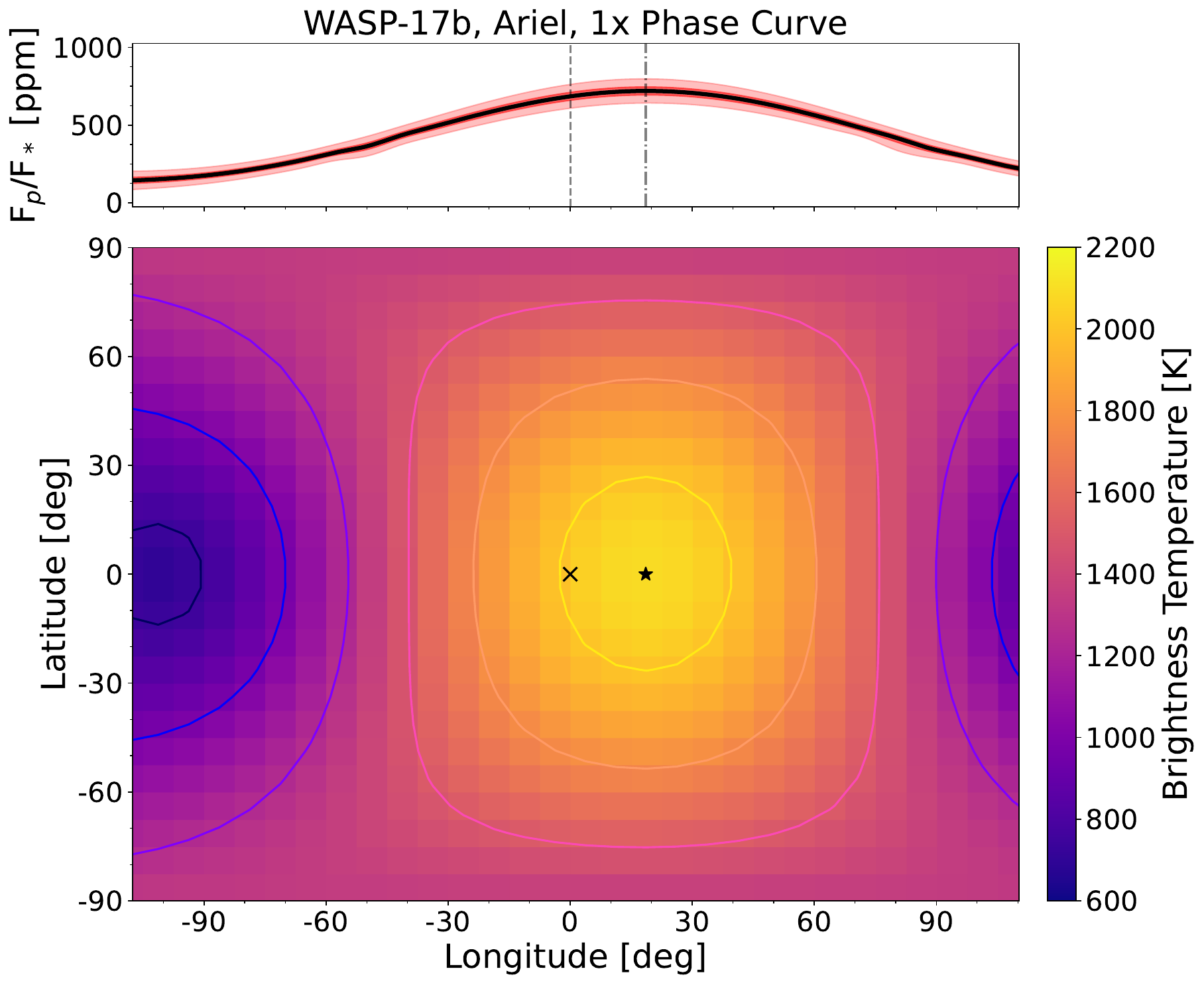}
    \includegraphics[width=0.335\linewidth]{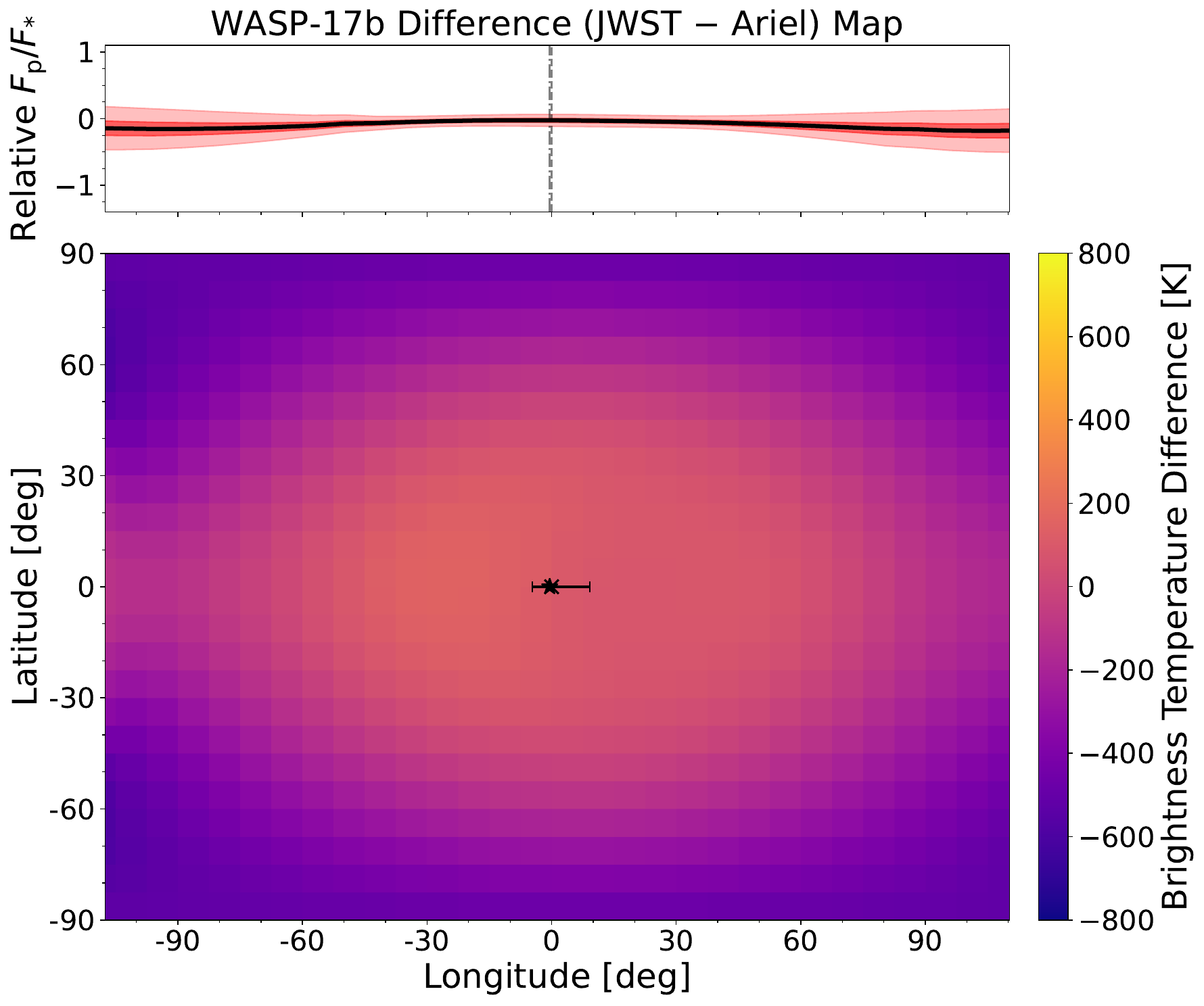}
    \caption{WASP-17b eclipse maps, {with flux profiles shown above. The cross and dashed line mark the location of the substellar point; the star and dash-dotted line mark the location of the hotspot}. \textit{Left:} Simulated version of the JWST MIRI/LRS map, derived from one eclipse observation \citep{wasp17eclipsemap}. \textit{Middle:} Equivalent Ariel AIRS Ch1 maps, derived from (top) five eclipse observations ($\sim$50 hrs), (middle) twenty eclipse observation ($\sim$ 200 hrs), and (bottom) one phase curve ($\sim$ 100 hrs). \textit{Right:} Difference maps (JWST$-$Ariel). Moving down the rows, we better recover the input thermal structure, and improve both the precision and accuracy of the recovered parameters (temperature profile and hotspot offset) with Ariel. The use of a full or at least partial phase curve is the most time-efficient way for Ariel to recover the input map for this dimmer, lower-ranking mapping target.}
    \label{fig:w17_maps}
\end{figure*}

\begin{figure}
    \centering
    \includegraphics[width=1\linewidth]{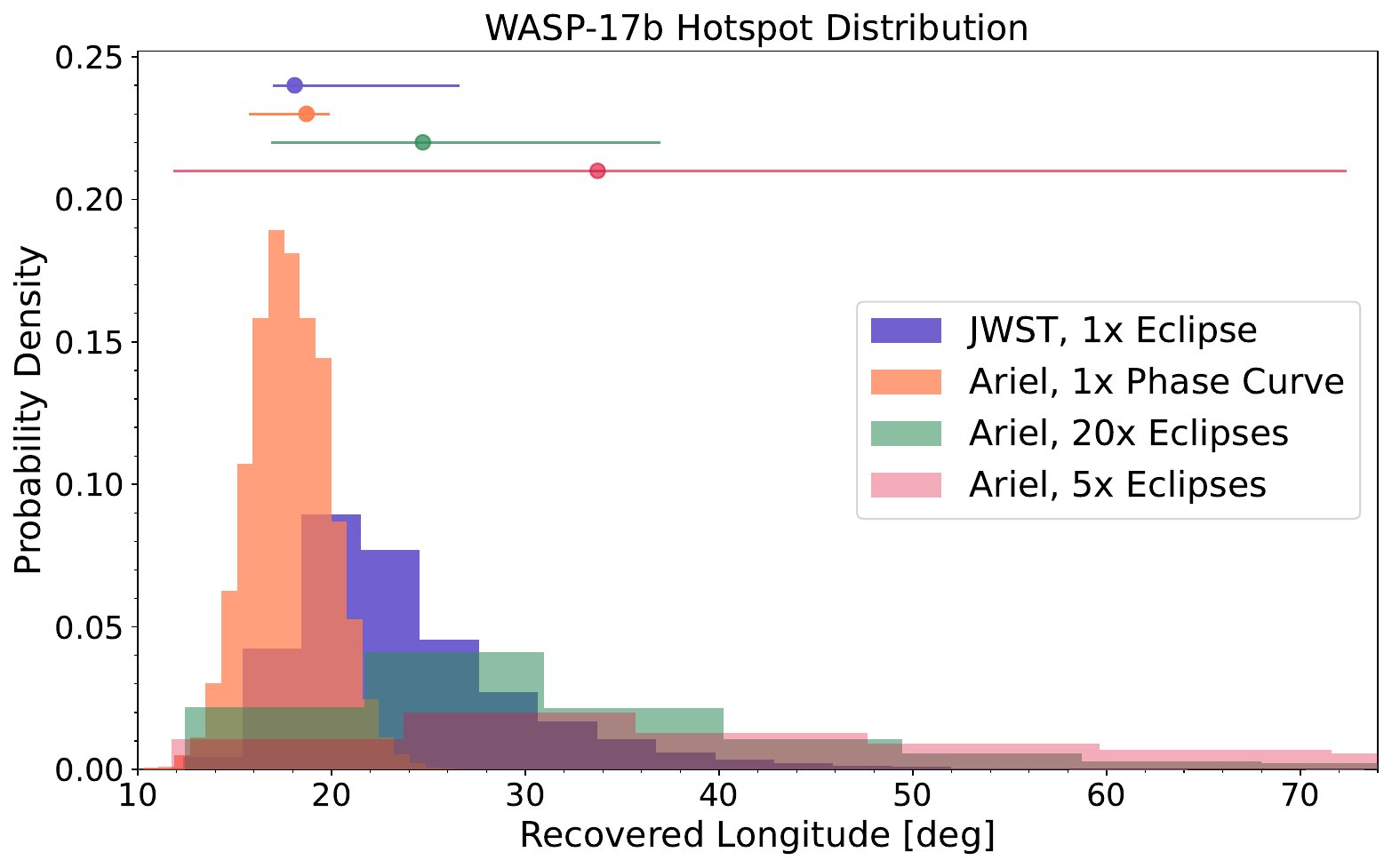}
    \caption{WASP-17b hotspot location posteriors. In blue is what we recover from the JWST MIRI/LRS map, and in red, green, and orange what we recover from the Ariel AIRS Ch1 maps derived from five eclipses, twenty eclipses, and one phase curve, respectively. Using eclipses only, we are unable to match the precision nor accuracy of the JWST map. Using a full phase curve, we recover the input hotspot offset with both high accuracy and precision.}
    \label{fig:w17_hotspot_hist}
\end{figure}

Despite being successfully eclipse mapped by JWST MIRI/LRS using only a single eclipse observation \citep{wasp17eclipsemap}, WASP-17b does not rank amongst the top 15 targets for any of our JWST spectrographs of interest \citep{boone2023}. Given that Ariel has shown the potential to map all of these high-ranking targets, we therefore elected to expand the analysis of \citet{boone2023} in order to identify where WASP-17b falls (see Section \ref{sec:ariel_jwst_comp} for further details). In the appendix, we tabulate the rankings of the 100 best eclipse mapping targets for JWST NIRISS/SOSS (Table \ref{tab:emm_list_soss}), NIRSpec/G395H (Table \ref{tab:emm_list_g395h}), and MIRI/LRS (Table \ref{tab:emm_list_miri}) that we derived in this way.

From this expanded analysis, we find that WASP-17b ranks as approximately the 100th best eclipse mapping target for JWST MIRI/LRS. Whilst it scores much higher in terms of its longitudinal metric (34th), a testament to the ability of \citet{wasp17eclipsemap} to recover the longitudinal hotspot shift using only a single eclipse observation, its overall EMM is brought down by its poorer latitudinal ranking (167th), a consequence of WASP-17b's low impact parameter of 0.35. This explains why no latitudinal structure was recovered in the JWST eclipse map, with a lower-order two-component (N2) model being preferred. This test case is therefore used to push Ariel to its limits by testing its ability to compete with JWST for a {much dimmer and} lower-ranking mapping target. Below we detail four cases: (1) using five eclipses, (2) using 20 eclipses, (3) an extended baseline method, and (4) a full orbital phase curve.

\subsubsection{Five Eclipses Case}
We take the L2N2 MIRI/LRS map of WASP-17b from \citet{wasp17eclipsemap}, post-process it into an AIRS Ch1 light curve {using the methods outlined in Section \ref{sec:sim_framework}}, and continue to scale the \texttt{ArielRad}-derived flux precision by $\sqrt{N}$ the number of observations until we recover the same model with Ariel. {As a dimmer target, the native cadence of the observation was too long to finely sample the dayside flux profile. We therefore used the methods outlined in Section \ref{sec:sim_framework} to rescale the cadence to the maximal value that ensures adequate spatial scanning, and degraded the flux precision accordingly.}

{Despite this reduced flux precision,} we find that, even for this {dimmer}, lower-ranking target, the same mapping model is recoverable at statistical significance \citep[$\Delta \mathrm{BIC}$ $\gtrsim$10, ][]{Raftery1995BIC} using as few as five eclipse observations; this map is shown in the top row of Figure \ref{fig:w17_maps}. However, unlike the previous test cases, where the same atmospheric structure was recovered {with similar constraints as the JWST map}, this five-eclipse Ariel map fails to adequately constrain either the hotspot offset {or temperature profile}.

{Whilst the median dayside temperature is recovered accurately, the contour profile in Figure \ref{fig:w17_maps} shows that the 2D structure is inaccurate, and the overhead flux profile shows that it is largely unconstrained, with a median dayside uncertainty of 180 K.} This leads to a discrepant and poorly constrained hotspot offset of $33.7^{+38.7 \circ}_{-21.9}$, the posterior of which is plotted in Figure \ref{fig:w17_hotspot_hist}. The Ariel map also exhibits a much shallower day-night temperature gradient {closer to 500 K than the steep 1000 K gradient of the JWST map. Whilst our previous test cases showed that we expect to recover a slightly shallower thermal profile with AIRS Ch1 than we do with MIRI/LRS, this level of discrepancy is too large to be explained by the different instrument bandwidths, and can rather be attributed to poor signal recovery}.

{Thus, despite Ariel recovering the correct mapping model, the map itself does not agree either qualitatively or quantitatively with the JWST map. This is converse to the WASP-18b case, where the threshold for signal identification also produced a map with similar quantitative constraints. This is because that model was much higher order, therefore it necessitated precise constraints in order to correctly resolve the small-scale signals. The lower order WASP-17b map, on the other hand, means that whilst Ariel is able to recover these higher-amplitude signals using only 5$\times$ as many observations, this does not necessarily mean that it is able to adequately constrain them.}

\subsubsection{Twenty Eclipses Case}
\label{sec:20_eclipse_case}
Due to the $\sqrt{N}$ scaling of photon noise, the greatest increase in light curve SNR comes from the stacking of the first few observations: many more are needed beyond this to increase the SNR enough to make substantial improvements. In this case, even by increasing the number of eclipse observations to twenty, at the upper limit even for the highest-ranking Ariel targets \citep{ariel_target_list}, Ariel is still unable to satisfactorily replicate the input map.

The hotspot offset, as evidenced by the posteriors in Figure \ref{fig:w17_hotspot_hist}, is better constrained in this case, with a measured value of $24.6^{+12.3 \circ}_{-7.8}$, but not to a satisfactory enough degree to warrant using 20$\times$ as many observations as is required with JWST. Similarly, the {temperature profile (middle row of Figure \ref{fig:w17_maps}) is better constrained to a median dayside precision of 100 K, and yields a more consistent contour profile with a better constrained flux profile, but the day-night gradient is still not satisfactorily reproduced, as evidenced by the residual structure of the difference map.}
% 750 K.

We therefore conclude that Ariel is unable to adequately derive the input map for this lower-ranking test case using a feasible number of repeated eclipse observations alone.

\subsubsection{Extended Baseline Mapping}
\label{sec:extended_baseline_mapping}
Instead of stacking more eclipses, which beyond $N$=20 would contribute less than 0.5\% improvement to the SNR of the light curve with each additional observation, we here test whether changes to the observing style may improve the constraints of the Ariel map.

Phase mapping uses the shape of the phase curve to measure the hotspot offset and thermal profile, whilst eclipse mapping uses the shape of ingress/egress. These techniques therefore provide two distinct and independent measurements of these parameters. By extending the pre-eclipse baseline of the light curve to measure the peak of the phase curve, we can perform both of these measurements simultaneously via joint phase and eclipse mapping. This allows us to use the constraints of the former to anchor the constraints of the latter and alleviate signal-versus-noise degeneracies. This strategy is being utilised in JWST GO 5687 \citep{5687_proposal}, and has shown successful application by combining a Spitzer partial phase curve with JWST eclipses of HD 189733b \citep{lally2025}.

{Whilst we only plot the regions of the planet visible during the observation, our models naturally extend over the entire planet. Whilst not formally valid over the unobserved regions because they are not data-driven, the nightside flux of these models are driven by the day-to-limb contrast of the eclipse map, and gives a first-order approximation of what we would expect to measure for the nightside. Hence, these models can be used to approximate full phase-curve observations, at least in the regime of simulations for an in-development observatory.}

By extending our simulations to a full phase curve observation in this way, we find that the peak of the phase curve for WASP-17b using the \citet{wasp17eclipsemap} input map falls approximately six hours before the start of eclipse. The JWST pre-eclipse baseline of three hours, which we also adopt for our Ariel simulated light curves, is therefore too short to measure this peak.

We first test extending the pre-eclipse baseline by 3$\times$, to nine hours, in order to capture the location of this peak with a few hours of baseline either side to characterise its morphology.
We find that we are now able to recover the mapping signal using only one observation, with a measured hotspot offset of $32.1^{+17.2 \circ}_{-12.4}$. Whilst the accuracy is still lacking, the precision in this case is $\sim$2$\times$ better than the five standard baseline eclipse case that was required before to derive this signal previously, and achieved in one third of the time (15 vs 45 hours, respectively). {The same is true for the derived thermal profile, which is now much more consistent with the steep day-night gradient of the input map at order 1000 K, and constrained at a more comparable precision (75 K median dayside precision for Ariel compared to 60 K for JWST)}.

\begin{figure}
    \centering
    \includegraphics[width=1\linewidth]{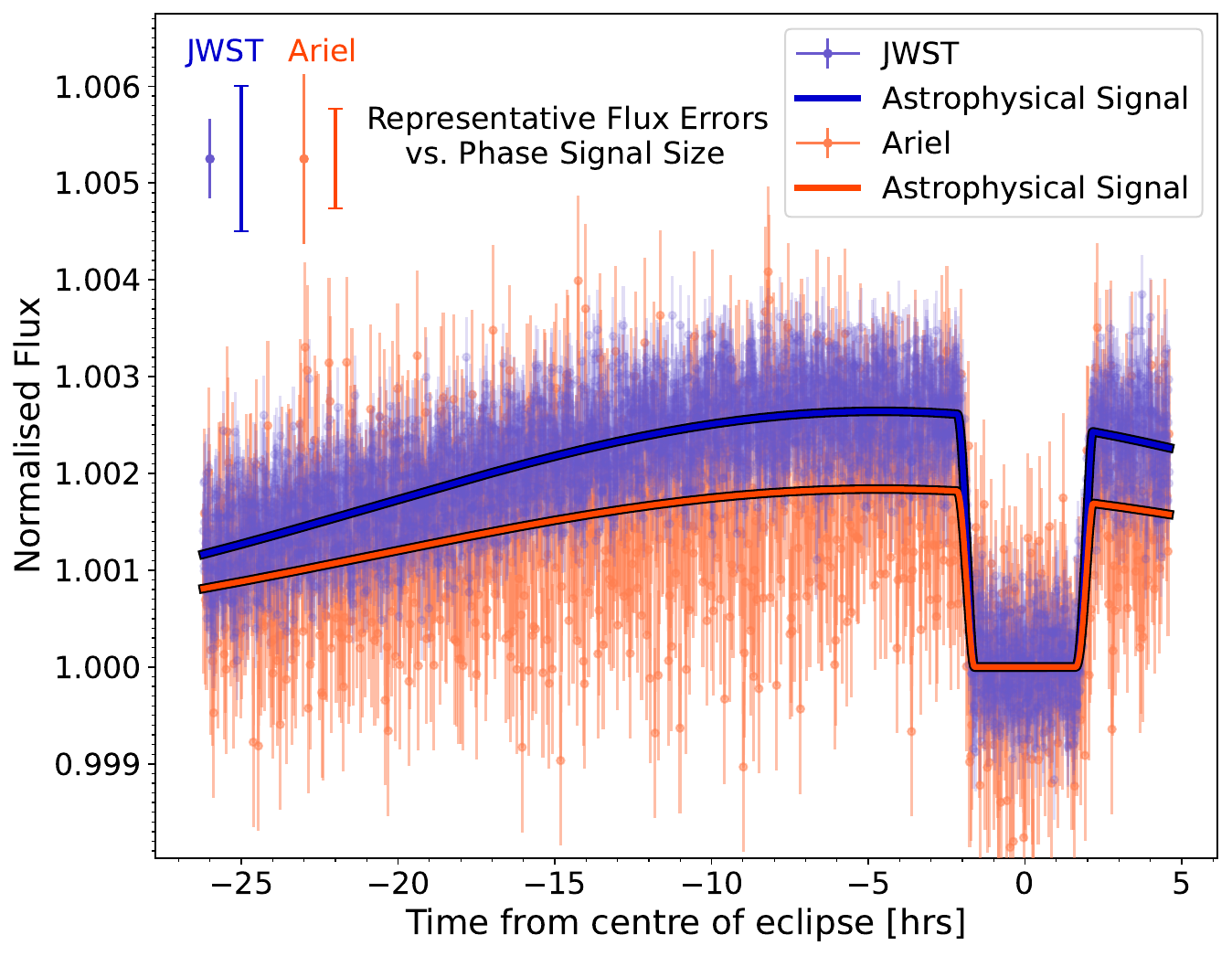}
    \caption{Partial phase curve of WASP-17b compared to the phase signal size, as observed using JWST MIRI/LRS compared to Ariel AIRS Ch1.}
    \label{fig:partial_phase_curve}
\end{figure}

By further extending the pre-eclipse baseline, we find further improvement in the accuracy and precision of {both} the hotspot offset {and the thermal profile}. With a 6$\times$ pre-eclipse baseline extension (18 hours, 24 total), we recover a hotspot offset of $22.6^{+2.6 \circ}_{-5.2}$ {and $T_{\mathrm{day-night}}\sim$1000 K with 50 K median dayside precision}. With an 8$\times$ extension (24 hours, 30 total), we recover a hotspot offset of $18.5^{+4.0 \circ}_{-3.9}$ {and $T_{\mathrm{day-night}}\sim$1000 K with 40 K median dayside precision}. The latter is the example plotted in Figure \ref{fig:partial_phase_curve}. With these baseline extensions, we now recover these key mapping parameters at even better precision than the standard baseline JWST eclipse ($\sim$10 hours), using only up to $\sim$3$\times$ as much observing time. 

The amount of pre-eclipse baseline extension required is longer for Ariel than what is expected for JWST to successfully employ this strategy, with the GO 5687 program finding from their simulations that only one to two additional hours prior to the phase curve peak were necessary in order to successfully characterise it.
This reason for this is evidenced in Figure \ref{fig:partial_phase_curve}, which shows that the phase signal is much more easily identifiable close to eclipse for JWST case than it is in for Ariel. {This is partially a result of the larger amplitude of the phase signal due to the longer wavelengths and wider bandwidth of MIRI/LRS compared to AIRS Ch1, but primarily due to the higher native flux precision of JWST for dimmer targets.}

Despite the longer baseline extension required, the time-efficiency gained compared to using repeated standard baseline eclipses still holds true. Even the minimal 5-eclipse case would require $\sim$50 hours of observing time compared to $\sim$30 hours for the scenario plotted in Figure \ref{fig:partial_phase_curve}. Hence, the most time-efficient way for Ariel to map a large population of planets, which include dimmer, lower-ranking mapping targets, is through the incorporation of phase information. The current Ariel strategy can dedicate up to twenty observations for high-ranking targets, and five for low-ranking targets \citep{ariel_target_list}. By comparing the observing times outlined above, we thus find that these extended-baseline observations are feasible.

However, there is no ``one strategy fits all'' way of facilitating this observing style. The position of the peak of the phase curve will vary from target-to-target depending on the magnitude of the hotspot offset. HD 209458b, for example, with its much larger offset compared to WASP-17b, has a peak position $\sim$10$-$11 hours prior to the beginning of eclipse \citep{zellem2014}. Conversely, targets like CoRoT-2b have been observed to have westward hotspot offsets \citep{dang2018}, meaning their peak in flux occurs post-eclipse, and therefore require a post-eclipse baseline extension instead. The brightness of the target and resultant flux precision will also determine how much baseline either side of the peak need be characterised in order to identify the phase signal and successfully apply this strategy, as Figure \ref{fig:partial_phase_curve} shows. Hence, the best way to facilitate this observing style for a large and diverse number of targets -- and produce a uniform final dataset, per the Ariel mission aims \citep{ariel_citation} -- is to observe full phase curves of each target.

\subsubsection{Full Orbit Phase Curves}
\label{sec:full_orbit_pc}
\begin{figure*}
    \centering
    \includegraphics[width=1\linewidth]{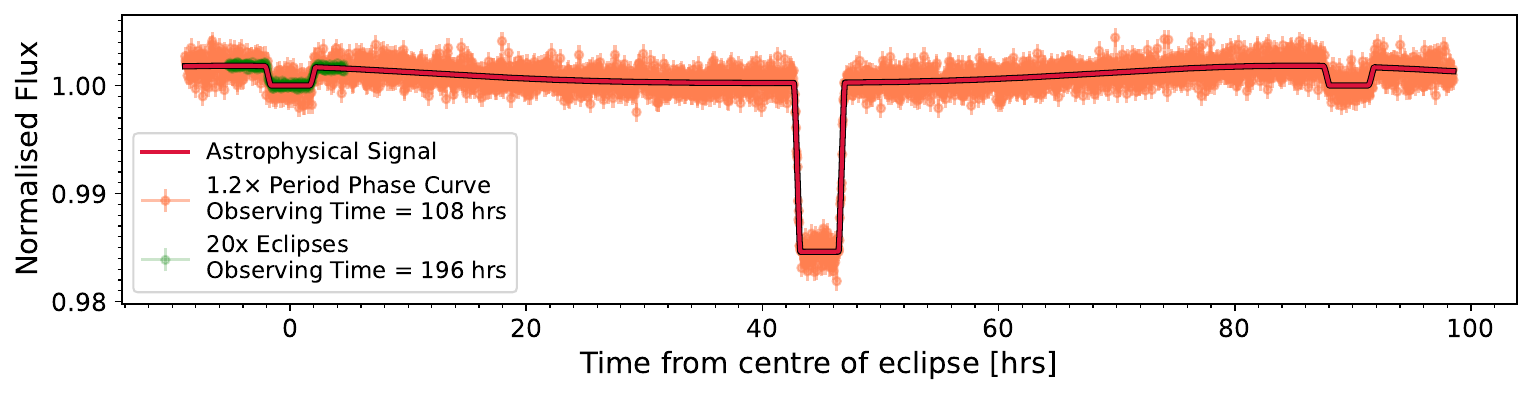}
    \caption{Proposed observing strategy to best facilitate mapping with Ariel. By observing 1.2$\times$period phase curves, a number of benefits are gained (see Section \ref{sec:full_orbit_pc}) which allow Ariel to derive better mapping results using less observing time than is feasible by stacking repeat eclipse-only observations. We use our WASP-17b AIRS Ch1 test case for illustrative purposes here, where a phase curve of this nature (orange) is measurable in almost half the observing time required to measure 20 eclipse-only observations (green), and achieves better results.}
    \label{fig:full_phase_curve}
\end{figure*}

By designing these phase curves as being of observing length 1.2$\times$period, centred on the transit with two eclipses bracketing the observation either side (see Figure \ref{fig:full_phase_curve}), we gain a number of advantages that would bolster Ariel's mapping abilities. We outline these qualitatively below before going on to show its quantitative applications to WASP-17b.

\begin{itemize}
    \item \textbf{Guarantee of Signal Characterisation and Data Uniformity:} With a full-orbit phase curve, the peak is guaranteed to be observed with enough baseline to characterise its morphology no matter how far it is shifted from mid-eclipse. This enables us to anchor the large-scale longitudinal profile and break ingress/egress degeneracies, regardless of the magnitude or direction of the hotspot offset. This uniform observing strategy facilitates the holistic inferences that form the core of the Ariel mission aims \citep{ariel_citation}.

    \item \textbf{Precise System Parameters:} The shape of the phase curve, including the transit, can be fit in order to improve our precision on key system parameters, which are unlikely to be known accurately for every target in such a large sample. High precision on system parameters, in particular the inclination and semi-major axis of the orbit, is imperative in order to correctly attribute the small-scale signals of eclipse mapping \citep{wasp43nirspec_eclipsemap}.

    \item \textbf{Centre-of-Transit Time and Ephemeris Constraints:} The transit can further be used to accurately measure the centre-of-transit time and attribute centre-of-eclipse timing offsets to a mapping signal \citep{eclipse_timing}. Measuring a concurrent transit affords greater accuracy to do this \citep{wasp17eclipsemap}, whereas ephemeris propagation from older observations has proved to hinder mapping efforts \citep{schlawin2024}.

    \item \textbf{SNR Increase and Variability Calibration:} By observing two eclipses, we gain a $\sqrt{2}$ SNR improvement during ingress/egress, which increases the EMM of any given target \citep{boone2023}. Conversely, any \textit{changes} to the eclipse depth and/or shape would be indicative of short-term variability, whether that be astrophysical or systematic, and allow us to diagnose this and correct for it. This is much more challenging using individual observations due to calibration differences \citep{wasp43miri_eclipsemap, lally2025}. 

    \item \textbf{Wealth of Ancillary Science:} Finally, whilst an eclipse baseline extension is helpful for eclipse mapping, it does not provide enough ancillary science to justify the cost for such a large sample size. With a full phase curve, on the other hand, a wealth of ancillary science becomes available. This includes our ability to map not only the dayside, but the \textit{global} emission profile, in addition to facilitating other mapping techniques such as transmission mapping \citep{harmonica}.
\end{itemize}

\noindent In order to show the quantitative improvement gained by adopting this observing strategy, we finally test simulating and recovering on a 1.2$\times$period phase curve Ariel observation of WASP-17b (shown in Figure \ref{fig:full_phase_curve}, compared to the 20-eclipse case from Section \ref{sec:20_eclipse_case}). The recovered map is
% {and flux profile are}
shown in the bottom row of Figure \ref{fig:w17_maps}, and the hotspot posterior is shown in Figure \ref{fig:w17_hotspot_hist}. With only this single observation now, the hotspot is the most accurately and precisely recovered at $18.7^{+1.2 \circ}_{-3.0}$. This high precision is primarily driven by the addition of the phase signal, with the measurement of the phase curve peak heavily anchoring this constraint. This demonstrates the advantages of combining phase and eclipse mapping information for deriving precise constraints on the atmospheric structure. We note that a four-component (N4) model is now required to fit the additional phase signal, but that the derived structure matches the complexity of the input N2 map derived from eclipse-only.

We also recover the input dayside temperature profile better than we could using a reasonable number of eclipses with Ariel, {driven by our characterisation of the shape of the phase curve}. For ease of comparison, we only show the same longitudes for this map as those observed from the original eclipse-only observation timings, but note that such an observation would in fact produce a map of the global profile. The looser contour structure around the hotspot now much more closely resembles that of the input map, and the steeper day-night temperature contrast of order 1000 K is now successfully recovered compared to the eclipse-only Ariel maps, which erroneously indicate a more uniform dayside. {The flux profile, and therefore so too the temperature profile, is also much more tightly constrained, with a median precision of 30 K across the dayside compared to 60 K for the JWST map.}

The difference map shows that there are still some discrepancies at high latitudes, primarily at the limbs. These are primarily artefacts introduced by the higher-order eigenmaps (N3+) required to fit the phase signal. However, as previously discussed, latitudinal information is not constrained by phase maps, whilst the limbs are not tightly constrained by eclipse maps. Thus, when comparing eclipse-only maps versus phase-and-eclipse maps, close agreement is only expected in the equatorial latitudes of the central dayside longitudes. The difference map shows this to be true, with close, high-precision agreement between the JWST and Ariel maps here. The high-latitude discrepancies at the limbs are therefore of least concern, particularly since they contribute negligible flux to the overall map regardless.

Comparing between the best-case eclipse-only scenario, which required 20 observations, and this phase curve scenario, it is evident that the latter is the more efficient mapping strategy. In terms of time efficiency, $\sim$200 hours of observing time would be required to observe 20 eclipses, without even accounting for the inordinate slewing and waiting time that these repeat observations would generate. The phase curve scenario, on the other hand, amounts to only $\sim$100 hours. Without even factoring in time efficiency gained from only having to perform one observation, this is equivalent to only 10 standard-baseline eclipses, which is well within the limit of 20 for high-ranking Ariel targets like WASP-17b \citep{ariel_target_list}.

Hence, the primary conclusion of this test case is that Ariel also has the potential to map dimmer, lower-ranking mapping targets, and that the most practical and efficient way to do this is through the use of phase curves. As such, we conclude at this point that Ariel is in fact capable of facilitating a population-level mapping survey.

\subsection{WASP-43b}

Here, we consider what parameters {Ariel would be most suited to constraining with a population-level mapping survey}. Our previous test cases have shown that Ariel is capable of mapping the longitudinal profile of targets spanning two orders of magnitude in EMM \citep[][and see Appendix]{boone2023}. However, we have not yet tested Ariel's ability to recover latitudinal signals.

Longitudinal signals are more observable for two reasons. First, they are present in the entire phase curve, whilst latitudinal signals are only measurable from ingress/egress.
Second, the amplitude of longitudinal signals are inherently larger for tidally locked gas giants: the atmospheric dynamics
% of hot Jupiters
tend to be dominated by rotational circulation regimes, resulting in substantial east-west hotspot offsets with large limb-to-limb temperature contrasts, whilst we expect a generally symmetric north-south profile with much shallower pole-to-pole temperature contrasts \citep{hammond_and_lewis, beltz2022}. For these reasons, the majority of JWST eclipse maps have not yet been able to robustly measure latitudinal signals \citep[e.g.,][]{wasp18eclipsemap, wasp17eclipsemap, lally2025}.

The MIRI/LRS eclipse map of WASP-43b constituted the first robust detection of the latitudinal profile of an exoplanet atmosphere \citep{wasp43miri_eclipsemap}, with tentative indications of a latitudinal hotspot offset of $\sim$10$^{\circ}$. This latitudinal offset was later confirmed with a 4$\sigma$ detection in a NIRSpec/G395H eclipse map of WASP-43b; the detection of this signal was attributed as only being made possible by the exquisite data quality of JWST \citep{wasp43nirspec_eclipsemap}. We therefore use these final two test cases to determine the limits of Ariel's eclipse mapping abilities by assessing its ability to also recover this latitudinal signal.

\begin{figure*}
    \centering
    \includegraphics[width=0.475\linewidth]{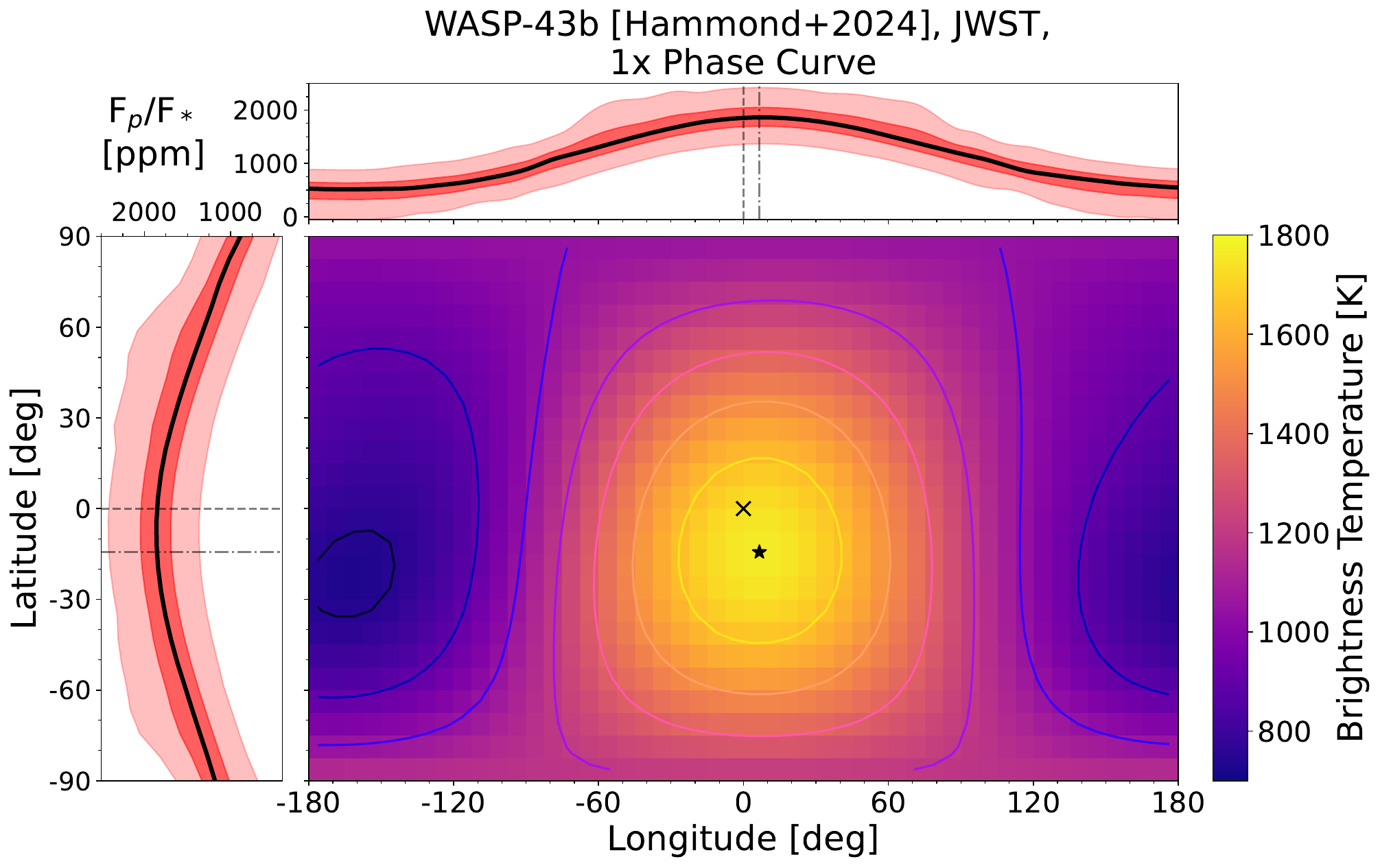}
    \includegraphics[width=0.475\linewidth]{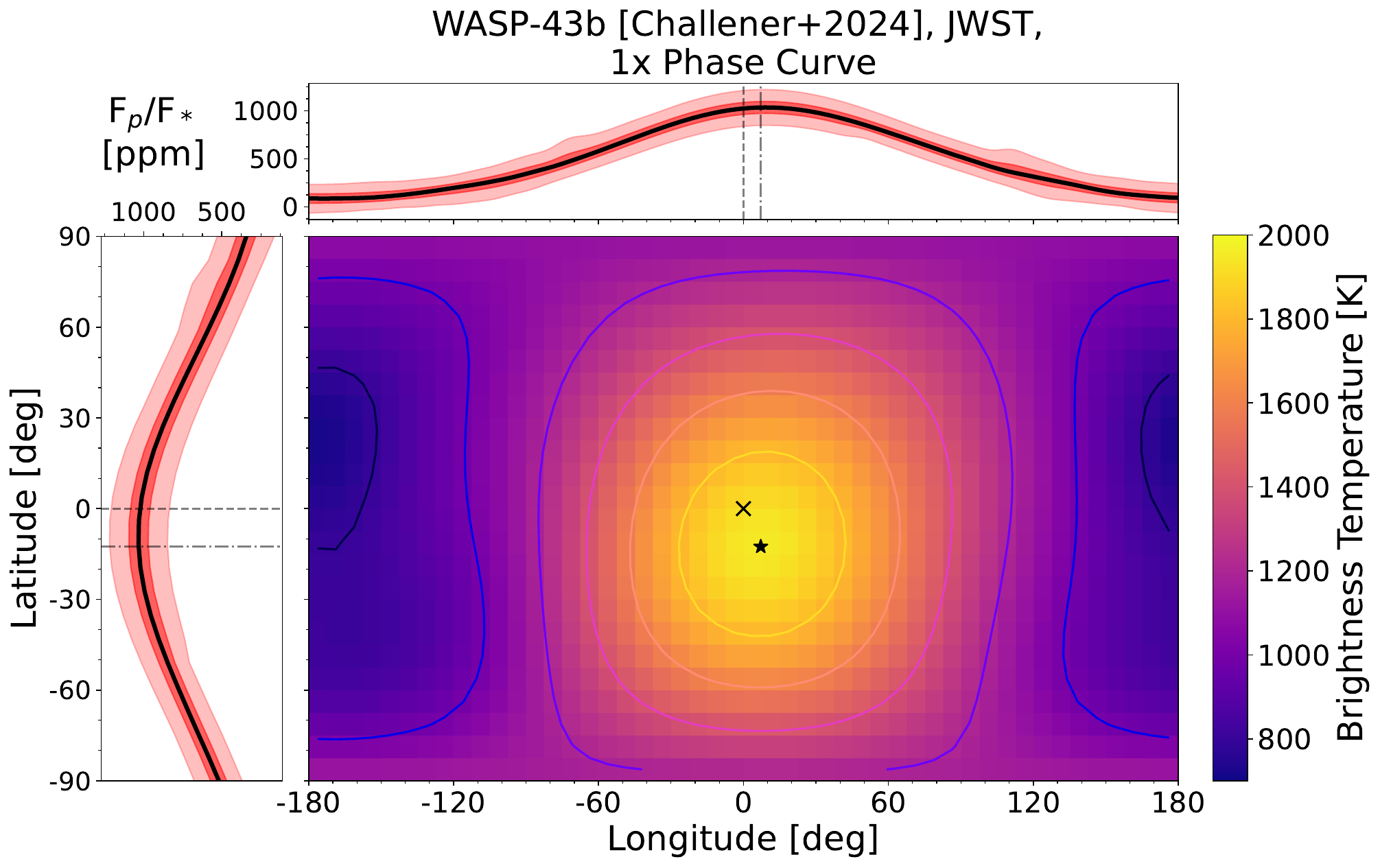}
    \includegraphics[width=0.475\linewidth]{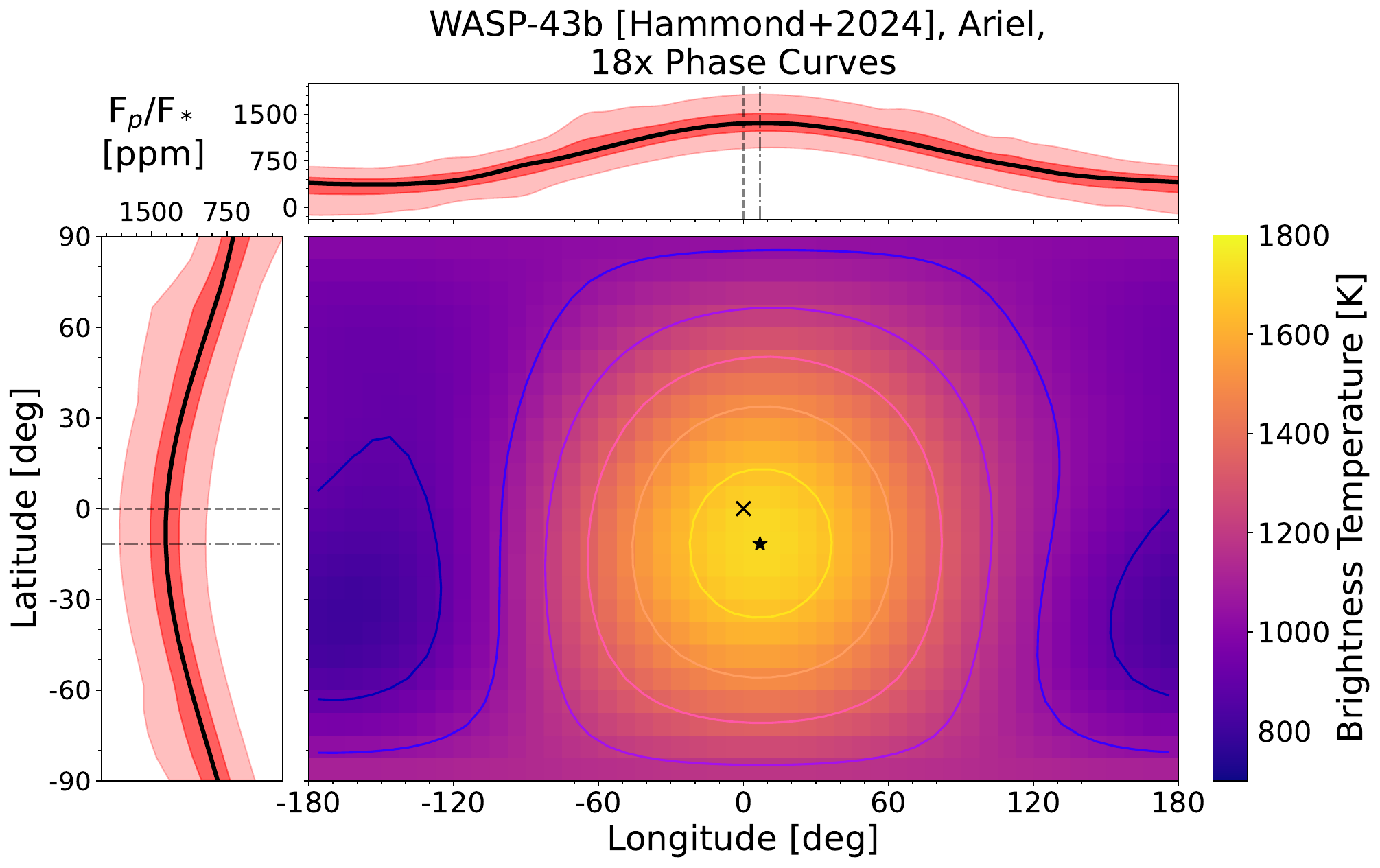}
    \includegraphics[width=0.475\linewidth]{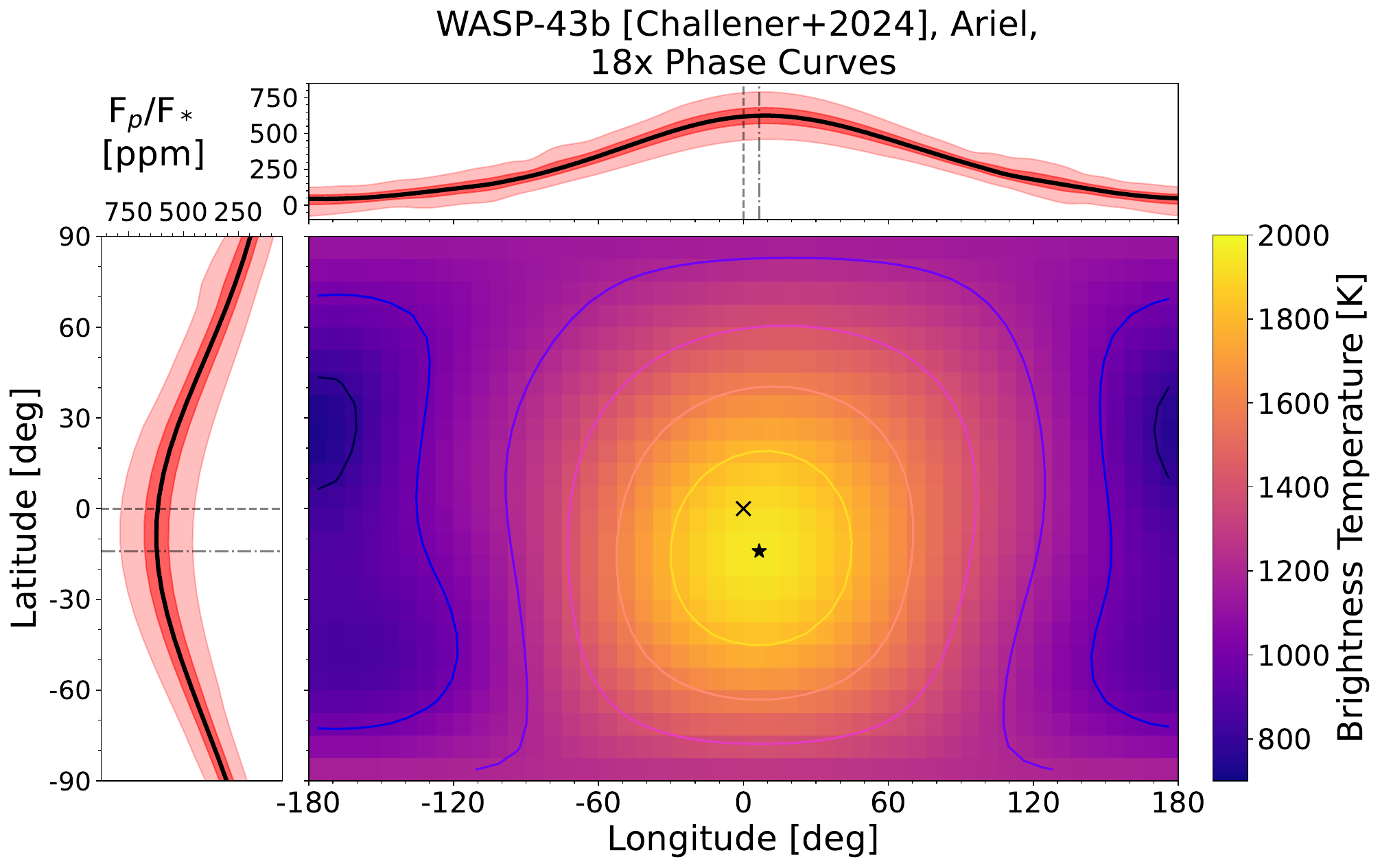}
    \includegraphics[width=0.475\linewidth]{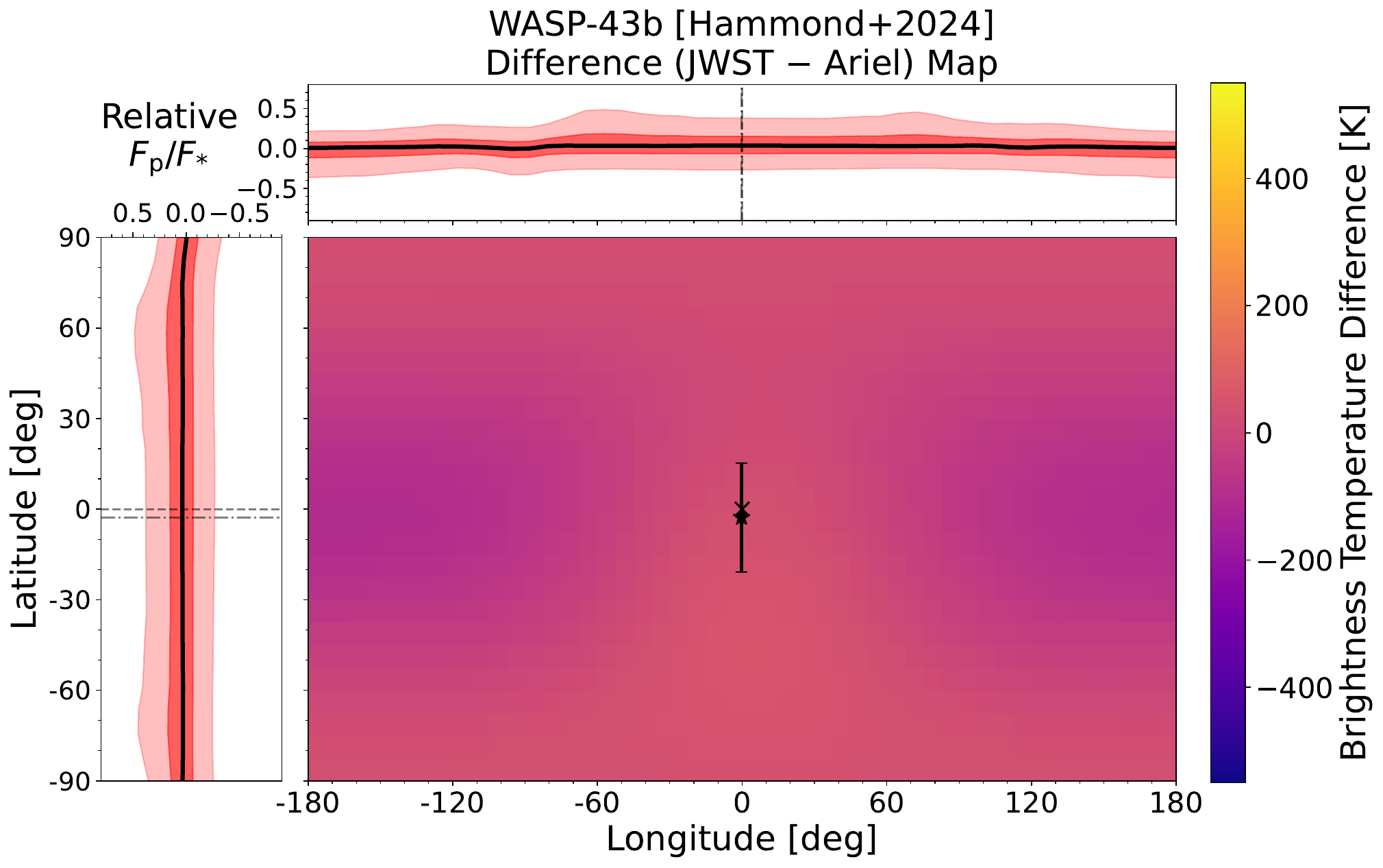}
    \includegraphics[width=0.475\linewidth]{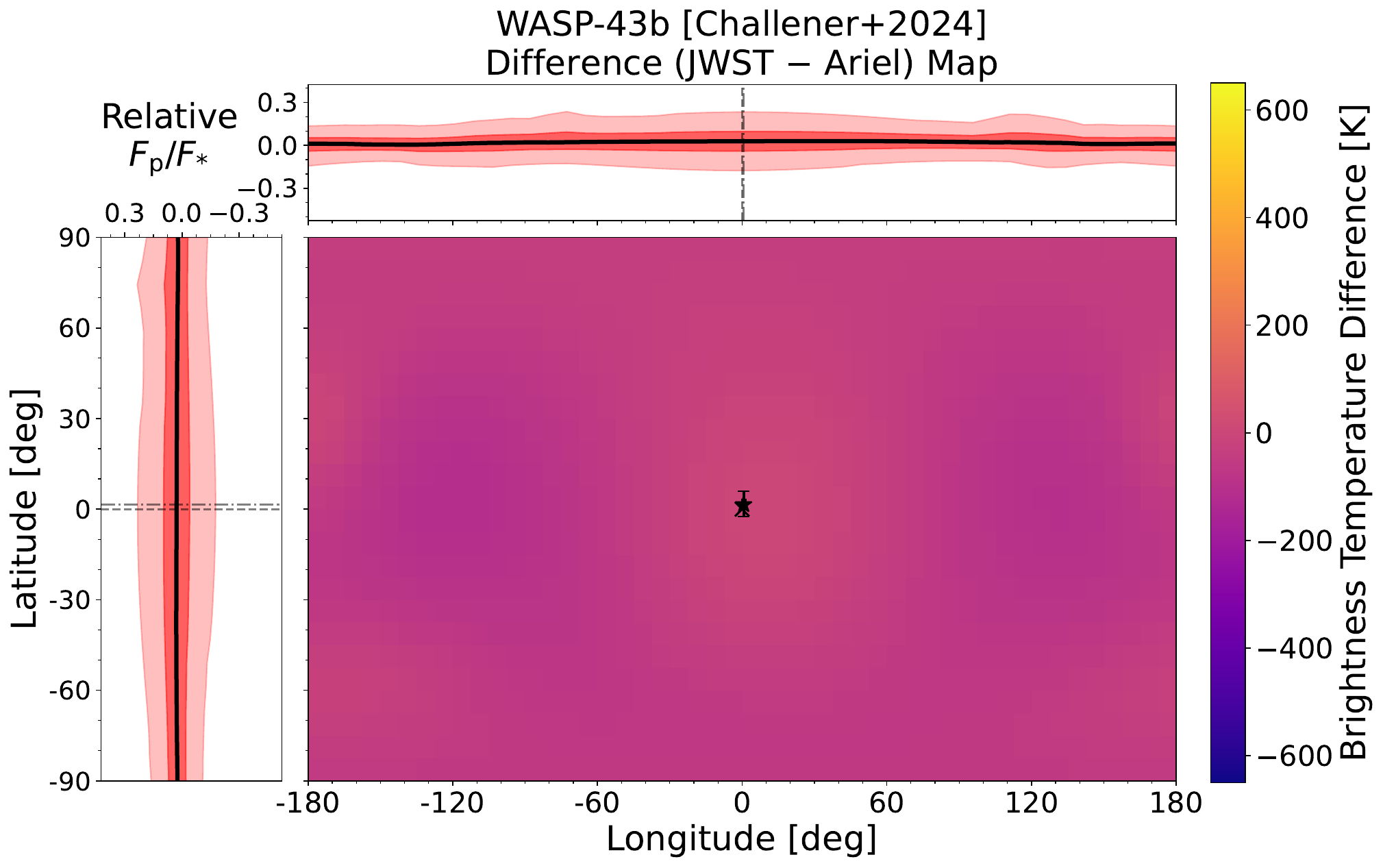}
    \caption{WASP-43b (\textit{left:}) \citet{wasp43miri_eclipsemap} and (\textit{right:}) \citet{wasp43nirspec_eclipsemap} eclipse maps, {with the longitudinal and latitudinal flux profiles shown above and on the left, respectively. The cross and dashed lines mark the location of the substellar point; the star and dash-dotted lines mark the location of the hotspot}. \textit{Top:} Simulated versions of the JWST (left) MIRI/LRS and (right) NIRSpec/G395H maps, derived from single phase curve observations. \textit{Middle:} Equivalent Ariel AIRS Ch1 and Ch0 maps, respectively, each derived from 18 simulated phase curve observations. \textit{Bottom:} Difference maps (JWST$-$Ariel). The flat structures show that we accurately recover the complex input profile in both cases. The overlap of the differenced hotspots with the substellar points show that we recover this parameter accurately in both longitude and latitude. However, latitudinal characterisation with Ariel is incredibly time costly.}
    \label{fig:w43_maps}
\end{figure*}

The primary WASP-43b MIRI/LRS analysis was conducted using spherical harmonic mapping, but an alternative eigenmapping analysis was also conducted with \texttt{ThERESA}, yielding an L2N6 mapping model \citep{wasp43miri_eclipsemap}. The large number of basis components required to fit this mapping signal is a testament to the data quality, resulting in a complex mapping profile with a high degree of asymmetric structure. A similar but higher-degree L3N6 model was used to fit the mapping signal of the NIRSpec/G395H phase curve \citep{wasp43nirspec_eclipsemap}, yielding a similarly complex morphology which is consistent between the JWST instruments (see the top row of Figure \ref{fig:w43_maps}). Hence, we are also testing the synergy of different Ariel spectrographs with this test case by establishing whether they can similarly derive consistent morphologies. Such synergies are even more important for Ariel since it simultaneously measures with all of its instruments at once for every observation \citep{ariel_citation}, {facilitating inherent 3D mapping for every target}.

\subsubsection{The Latitude Problem}

We post-process the JWST MIRI/LRS and NIRSpec/G395H maps of WASP-43b into Ariel AIRS Ch1 and Ch0 phase curves, respectively, {using the methodology outlined in Section \ref{sec:sim_framework}}. {The same as for our other WASP cases, the saturation times surpassed the maximal cadence for adequate spatial scanning of the dayside atmosphere for both spectrographs. We therefore rescaled the cadence per the methods of Section \ref{sec:sim_framework}, thereby decreasing the flux precision of the light curves.} Given the small-scale nature of latitudinal signals, it is perhaps then not surprising that with single observation Ariel phase curves, we are unable to recover these signals.

Whilst we recover {a consistent temperature range (see Table \ref{tab:jwst_vs_ariel}), with the same median dayside precision achieved with Ariel as with JWST}, we recover only four-component (N4) mapping models, with eastward hotspot offsets of $10.9\pm0.{8}^{\circ}$ and $11.3^{+0.7\circ}_{-0.6}$ for AIRS Ch1 and Ch0, respectively. The same as in the WASP-17b phase curve case, an N4 mapping model corresponds to no constrained latitudinal structure for maps derived from full phase curves. Whilst these longitudinal hotspot offsets are discrepant from the eclipse mapping values due to the small uncertainties, they interestingly align with the prior phase-mapping only results of \citet{stevenson2014} and \citet{murphy2023}, who find offsets of $12.3\pm1.0^{\circ}$ and $10.4\pm1.8^{\circ}$ from HST/WFC3 and Spitzer/IRAC phase curves, respectively\footnote{Note that hotspot offsets can differ as a function of pressure, but the high gravity of WASP-43b means that we expect to see similar offsets across all observable pressures for this planet \citep{katariawasp43}}. \citet{wasp43nirspec_eclipsemap} also post-processed from their eclipse mapping posteriors the phase mapping-only value they would have measured (i.e., assuming no eclipse mapping signal), and found a similarly consistent phase-curve offset of $10.0\pm0.8^{\circ}$.

This discrepancy between hotspot offsets derived from phase maps versus eclipse maps signals is a known effect. \citet{nullspace} showed through numerical simulations that the values can differ due to the different observing methods used to derive them: as discussed in Section \ref{sec:extended_baseline_mapping}, phase mapping exploits hemispherically-averaged rotational measurements and therefore constrains only large-scale, longitudinal structure, whereas eclipse mapping finely samples the dayside profile in both longitude and latitude. For WASP-17b, these two measurements (i.e., phase mapping versus eclipse mapping offsets) give equivalent results both in our simulations and the original data \citep{wasp17eclipsemap} because the eclipse map contains no small-scale or latitudinal structure. The WASP-43b maps, on the other hand, contain a degree of small-scale structure, and the high impact parameter of $b=0.66$ \citep{wasp43miri_eclipsemap} means that the latitudinal structure is robustly measured. The phase mapping signal is blind to both this small-scale and latitudinal information, therefore for higher-order maps like these, the two methods may begin to diverge. As such, they no longer anchor each other as they did for WASP-17b, with the longer-duration and higher-amplitude phase mapping signal beginning to dominate.

We simulate increasing our number of phase curve observations in both bandpasses until we recover the higher-component, latitudinally-inclusive models. {We find that in both cases, at least 10 repeated phase curve observations are required in order to recover the latitudinal signal, and that the exact number is dependent on the random scatter of the data, varying upwards to 25 repeat observations. The same mapping model and thermal structure is derived in each case, verifying that the astrophysical signal is indeed being correctly identified; therefore, the random scatter of the data is not affecting our ability to characterise the signal, but rather the noise threshold at which point we are able to resolve it.}

\begin{figure}
    \centering
    \includegraphics[width=1\linewidth]{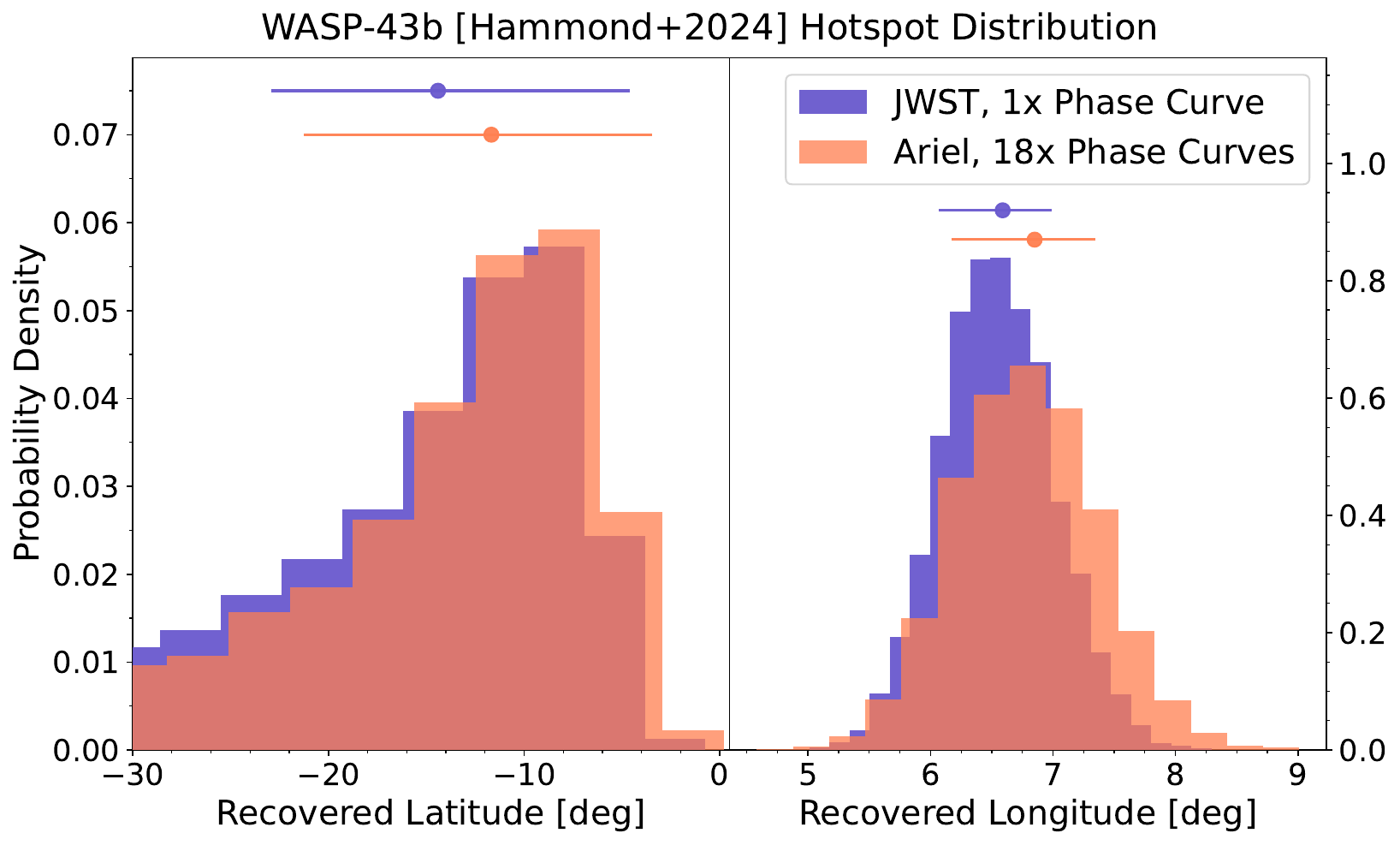}
    \includegraphics[width=1\linewidth]{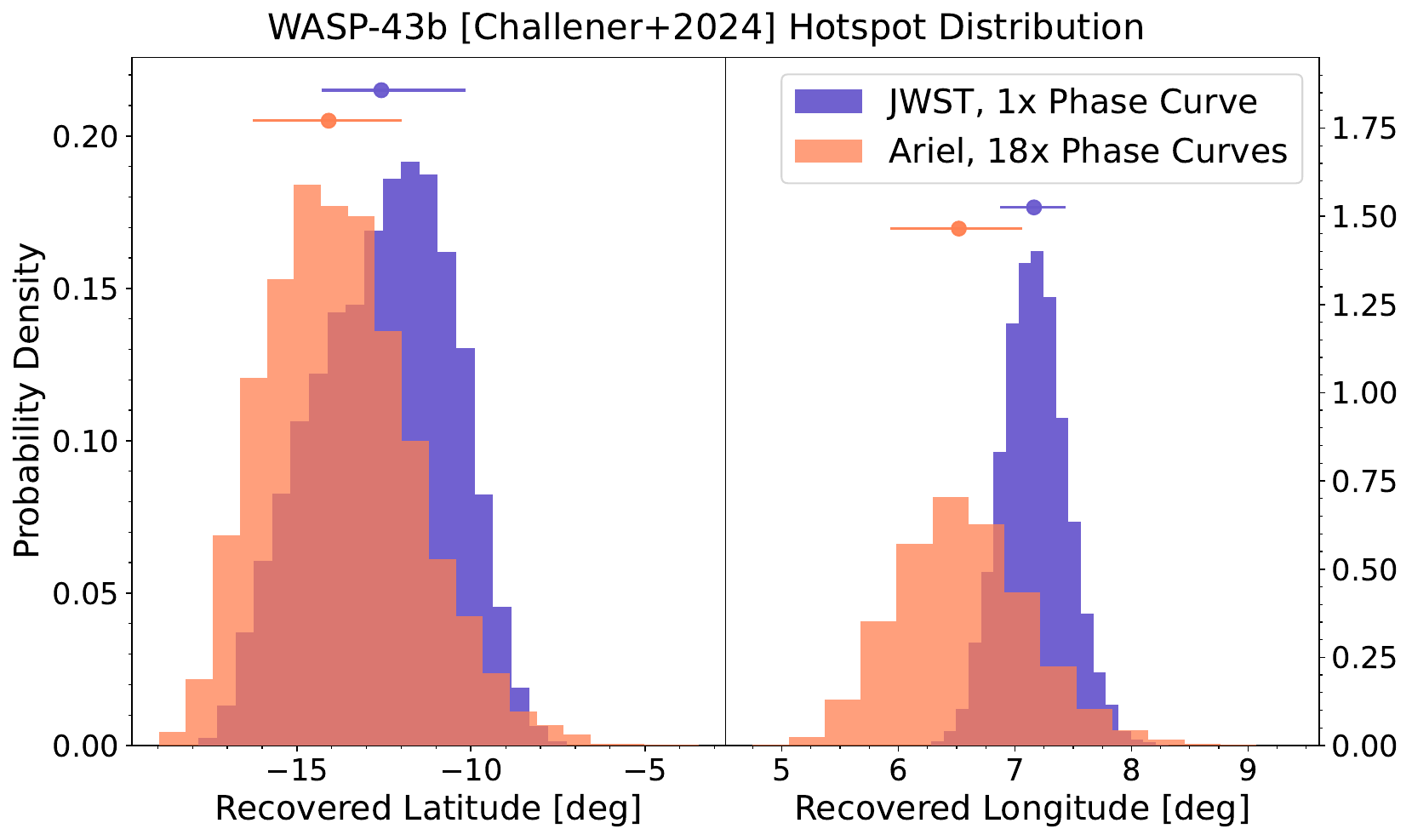}
    \caption{WASP-43b (\textit{top}:) \citet{wasp43miri_eclipsemap} and (\textit{bottom}:) \citet{wasp43nirspec_eclipsemap} hotspot location posteriors. In blue is what we recover from the JWST (\textit{top}:) MIRI/LRS and (\textit{bottom}:) NIRSpec/G395H single phase curve maps, and in orange is what we recover from the 18 Ariel (\textit{top}:) AIRS Ch1 and (\textit{bottom}:) Ch0 phase curve maps, respectively. We recover consistent results, even in terms of latitude, but this test case highlights the time-costly nature of measuring such signals with a smaller observatories like Ariel.}
    \label{fig:w43_hotspot_hist}
\end{figure}

The much higher data quantity required for Ariel to compete with JWST here can be attributed to the fact that, whilst JWST light curves far supersede the quality threshold for \textit{longitudinal} eclipse mapping, they do not for \textit{latitudinal} eclipse mapping, due to the inherently greater challenge of measuring the latter. Hence, we must stack far more Ariel observations in order to achieve the same latitudinal constraints as JWST compared to the number required to achieve the same longitudinal constraints.

{The reason that the number of observations required is so heavily dependent on the random noise properties of the data is twofold. Firstly, going from 10 to 25 observations only in fact increases the SNR by 11.6\%, which is comparable to what is achieved by going from 2 to 3 observations (13.0\%). This is due to the inherent quadratic drop-off of SNR increase with repeat $N$ observations, resulting in a wider spread in the number of required observations in this high-$N$ regime that seemingly exacerbates the issue.}

{Secondly, Ariel's lower collecting area results in longer integration times than JWST, and therefore fewer data points during ingress/egress. Since latitudinal signals are small-scale and localised to ingress/egress, this means that small modulations to individual data points here has a far greater impact on our ability to measure them with Ariel than it does with JWST. We tested whether increasing our sampling of ingress/egress beyond the optimised cadence could alleviate this issue, but found that the sacrifice in flux precision led to the same results. Conversely, when using the native integration times rather than the optimised cadence, the enhanced flux precision is unable to compensate for the reduced spatial scanning. We thus conclude that this is an inherent problem associated with latitudinal eclipse mapping using smaller observatories.}

\subsubsection{Comparison between Ariel and JWST}

We discuss the median Ariel phase curve scenario for each test case, comprising 18 repeated observations. In both cases, we recover highly consistent latitudinal hotspot offsets with Ariel as we do with JWST, as evidenced by the overlapping posteriors in Figure \ref{fig:w43_hotspot_hist}, with values of $-11.8^{+8.2\circ}_{-9.6}$ and $-14.1^{+2.1\circ}_{-2.2}$ for the AIRS Ch1 and Ch0 maps, respectively. We also recover longitudinal hotspot offsets which are now consistent with the eclipse mapping values of $6.9^{+0.5\circ}_{-0.7}$ and $6.5^{+0.5\circ}_{-0.6}$, respectively, indicating that our modelling framework is now correctly identifying both the phase and eclipse mapping signals. The flat difference maps
% {and flux profiles in the bottom row of}
in the bottom row of Figure \ref{fig:w43_maps} support this, showing that we recover highly consistent thermal profiles now with the same degree of small-scale structure.

{The median dayside temperature is constrained at comparable precision with Ariel as it is with JWST, at 50 K for AIRS Ch1 and 25 K for AIRS Ch0. Interestingly, the precision is the same for these Ariel maps derived from 18 phase curves as it is for the Ariel maps derived from single phase curves, despite the increased flux precision. This is due to the added uncertainty that the additional basis components (N5+) add to the higher-order models; the eigenmaps are ranked in order of visibility, therefore higher-order basis maps correspond to smaller-scale signals which will naturally be less constrained \citep{theresa}. The four-component mapping models derived in the single phase curve cases, on the other hand, are made up only of the highest magnitude signals, resulting in precise constraints even at lower flux precision. For more nuances of eclipse mapping temperature uncertainties, see \citet{nullspace}.}

{Comparing between JWST and Ariel in Figure \ref{fig:w43_maps} and Table \ref{tab:jwst_vs_ariel}, the AIRS Ch1 map shows the same trend as observed for previous test cases: a consistent median dayside temperature to what is observed with MIRI/LRS, but over a narrower temperature range. AIRS Ch0 has a more comparable bandwidth to its JWST equivalent, NIRSpec/G395H (see Figure \ref{fig:jwst_vs_ariel_wvs}), therefore we find not only a comparable median dayside temperature in this case, but also a comparable temperature range.}

\subsubsection{Phase Curves vs. Eclipses}

In an attempt to overcome the prohibitive observing time of {$\geq$10 phase curves}, we tested whether the same number of eclipse-only observations would be sufficient to recover the input mapping model in this case. WASP-43b is one of the top 10 highest-ranking mapping targets for both MIRI/LRS and NIRSpec/G395H \citep{boone2023}, therefore we would expect to be able to map it from eclipse-only observations. Critically, latitudinal signals are only present during ingress/egress, therefore the phase information, which only constrains longitudinal signals, may not be necessary in order to measure them. We simulated these eclipses with standard 1:1 baseline, with an additional one hour pre-eclipse for systematic characterisation, and a buffer of half an hour post-eclipse for a scheduling window.

We found that with a similar number (i.e., 10--25) of eclipse-only observations of each respective test case, Ariel is only able to recover the large-scale longitudinal profile through two-component maps, which is equivalent to what was found from single phase curves (but with N4 to also fit the phase signal). {For comparison with the maps derived from 18 phase curves, we discuss the median case of 18 eclipse-only observations.}

{We recover a similar median dayside temperature and range in these eclipse-only maps as we do in the phase curve maps; note that the minimum temperatures quoted in Table \ref{tab:jwst_vs_ariel} are higher than the phase curve cases because with eclipse-only observations, we no longer measure the entirety of the nightside. Whilst the median dayside temperatures agree between the maps, they are constrained at higher precisions in these eclipse-only maps of 15 K and 10 K for AIRS Ch1 and Ch0, respectively, compared to 50 K and 25 K for the phase curve maps. The higher precision compared to the comparably low-order maps derived from single phase curves (N4, with the N3+ fitting the phase signal in these cases) can be attributed to the $\sqrt{18}$ SNR increase winning out over the added phase signal on the dayside. The better precision compared to the higher-order maps derived from 18 phase curves (N6) can again be attributed to the higher-order (N5+) basis maps contributing larger uncertainties.}

Interestingly, we also recover consistent hotspot offsets in these eclipse-only maps as was recovered from the single phase curve cases, of $9.6\pm1.0^{\circ}$ and $10.9\pm1.0^{\circ}$, respectively. This indicates that, even for this high-ranking target, the phase signal is in fact imperative in order to soundly anchor the large-scale, longitudinal structure, break ingress/egress signal degeneracies, and thus constrain the latitudinal structure. This same conclusion was reached in the original MIRI/LRS analysis by removing the phase information and fitting only the two eclipses \citep{wasp43miri_eclipsemap}. For Ariel, without the phase information, we only recover the large-scale longitudinal structure, bringing the eclipse mapping hotspot offsets in-line with the phase mapping offsets. This test case is therefore another advocate for the need to use full phase curves in order to holistically map a large sample of planets.\\

The primary conclusion of this test case is therefore that latitudinal signal characterisation is better left to JWST for the most optimal, individual case targets, {since it requires an observatory with a large enough collecting area to simultaneously facilitate high flux precision and high cadence}. Although these test cases have shown that Ariel can do the same, the data quantity needed to do so compared to JWST is unjustifiable.
We recommend that what Ariel focus on is leveraging its greater dedicated observing time to model the longitudinal profiles of a large number of targets spanning a range of EMMs.

As proven by our test cases, Ariel has the capability to do this using single phase curve observations. In this regime of large-scale longitudinal signals, phase mapping and eclipse mapping offsets are equivalent, facilitating this observing strategy \citep{nullspace}. In particular, the day-night temperature contrast and east-west hotspot offset are measurable from such maps, which, as proven by a number of recent studies \citep[e.g.,][]{wasp17eclipsemap, lally2025}, are enough to constrain the first-order atmospheric dynamics. We therefore recommend that a population-level mapping survey with Ariel focus primarily on characterising these signals.

\subsection{Summary of Test Cases}
\label{sec:test_case_summary}
We used these test cases in order to provide an assessment of Ariel's eclipse mapping capabilities. Below, we group the questions that we set out to address, and the test cases that answered them.

\begin{enumerate}
  \item \textbf{Does Ariel have eclipse mapping capabilities?} The HD 189733b and HD 209458b test cases showed that Ariel has at least baseline eclipse mapping capabilities, since we were able to eclipse map these planets using all three of the Ariel spectrographs using only 1--2 eclipse observations.
  \item \textbf{How does Ariel compare against JWST?} The WASP-18b test case showed that, whilst more data may be required to map even high-ranking targets due to Ariel's lower collecting area, the amount is not exorbitant and in fact perfectly feasible for a dedicated exoplanet mission. Using as few as three eclipse observations, Ariel could eclipse map many of the highest-ranking mapping targets.
  \item \textbf{What is the best way to design these observations?} The WASP-17b test case showed that the data quantity required to map
  % a large population of planets, including those
  planets with lower EMMs does in fact start to become unfeasible using eclipse-only observations. However, the incorporation of phase information can help anchor our longitudinal signal in order to better identify signals in ingress/egress, and recover the mapping signal. For lower-ranking targets, the use of phase curves is in fact the most time-efficient way to conduct these observations, and allows for a homogeneous dataset from which the signal can always be robustly identified no matter the specific morphology of the phase curve.
  \item \textbf{What should be Ariel's primary mapping focus?} The WASP-43b test cases showed that, whilst Ariel can compete with JWST for longitudinal signals using a reasonable amount of data, it struggles to recover the more notorious latitudinal signals, requiring a much larger dataset. Hence, we conclude that latitudinal studies, which require {simultaneously} high SNR {and cadence}, be left to the likes of JWST. What {we recommend} Ariel {would be better focusing} on is its ability to dedicate a great deal of observing time to a large population of planets.
  \end{enumerate}

By observing phase curves of a large and diverse sample, we have shown that Ariel has the ability to map all of their longitudinal profiles, from which their east-west hotspot offsets and day-night temperature contrasts are measurable. These parameters are enough to diagnose the first-order atmospheric dynamics \citep{komacek2017, roth2024} and facilitate the largest-scale assessment of exoplanet atmospheric dynamics to date.

% \section{Survey Design}
\section{Ranking the Best Mapping Targets for Ariel}
\label{sec:emm_ranking}

Based on the results of our test cases, we now repeat the methodology that \citet{boone2023} applied to determine the best eclipse mapping targets for JWST in order to do the same for Ariel. In this way, we provide a target list {of the most advantageous targets for a} population-level mapping survey {with Ariel,} designed to map the large-scale longitudinal profile and constrain the first-order atmospheric dynamics of these targets.

\citet{boone2023} used analytic methods to derive a quantitative metric that assesses the eclipse mapping potential of a target observed using a specific instrument, based on the orbital parameters of the target and the performance of the instrument. This metric is therefore generalisable to any target observed using any observatory. Two quantities of interest are derived that encapsulate the eclipse mapping potential of a target.

The first quantity is the eclipse mapping metric (EMM) itself. This corresponds to the minimum resolution element observable with eclipse mapping for a given planet observed using a given instrument, which is essentially the maximum resolution to which we can spatially resolve the planet, in units of degrees. This quantity is first calculated in a decomposed manner for the longitudinal and latitudinal components, EMM$_x$ and EMM$_y$, respectively, which are defined as

\begin{equation}
\label{eq:emmx}
    \mathrm{EMM_x} = 180^\circ \ \times \left[\frac{2\sqrt{6\ln2}}{\pi^2\sqrt{1-b^2}} \left( \left( \frac{F_0}{\sigma} \frac{1}{\pi} \sqrt{\frac{2 R_{\mathrm{p}}}{\pi a \sqrt{1-b^2}}}  \right)^{1/3} + \frac{1}{2} \right)^{-1} \right],
\end{equation}

\noindent and

\begin{equation}
\label{eq:emmy}
    \mathrm{EMM_y} = 180^\circ \ \times \left[\frac{2\sqrt{6\ln2}}{\pi^2 b} \left( \left( \frac{F_0}{\sigma} \frac{1}{\pi} \sqrt{\frac{2 R_{\mathrm{p}}}{\pi a \sqrt{1-b^2}}}  \right)^{1/3} + \frac{1}{2} \right)^{-1} \right],
\end{equation}

\noindent with the overall EMM defined as the root sum of their quadrature,

\begin{equation}
    \mathrm{EMM} = \sqrt{\mathrm{EMM_x}^2 + \mathrm{EMM_y}^2}.
\end{equation}

\noindent Here, $F_0$ is the phase-averaged flux from the planet (carrying the system brightness and planetary temperature dependence), $\sigma$ is the full-orbit integrated error on the phase-averaged flux (carrying the precision and cadence dependence), $b$ is the impact parameter, $a$ is the semi-major axis of the orbit, and $R_{\mathrm{p}}$ is the planetary radius (the combination of which carry the dependence of the stellar-edge angle and ingress/egress duration).

The second parameter of interest is $N_{\max}$, which is the highest spherical harmonic order detectable. This parameter essentially defines the complexity of the signal that we can recover for a specific planet measured with a specific instrument using eclipse mapping. It is defined as

\begin{equation}
\label{eq:nmax}
    N_{\max} = \left[ \frac{F_0}{\sigma} \frac{1}{\pi} \sqrt{\frac{2 R_{\mathrm{p}}}{\pi a \cos{\theta_0}}} \right]^{\frac{1}{\gamma+1}}|_{\gamma=2},
\end{equation}

\noindent where the already defined parameters carry the same dependence as previous, and $\cos\theta_0=\sqrt{1-b^2}$ carries the impact parameter dependence in this case.

Because this method is purely analytical, no input planetary models are used in order to calculate these metrics; however, some assumption must be made on the expected morphology of the planetary emission profiles in order to derive a quantitative metric, in particular the expected feature sizes. This dependence is incorporated via the $\gamma$ factor, which carries the scaling of the assumed spatial power spectrum of these profiles as $F_N$$\sim$$N^{-\gamma}$. Similarly to \citet{boone2023}, we assume $\gamma=2$ based on the predictions of numerous benchmarked GCM simulations of hot Jupiters (see \citet{showman2020} for a comprehensive summary), which are our primary targets of interest for eclipse mapping purposes. Note that $N_{\max}$ is used in the derivation of the EMM quantities, therefore this feature-scale assumption is folded into those metrics.

\begin{figure*}
    \centering
    \includegraphics[width=0.33\linewidth]{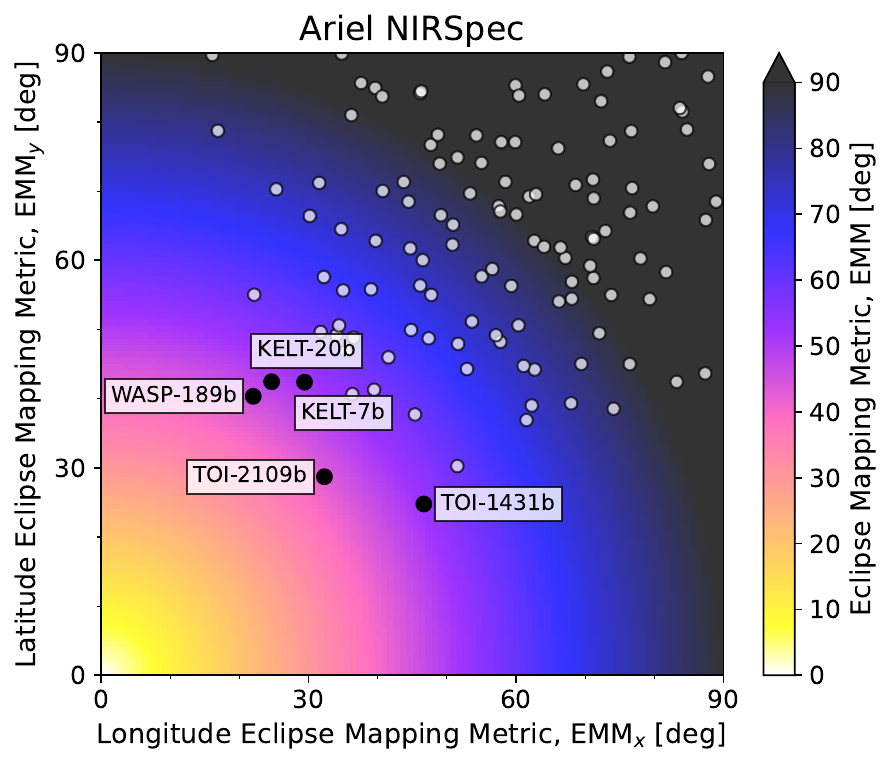}
    \includegraphics[width=0.33\linewidth]{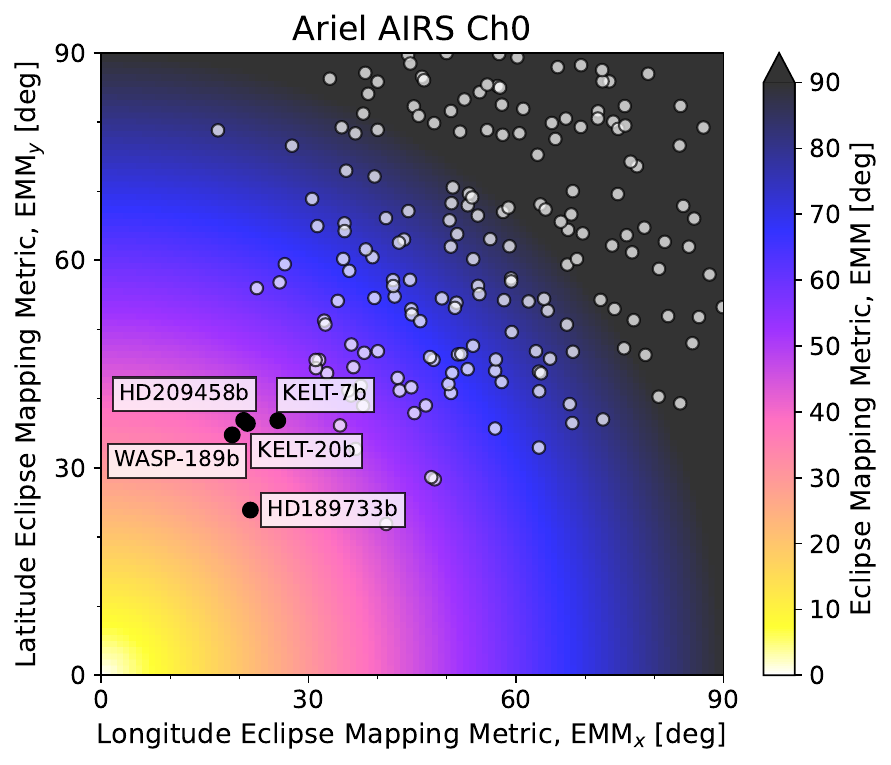}
    \includegraphics[width=0.33\linewidth]{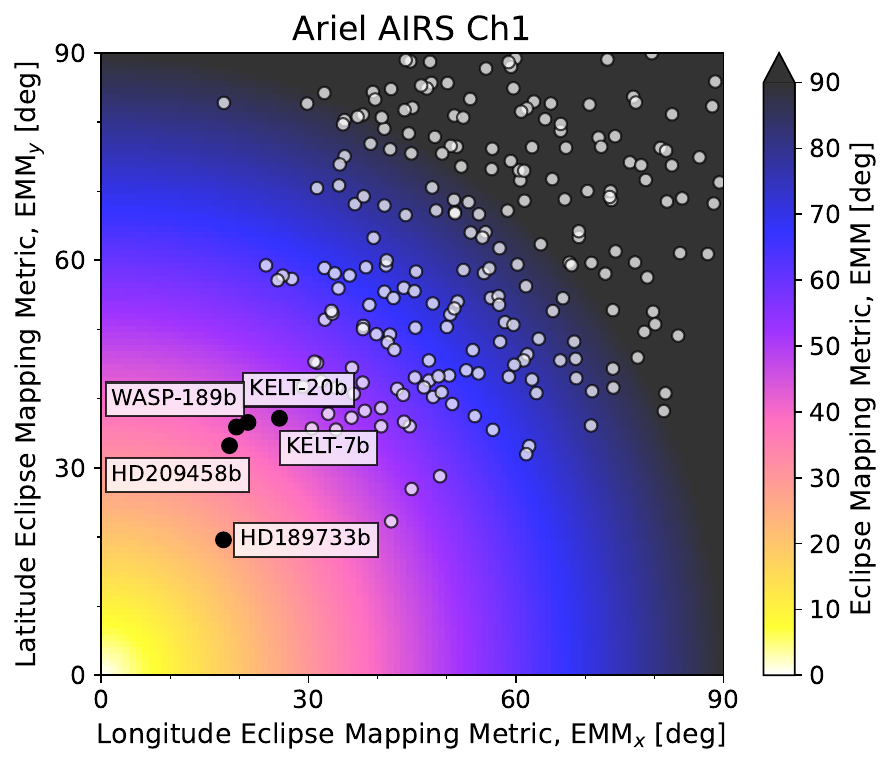}
    \caption{The eclipse mapping metric (EMM) parameter space for the three Ariel spectrographs: (left) NIRSpec, (middle) AIRS Ch0, and (right) AIRS Ch1. Targets increase in ranking going to the bottom left of the plots. The top five highest-ranking targets for each instrument are labelled.}
    \label{fig:emm_plots}
\end{figure*}

We calculate these quantities for all planets in the Ariel Mission Candidate Sample\footnote{\url{https://github.com/arielmission-space/Mission_Candidate_Sample}, v. 2024-07-09.} \citep[MCS,][]{ariel_target_list} for the three Ariel spectrographs: NIRSpec, AIRS Ch0, and AIRS Ch1. We define a planet to be ``eclipse mappable'' if it scores EMM$<90^{\circ}$, meaning it can be mapped at better than hemispheric resolution. We also require that $N_{\max} \geq1$, meaning that we can resolve higher-order features than would be possible from only a phase signal.

Because these metrics are dependent on the precision and cadence of the observing instrument, the observations must be simulated in order to predict these quantities.
We again use \texttt{ArielRad} \citep{arielrad} to simulate these observations, using the same methodology as that outlined in Section \ref{sec:sim_framework}: we adopt the minimum saturation time within a spectrograph's wavelength channels as the cadence, unless this saturation time exceeds the time to obtain 32 samples across ingress/egress, which is required to maximise the spatial information content of the maps \citep{nullspace}. In such cases, we adopt a shorter cadence value that ensures this criterion, and degrade the flux precision accordingly by substituting this value back into \texttt{ArielRad}. For all observations, we then increase the flux precision
by $\sqrt{2}$ as we are assuming 1.2$\times$period phase curve observations of each target bracketed by two eclipses either side, motivated by the results of our test cases (see Section \ref{sec:test_case_summary}). 

We note that these mapping metrics account only for ingress/egress signals, and therefore do not account for the additional phase information in the phase curve outside of eclipse. However, this would not increase either the EMM nor the $N_{\max}$ since phase information cannot provide smaller-scale information than is already possible with eclipse mapping. The addition of this phase information therefore may seek to include more targets in our rankings, but would not increase the rankings of those that already meet our criteria. Since we are already dealing with large quantities of targets which we will eventually seek to limit to only the most optimal for our survey criteria, the addition of these lower-ranking targets is therefore unnecessary.

\subsection{Ariel--JWST Comparison}
\label{sec:ariel_jwst_comp}

We tabulate all of the Ariel ``eclipse mappable'' targets identified by our calculations in Tables \ref{tab:emm_list_nirspec}, \ref{tab:emm_list_airs_ch0}, and \ref{tab:emm_list_airs_ch1} for Ariel NIRSpec, AIRS Ch0, and AIRS Ch1, respectively, listed in order of their EMM. In total, we find that 66 targets are mappable with Ariel NIRSpec, 113 with AIRS Ch0, and 136 with AIRS Ch1. The preceding tables are essentially subsets of one another, with every target but one in both NIRSpec (HATS-67b) and AIRS Ch0 (K2-107b) also contained within AIRS Ch1, totalling 138 unique targets overall.

As previously mentioned in Section \ref{sec:ariel_sims_w17}, we also repeated the JWST analysis of \citet{boone2023} in order to expand their rankings and encourage future eclipse mapping studies of more targets. Like \citet{boone2023}, we used \texttt{PandExo} \citep{pandexo} to simulate the flux precision and cadence of each JWST observation. We tabulate the top 100 eclipse mapping targets for JWST NIRISS/SOSS, NIRSpec/G395H, and MIRI/LRS in Tables \ref{tab:emm_list_soss}, \ref{tab:emm_list_g395h}, and \ref{tab:emm_list_miri}, but here assuming the SNR of one eclipse since JWST has a much larger collecting area and time-constrained nature than Ariel. For comparison with our Ariel findings, we find 200 targets to be mappable with JWST NIRISS/SOSS, 362 with NIRSpec/G395H, and 296 with MIRI/LRS (the full rankings of these instruments can be found on Zenodo at \href{https://doi.org/10.5281/zenodo.17245372}{10.5281/zenodo.17245372}). These lists are again subsets of one another, with nine targets unique to MIRI/LRS, 66 to NIRISS/SOSS, and none to G395H, for a total of 437 unique targets overall. With our proposed observational design, Ariel therefore has the capabilities to map up to a third of the total number of targets that JWST could. {However, we highlight that a single Ariel observation would measure these planets across its entire wavelength range ($0.5-7.8 \ \upmu$m), whereas JWST can only observe using one instrument at a time.}

These rankings for each of the Ariel spectrographs contain many similar targets to their JWST equivalents, with the top-ranking targets being largely the same between the equivalent instruments (see Figure \ref{fig:jwst_vs_ariel_wvs}) of each observatory. This is to be expected since the overall sample selection is dominated by the system parameter dependence, which carry the greatest weighting in the metric calculations (see Equations \ref{eq:emmx}, \ref{eq:emmy}, and \ref{eq:nmax}). The specific scaling of where planets rank, on the other hand, is more dependent on the wavelength range, precision, and cadence of the instrument, hence why we see targets fall in different positions between the Ariel and JWST results (see Appendix tables).

The most notable difference between the JWST and Ariel rankings is not in the targets themselves, but in the distribution that they span in EMM parameter space. Figures 7, 8, and 9 of \citet{boone2023} show the EMM parameter space of the top 100 targets for JWST NIRISS/SOSS, NIRSpec/G395H, and MIRI/LRS. These plots show a clustering of targets at small $\mathrm{EMM_x}$ (increased longitudinal resolution), whilst being more widely distributed in $\mathrm{EMM_y}$. In Figure \ref{fig:emm_plots}, we plot the equivalent EMM parameter space of all the Ariel eclipse mappable targets for its three spectrographs. Conversely to the JWST parameter space, these plots show a similarly broad distribution of targets in both $\mathrm{EMM_x}$ and $\mathrm{EMM_y}$.

This can be explained by the inherent observability of each signal. The assumed power law used to derive the EMM carries the higher-amplitude nature of longitudinal signals compared to the lower-amplitude latitudinal signals, which is consistent with what we see from our test cases. Hence, longitudinal signals are naturally easier to measure. As such, the longitudinal eclipse mapping signals of even lower-ranking targets are easily identifiable by both JWST and Ariel. With increasingly higher precision, increasingly smaller-scale longitudinal signals therefore become easily observable for many targets.
Because \citet{boone2023} only plot the top 100 targets for each JWST spectrograph, we therefore observe a clustering of these targets at small, high-ranking EMM$_x$ below $\sim$40--50$^\circ$. Due to the lower collecting area and therefore flux precision of Ariel, this limit is not as easily reached for as many of the top-ranking targets, resulting in a wider distribution of targets in EMM$_x$ for approximately the same number of plotted targets as JWST, extending to $\mathrm{EMM}_x$$\sim$80--90$^\circ$. To showcase this, we re-plot the EMM parameter space of the JWST spectrographs in Figure \ref{fig:emm_plots_jwst}, including all targets with EMM$<90^{\circ}$. When additionally including these lower-ranking targets, we see that the EMM$_x$ target distribution is now as widely dispersed as that of Ariel, confirming that this apparent $\mathrm{EMM}_x$ bias arises from the primary dependence of longitudinal signal observability being tied to the precision of the observing instrument.

The geometry of eclipse also means that the longitudinal profile itself is optimally scanned over for most targets, whereas the latitudinal profile is only well sampled for planets with high impact parameters. Due to the cosine relationship between inclination and impact parameter ($b$$\propto$$\cos{i}$), this skews the observability of the latitudinal profile to the small fraction of planets that orbit far from the stellar equator. Hence, even with infinitely high flux precision, latitudinal signals are still unrecoverable if the orbital geometry is suboptimal. 
The observability of latitudinal signals is therefore set primarily not by the flux precision of the instrument, but by the orbital geometry of the system, particularly the impact parameter of eclipse. Hence, the distribution of targets in $\mathrm{EMM}_y$ is similarly broad for both JWST and Ariel. 

\subsection{Target Selection}
\label{sec:target_selection}

\begin{figure*}
    \centering
    \includegraphics[width=1\linewidth]{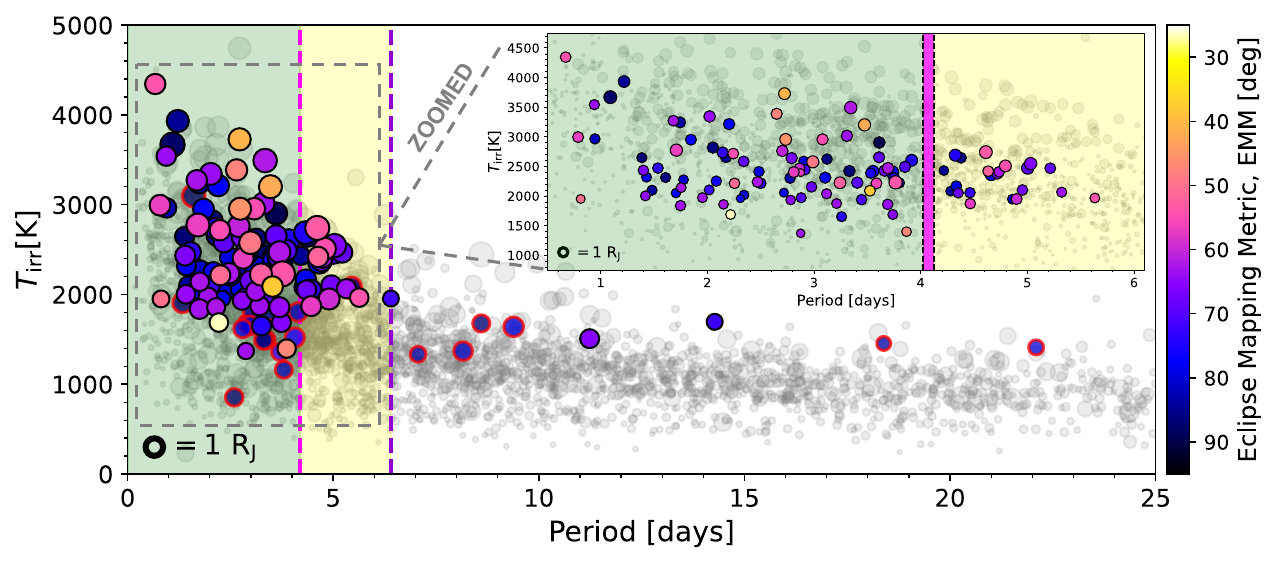}
    \caption{Ariel ``eclipse mappable'' targets. The data points are scaled according to the radius of the planet, and shaded according to their AIRS Ch1 EMM values {(except HATS-67b and K2-107b, for which we use their NIRSpec and AIRS Ch0, values, respectively - see Section \ref{sec:target_selection} for details)}. {Outlined in red} are targets which are only mappable with 1/3 of the Ariel spectrographs{, which we exclude from our survey design due to their lower scientific contribution}. In grey is the total population for a sense of demographics. The dotted lines mark the different period cut-offs discussed in the text, and the inset shows a zoom-in of these shorter period targets. Targets {outlined in black} in the green shaded region $(P\leq4.2 \ \mathrm{days})$ are the ones that we ultimately adopt.}
    \label{fig:survey_P-T_param_space}
\end{figure*}

Figure \ref{fig:survey_P-T_param_space} shows the period-temperature parameter space of the Ariel ``eclipse mappable'' targets. They span irradiation temperatures between 850$-$4350 K, periods up to $\sim$22 days, and radii from 0.65$-$2.1 $\mathrm{R}_{\mathrm{J}}$. A population-level eclipse mapping survey of irradiated gas giants, ranging from warm to ultra-hot Jupiters, is therefore possible using Ariel. However, we calculate that to observe each of these 138 eclipse mappable targets using {a} $1.2\times$period phase curve observing strategy would require 600 days of observing time. With a 4-year primary mission lifetime and a projected $\sim$70\% observing efficiency \citep{ariel_target_list}, Ariel is expected to have 24,800 hours, or $\sim$1033 days, of science time available. Hence, approximately 60\% of this science time would be needed to adopt this strategy. Given that Ariel aims to observe $\sim$1000 targets, this amount of science time is clearly unfeasible. To reduce this overly large fraction, we therefore apply selection cuts in order to craft the most time-effective survey which would maximise the scientific returns whilst still maintaining the diversity of the sample.

We begin by excluding targets that are mappable using only one of the Ariel spectrographs, as this would not maximise the scientific returns of the mission. There is only one target each that is mappable with the NIRSpec and AIRS Ch0 instruments that are not mappable with AIRS Ch1: these are HATS-67b and K2-107b, respectively. Additionally, there are 24 targets that are mappable with AIRS Ch1 but not the other two spectrographs. After removing all of these, we are left with 112 targets which are mappable in at least 2/3 of the Ariel spectrographs. This brings the science observing time down to 432 days, or 42\% of the projected available science time.

\begin{figure}
    \centering
    \includegraphics[width=1\linewidth]{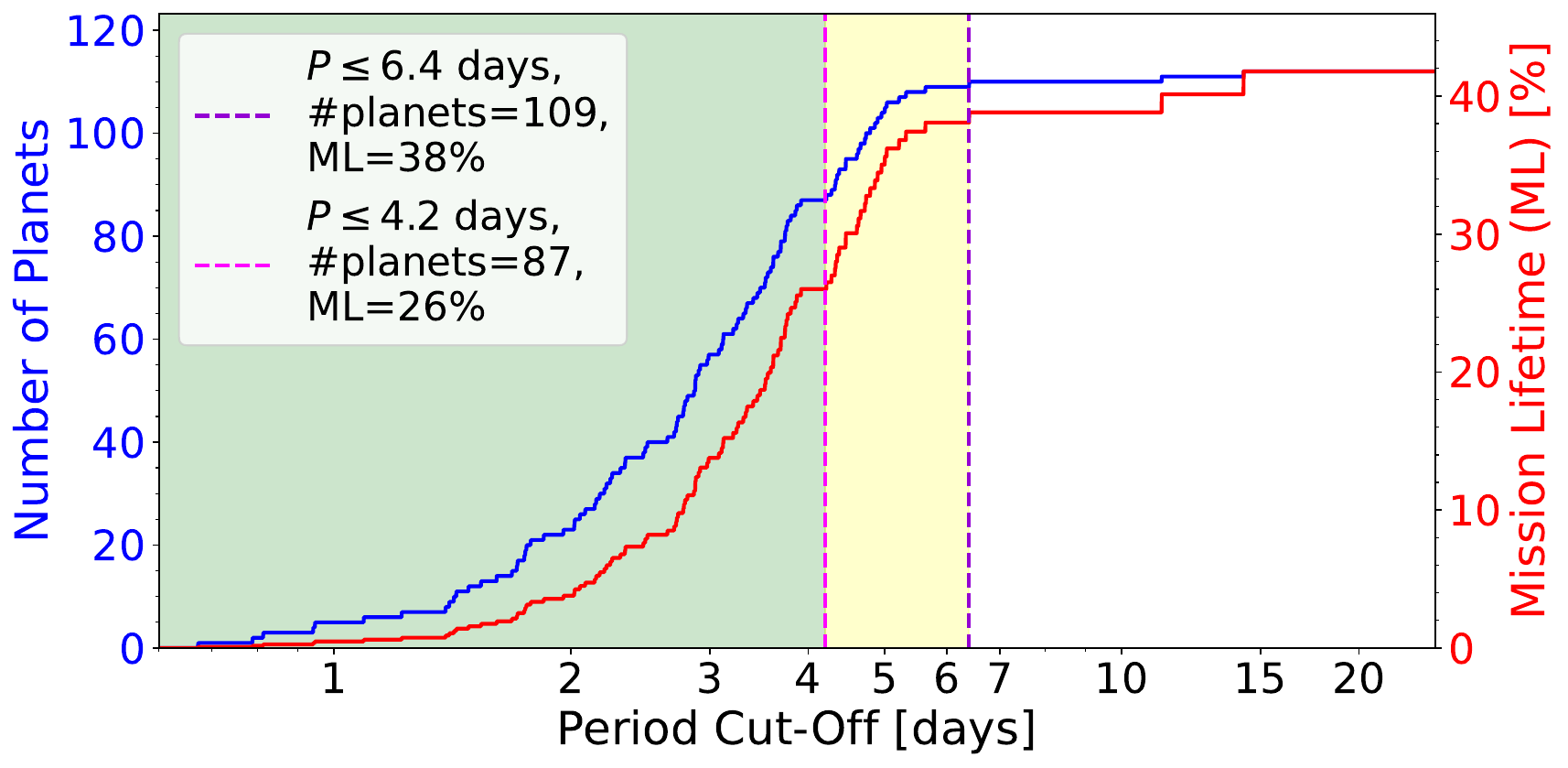}
    \caption{Number of ``eclipse mappable'' targets that are mappable in at least 2/3 Ariel spectrographs, versus the percentage of projected Ariel science time available \citep[24,800 hours,][]{ariel_target_list} that would be required to observe them, as a function of different period cut-offs. We highlight with dotted lines two plateaus of these curves at $P=4.2$ and $P=6.4$ days.}
    \label{fig:survey_time_vs_planets}
\end{figure}

In Figure \ref{fig:survey_time_vs_planets}, we plot the number of these potential targets and the percentage of the projected Ariel science time that would be required to observe them, as a function of different maximum period cut-offs. At $P>1$ day, we see a steadily increasing rate in the number of targets in the survey for small added increments in period cut-off, which fully plateaus by $P$$\sim$15 days, at which point we include all 112 of the targets that are mappable in at least 2/3 of the Ariel spectrographs. This plateau begins at $P$$\sim$6.4 days (purple dashed line), signifying a reduction in time efficiency to include targets with periods longer than this. Excluding such targets leaves us with 109 planets observable in 394 days, or $\sim38\%$ of the projected science time of the mission. The additional three targets above this cut-off therefore require a substantial 38 days, or $\sim$4\% of the total available science time, to be observed. This highlights the greater cost of including longer period planets, and that a period cut-off is required to prevent this.

\begin{table*}
	\centering
	\caption{{Suggested targets} for a population-level mapping survey with Ariel. These have $\mathrm{EMM}<90^{\circ}$ and $N_{\max} \geq 1$ in at least 2/3 of the Ariel spectrographs, and $P \leq 4.2$ days. We list the targets alphabetically, with relevant system parameters, and provide their eclipse mapping ranking for each of the Ariel spectrographs.}
	\label{tab:survey_targets}
	\begin{tabular}{lcccccccccr} % four columns, alignment for each
		\hline
        Planet Name & $T_{\mathrm{eq}}$ [K] & $R_{\mathrm{p}} \ [R_{\mathrm{J}}]$ & $M_{\mathrm{p}} \ [M_{\mathrm{J}}]$ & $P$ [days] & $b$ & $K$ (mag) & NIRSpec Ranking & AIRS Ch0 Ranking & AIRS Ch1 Ranking\\
		\hline

        GPX-1b & 2293 & 1.47 & 19.7 & 1.74 & 0.82 & 11.18 & 38 & 79 & 109\\
        
        Gaia-2b & 1314 & 1.32 & 0.82 & 3.69 & 0.79 & 9.74 & / & 55 & 45\\
        
        HAT-P-13b & 1727 & 1.27 & 0.85 & 2.92 & 0.62 & 8.98 & 49 & 59 & 66\\
        
        HAT-P-16b & 1627 & 1.29 & 4.19 & 2.78 & 0.42 & 9.55 & / & 104 & 106\\
        
        HAT-P-20b & 969 & 0.87 & 7.25 & 2.88 & 0.63 & 8.6 & / & 88 & 38\\
        
        HAT-P-22b & 1324 & 1.15 & 2.47 & 3.21 & 0.43 & 7.84 & / & 57 & 41\\
        
        HAT-P-30b & 1699 & 1.44 & 0.83 & 2.81 & 0.77 & 9.15 & 22 & 20 & 23\\
        
        HAT-P-50b & 1855 & 1.29 & 1.35 & 3.12 & 0.63 & 10.5 & / & 101 & 122\\
        
        HAT-P-56b & 1866 & 1.51 & 2.31 & 2.79 & 0.85 & 9.83 & 29 & 39 & 53\\
        
        HAT-P-6b & 1761 & 1.48 & 1.32 & 3.85 & 0.54 & 9.31 & 47 & 56 & 64\\
        
        HAT-P-7b & 2272 & 1.51 & 1.84 & 2.2 & 0.47 & 9.33 & 28 & 64 & 85\\
        
        HATS-1b & 1370 & 1.3 & 1.86 & 3.45 & 0.7 & 10.58 & / & 74 & 65\\
        
        HATS-65b & 1633 & 1.5 & 0.82 & 3.11 & 0.67 & 11.1 & / & 84 & 89\\
        
        HD149026b & 1693 & 0.74 & 0.38 & 2.88 & 0.63 & 6.82 & 43 & 23 & 25\\
        
        HD189733b & 1193 & 1.13 & 1.13 & 2.22 & 0.67 & 5.54 & 16 & 1 & 1\\
        
        HD202772Ab & 2133 & 1.54 & 1.02 & 3.31 & 0.41 & 6.96 & 27 & 31 & 39\\
        
        HD209458b & 1477 & 1.39 & 0.73 & 3.52 & 0.49 & 6.31 & 18 & 4 & 2\\
        
        K2-29b & 1167 & 1.19 & 0.73 & 3.26 & 0.62 & 10.06 & / & 102 & 75\\
        
        KELT-14b & 1961 & 1.74 & 1.28 & 1.71 & 0.86 & 9.42 & 12 & 17 & 22\\
        
        KELT-17b & 2090 & 1.52 & 1.31 & 3.08 & 0.57 & 8.65 & 13 & 15 & 19\\
        
        KELT-20b & 2263 & 1.74 & 3.38 & 3.47 & 0.5 & 7.42 & 3 & 3 & 4\\
        
        KELT-21b & 2053 & 1.59 & 3.91 & 3.61 & 0.42 & 10.09 & / & 111 & 135\\
        
        KELT-23Ab & 1565 & 1.32 & 0.94 & 2.26 & 0.57 & 8.9 & 21 & 9 & 10\\
        
        KELT-3b & 1952 & 1.56 & 1.94 & 2.7 & 0.53 & 8.66 & 17 & 26 & 28\\
        
        KELT-4Ab & 1822 & 1.7 & 0.9 & 2.99 & 0.69 & 8.69 & 7 & 8 & 8\\
        
        KELT-7b & 2089 & 1.6 & 1.39 & 2.73 & 0.57 & 7.54 & 4 & 5 & 5\\
        
        KELT-8b & 1571 & 1.62 & 0.66 & 3.24 & 0.86 & 9.18 & 24 & 16 & 12\\
        
        KPS-1b & 1448 & 1.03 & 1.09 & 1.71 & 0.75 & 10.93 & / & 89 & 86\\
        
        Qatar-1b & 1416 & 1.14 & 1.29 & 1.42 & 0.65 & 10.41 & 64 & 46 & 46\\
        
        TOI-1194b & 1382 & 0.78 & 0.38 & 2.31 & 0.8 & 9.34 & / & 97 & 84\\
        
        TOI-1431b & 2395 & 1.49 & 3.12 & 2.65 & 0.88 & 7.44 & 5 & 6 & 7\\
        
        TOI-157b & 1591 & 1.29 & 1.18 & 2.08 & 0.8 & 10.89 & / & 78 & 82\\
        
        TOI-1937Ab & 2096 & 1.25 & 2.01 & 0.95 & 0.87 & 11.23 & 40 & 76 & 103\\
        
        TOI-2109b & 3072 & 1.35 & 5.02 & 0.67 & 0.75 & 9.07 & 1 & 7 & 16\\
        
        TOI-2236b & 1687 & 1.28 & 1.58 & 3.53 & 0.77 & 9.96 & 63 & 73 & 80\\
        
        TOI-3331Ab & 1487 & 1.16 & 2.27 & 2.02 & 0.57 & 9.91 & / & 69 & 70\\
        
        TOI-4137b & 1570 & 1.21 & 1.44 & 3.8 & 0.58 & 10.0 & / & 105 & 105\\
        
        TOI-4463Ab & 1394 & 1.18 & 0.79 & 2.88 & 0.88 & 9.4 & / & 72 & 63\\
        
        TOI-905b & 1192 & 1.17 & 0.67 & 3.74 & 0.82 & 9.45 & / & 70 & 50\\
        
        TrES-2b & 1581 & 1.36 & 1.49 & 2.47 & 0.77 & 9.85 & 41 & 35 & 34\\
        
        TrES-4b & 1708 & 1.61 & 0.78 & 3.55 & 0.82 & 10.33 & 56 & 71 & 81\\
        
        TrES-5b & 1483 & 1.19 & 1.79 & 1.48 & 0.61 & 11.59 & / & 107 & 113\\
        
        WASP-101b & 1566 & 1.43 & 0.51 & 3.59 & 0.72 & 9.07 & 30 & 21 & 20\\
        
        WASP-104b & 1514 & 1.14 & 1.27 & 1.76 & 0.73 & 9.88 & 48 & 37 & 32\\
        
        WASP-119b & 1566 & 1.4 & 1.23 & 2.5 & 0.57 & 10.54 & / & 68 & 72\\
        
        WASP-120b & 1876 & 1.47 & 4.85 & 3.61 & 0.77 & 9.88 & 33 & 44 & 58\\
        
        WASP-123b & 1522 & 1.32 & 0.9 & 2.98 & 0.53 & 9.36 & / & 58 & 57\\
        
        WASP-12b & 2594 & 1.94 & 1.46 & 1.09 & 0.34 & 10.19 & 32 & 86 & 130\\
        
        WASP-135b & 1720 & 1.3 & 1.9 & 1.4 & 0.76 & 11.04 & 34 & 38 & 52\\
        
        WASP-142b & 1992 & 1.53 & 0.84 & 2.05 & 0.77 & 11.44 & 54 & 95 & 123\\
        
        WASP-14b & 1920 & 1.38 & 8.84 & 2.24 & 0.56 & 8.62 & 15 & 14 & 18\\
        
        WASP-153b & 1712 & 1.55 & 0.39 & 3.33 & 0.61 & 11.05 & / & 106 & 125\\
        
        WASP-15b & 1626 & 1.41 & 0.54 & 3.75 & 0.53 & 9.69 & / & 75 & 76\\
        
        WASP-163b & 1634 & 1.2 & 1.87 & 1.61 & 0.45 & 10.15 & / & 108 & 121\\
        
        WASP-164b & 1608 & 1.13 & 2.13 & 1.78 & 0.82 & 10.96 & / & 93 & 99\\
        
        WASP-167b & 2363 & 1.58 & 8.0 & 2.02 & 0.77 & 9.76 & 10 & 28 & 44\\
        
        WASP-16b & 1440 & 1.22 & 1.24 & 3.12 & 0.65 & 9.59 & 58 & 42 & 40\\
        
        WASP-170b & 1426 & 1.1 & 1.6 & 2.34 & 0.69 & 10.72 & / & 100 & 98\\
        
        WASP-173Ab & 1871 & 1.2 & 3.69 & 1.39 & 0.4 & 10.0 & / & 103 & 127\\
        
        WASP-178b & 2469 & 1.81 & 1.66 & 3.34 & 0.54 & 9.7 & 9 & 24 & 29\\
        
        WASP-17b & 1699 & 1.93 & 0.48 & 3.74 & 0.35 & 10.22 & / & 85 & 91\\
        
        WASP-189b & 2636 & 1.62 & 1.99 & 2.72 & 0.48 & 6.06 & 2 & 2 & 3\\
        
    \hline
	\end{tabular}
\end{table*}

\setcounter{table}{3} 
\begin{table*}
	\centering
	\caption{Targets for proposed survey - continued.}
	\label{tab:survey_targets}
	\begin{tabular}{lccccccccc} % four columns, alignment for each
		\hline
        Planet Name & $T_{\mathrm{eq}}$ [K] & $R_{\mathrm{p}} \ [R_{\mathrm{J}}]$ & $M_{\mathrm{p}} \ [M_{\mathrm{J}}]$ & $P$ [days] & $b$ & $K$ (mag) & NIRSpec Ranking & AIRS Ch0 Ranking & AIRS Ch1 Ranking\\
		\hline
        WASP-18b & 2504 & 1.24 & 10.2 & 0.94 & 0.37 & 8.13 & 11 & 25 & 43\\
                
        WASP-19b & 2117 & 1.42 & 1.15 & 0.79 & 0.67 & 10.48 & 6 & 10 & 21\\

        WASP-24b & 1827 & 1.38 & 1.24 & 2.34 & 0.61 & 10.15 & 36 & 45 & 60\\

        WASP-2b & 1314 & 1.08 & 0.93 & 2.15 & 0.75 & 9.63 & / & 49 & 31\\
        
        WASP-31b & 1574 & 1.55 & 0.48 & 3.41 & 0.78 & 10.65 & 55 & 60 & 62\\
        
        WASP-32b & 1454 & 0.96 & 2.63 & 2.72 & 0.76 & 10.16 & / & 96 & 87\\
        
        WASP-33b & 2781 & 1.59 & 2.09 & 1.22 & 0.21 & 7.47 & 44 & 83 & 116\\
        
        WASP-36b & 1747 & 1.33 & 2.36 & 1.54 & 0.69 & 11.29 & 46 & 63 & 74\\
        
        WASP-3b & 2086 & 1.42 & 2.43 & 1.85 & 0.41 & 9.36 & 39 & 66 & 83\\
        
        WASP-43b & 1379 & 0.93 & 1.78 & 0.81 & 0.66 & 9.27 & 25 & 12 & 9\\
        
        WASP-46b & 1639 & 1.17 & 1.91 & 1.43 & 0.73 & 11.4 & / & 80 & 90\\
        
        WASP-48b & 1933 & 1.5 & 0.8 & 2.14 & 0.81 & 10.37 & 35 & 53 & 73\\
        
        WASP-49b & 1366 & 1.11 & 0.37 & 2.78 & 0.75 & 9.75 & / & 54 & 47\\
        
        WASP-50b & 1394 & 1.17 & 1.47 & 1.96 & 0.67 & 9.97 & 66 & 43 & 37\\
        
        WASP-52b & 1299 & 1.27 & 0.46 & 1.75 & 0.6 & 10.09 & / & 50 & 36\\
        
        WASP-54b & 1747 & 1.58 & 0.59 & 3.69 & 0.53 & 9.04 & 31 & 32 & 30\\
        
        WASP-69b & 988 & 1.11 & 0.29 & 3.87 & 0.65 & 7.46 & / & 30 & 6\\
        
        WASP-71b & 1830 & 1.18 & 1.39 & 2.9 & 0.55 & 9.32 & 61 & 82 & 96\\
        
        WASP-75b & 1705 & 1.27 & 1.07 & 2.48 & 0.89 & 10.06 & / & 87 & 95\\
        
        WASP-79b & 1716 & 1.53 & 0.85 & 3.66 & 0.5 & 9.06 & 45 & 48 & 55\\
        
        WASP-87b & 2313 & 1.38 & 2.18 & 1.68 & 0.6 & 9.55 & 8 & 22 & 35\\
        
        WASP-90b & 1840 & 1.63 & 0.63 & 3.92 & 0.84 & 10.25 & 51 & 67 & 79\\
        
        WASP-92b & 1879 & 1.46 & 0.8 & 2.17 & 0.61 & 11.52 & 57 & 94 & 112\\
        
        XO-6b & 1577 & 2.07 & 4.4 & 3.77 & 0.63 & 9.25 & 26 & 19 & 14\\
        
        XO-7b & 1744 & 1.37 & 0.71 & 2.86 & 0.73 & 9.24 & 23 & 29 & 27\\
      
    \hline
	\end{tabular}
\end{table*}

There is an earlier plateau in the time efficiency curve in the form of a step function which extends from $4\lesssim$$\ P \ $$\lesssim4.2$ days (pink dashed line), which is also in the inset of Figure \ref{fig:survey_P-T_param_space}. Establishing a period cut-off of 4.2 days at this first point of reduction in time efficiency leaves us with 87 targets, for which full 1.2$\times$ period phase curves would be observable using only 26\% (269 days) of the projected mission science time, a substantial improvement over that acquired using the later period cut-off.
The targets with periods below this cut-off span $T_{\mathrm{irr}}\sim$1350$-$4400 K, which still well samples the boundaries of the warm to ultra-hot Jupiter parameter space. The additional 22 targets with $4.2\lesssim P \lesssim 6.4$ days span a narrower range of $T_{\mathrm{irr}}$$\sim$2000$-$ $3000\ \mathrm{K}$, with similar radii and other non-plotted parameters, including mass and gravity, and therefore provide no significant additional diversity to our sample beyond period.
Hence, we elect to exclude these less time-efficient, longer periods targets, and adopt a period cut-off of 4.2 days. However, we note that we explore the benefits of including select long-period cases to test extreme case effects on the atmospheric dynamics in Section \ref{sec:survey_demographics}.

We tabulate the 87 targets that ultimately meet our {suggested} survey criteria in Table \ref{tab:survey_targets}, listed in alphabetical order. The Ariel MCS is divided into a tiered system from Tiers 1$-$3, with increasingly in-depth characterisation and observing time for higher-tier targets \citep{ariel_target_list}. A fourth tier is currently dedicated to ancillary science, including phase curves, using $\sim$10\% of the mission lifetime \citep{charnay2022}. Most of the targets in our curated survey are already designated as high-ranking Tier 3 targets \citep{ariel_target_list}, designated for in-depth characterisation using increased observing time, to which our proposed observing strategy is highly conducive.

With phase curve observations, both the transit and eclipse are observed, which facilitate transmission and emission spectroscopy; hence, while phase curves formally fall in Tier 4 of the current Ariel formulation, these observations would also meet Tier 3 criteria, enabling a hybridised approach between the two and thus sharing of designated observing time. A current Ariel phase curve working group report (priv. comm) estimates that by following this hybrid-tier approach, $\sim$25\% of the primary mission lifetime could be dedicated to phase curve observations, which would be enough to observe all of our Table \ref{tab:survey_targets} targets. The use of phase curves for observations also reduces slewing and waiting time compared to repeat transits and/or eclipse observations, which increases the mission efficiency and thus science time available \citep{charnay2022}, further facilitating the practicality and advantages of this strategy.

We note that an alternative analysis focusing only on the EMM$_x$ metric could be conducted since we advocate that Ariel should primarily focus on characterising the longitudinal profile rather than the latitudinal. However, when dealing with such a large quantity of targets, whilst the ranking order would change, the overall target list would remain largely invariant, therefore our overall survey and the parameter space it samples would not significantly change.

\subsection{Survey Demographics}
\label{sec:survey_demographics}
\begin{figure*}
    \centering
    \includegraphics[width=1\linewidth]{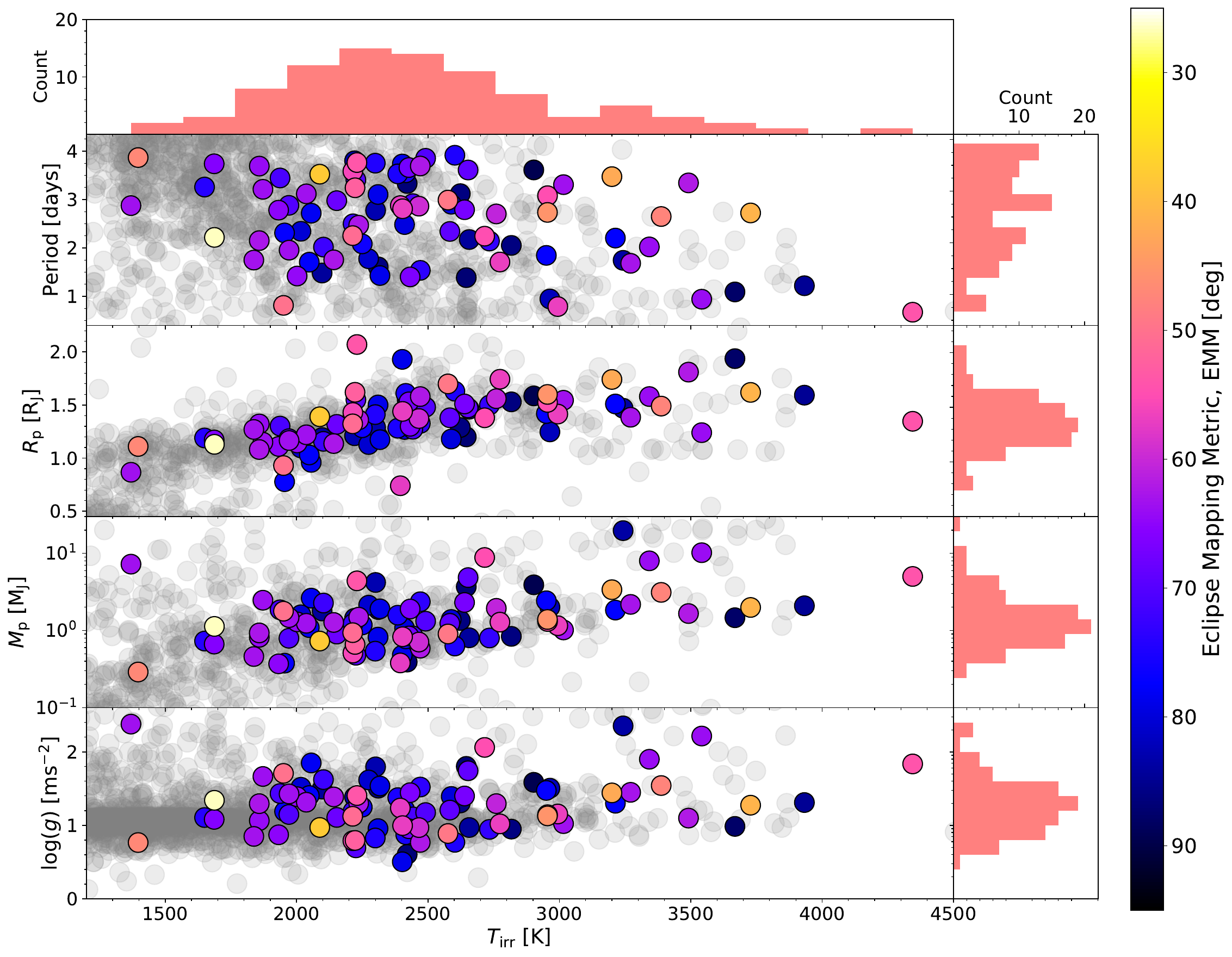}
    \caption{Population demographics of our 87 {selected} targets. We shade the targets according to their AIRS Ch1 EMM values, and show the total population in the background in grey. We plot the periods, radii, masses, {and gravities} of these targets against their irradiation temperatures. {A survey such as this would} well sample the parameter space of irradiated gas giants, enabling a comprehensive assessment of their thermal structures and atmospheric dynamics.}
    \label{fig:survey_total_param_space}
\end{figure*}

Figure \ref{fig:survey_total_param_space} shows the demographics of our {suggested} survey targets compared to the total population, in particular the periods, masses, radii, {gravities}, and irradiation temperatures that they span. Here, we explore the five parameters that are expected to dominate atmospheric dynamics at first order \citep{roth2024}, and how this large-scale survey with Ariel would diagnose their exact influence.

\begin{enumerate}
    \item \textbf{Irradiation Temperature:} This is one of the dominant factors that controls atmospheric dynamics, affecting both the heat circulation regime and resultant shaping of the temperature profile through factors like the hotspot offset \citep{showman2020}. The irradiated gas giant population transitions from a rotationally-dominated regime with efficient recirculation at the cooler end of the spectrum (i.e., for warm Jupiters), where advective timescales win out over radiative, to a divergence-dominated, inefficient recirculation regime at the hotter end, where radiative timescales win out over advective (i.e., for ultra-hot Jupiters). Both of these effects result in the lack of a hotspot offset, due to the lack of a hotspot for the former and a localisation of the hotspot at the substellar point for the latter. In the regime between these two, there exists a complex, non-monotonic relationship between the irradiation temperature and hotspot offset due to the competing influence of both rotational and divergent components \citep{beatty2019}. Figure \ref{fig:survey_total_param_space} shows that our survey targets span a range of irradiation temperatures, with a median value of 2400\,K ($T_{\mathrm{eq}}=1700$\,K) between a range of 1370\,K and 4350\,K. Hence, with this survey, we would be able to robustly diagnose the impact of irradiation temperature on the atmospheric dynamics in the critical transition regime between advective- and radiative-dominated environments.
    
    \item \textbf{Period and Rotation Rate:} All of our targets of interest are on short orbits and are therefore expected to be both highly irradiated and tidally locked, often leading to the formation of a super-rotating equatorial jet. Whether this jet dominates the circulation regime is influenced by the rotation rate \citep{kataria2016}. Faster rotation rates can seek to narrow jet widths, forming decoupled latitudinal banding in the atmosphere akin to the solar system gas giants, and lead to the formation of additional high-latitude jets. In this regime, latitudinal recirculation over the poles is expected to become the dominant circulation regime, with expected shallow longitudinal temperature gradients. Conversely, slower rotation rates lead to a single wide equatorial jet dominating the circulation regime, resulting in shallow latitudinal temperature gradients and heat primarily being recirculated around the equator \citep{showman2015}. The exact quantitative balance and transition between these regimes as a function of rotational and orbital period have not been observationally benchmarked, and their effect on observables like the hotspot offset is expected to be non-trivial. Figure \ref{fig:survey_total_param_space} shows that with our proposed survey, which samples planets with periods between 0.7 and 3.9 days, such an analysis could be facilitated by measuring the longitudinal temperature gradients.
    
    As a dedicated exoplanet mission, Ariel also has the potential to test more extreme cases in this rotational period regime by observing planets with periods too long to be feasible with the likes of JWST. Beyond our discussed period cut-offs of 4.2 and 6.4 days, there are two longer period planets that meet the criterion of being mappable in at least 2/3 Ariel spectrographs (see Figure \ref{fig:survey_P-T_param_space}): these are TOI-677b and HD 221416b, with periods of 11.2 and 14.3 days, respectively. Including targets such as this in the sample would allow Ariel to push the boundaries of our data-driven knowledge of how rotational and orbital periods affect exoplanet atmospheric dynamics, including testing the tidal locking radius itself \citep[see][]{wakeford2025}.
    
    \item \textbf{Gravity:} We expect higher gravity planets to have weaker day-night recirculation regimes as a consequence of damped circulation patterns \citep{roth2024}, which would manifest itself as a smaller hotspot offset, as was observed for WASP-43b \citep{wasp43miri_eclipsemap, wasp43nirspec_eclipsemap}. However, there is no general consensus on the magnitude or dominance of this effect. Mapping and comparing planets with similar characteristics, but different gravities, would therefore allow us to diagnose the quantitative impact of gravity on the atmospheric dynamics.
    Gravity also contributes in setting the opacity and pressure level of the photosphere, which influences the balance between radiative and advective timescales, and therefore the nature of the circulation regime \citep{showman2020}.
    
    Figure \ref{fig:survey_total_param_space} shows that our proposed survey spans planetary radii of $0.74-2.07 \ \mathrm{R_J}$ (median=1.36 $\mathrm{R_J}$) and planetary masses of 0.3--19.7 $\mathrm{M_J}$ (median=1.31 $\mathrm{M_J}$), and therefore also probes a wide range of surface gravities, {from log$(g)$=0.5--2.4 $\mathrm{ms}^{-2}$} (median=1.3 $\mathrm{ms}^{-2}$), including targets in the brown dwarf mass regime. These can effectively be treated as high-mass hot Jupiters, allowing Ariel to probe the extreme effects of gravity on the atmospheric dynamics. We note that mass does not factor into our EMM calculations, and therefore the upper mass of our survey is largely a consequence of the artificial 30 $\mathrm{M_J}$ cut-off of the NASA Exoplanet Archive \citep{nasa_exoplanet_archive}, from which the MCS is derived. Higher-mass brown dwarfs may therefore also be mappable; such work is being explored in D. Lewis et al. (in prep).
        
    \item \textbf{Metallicity:} We have thus far discussed the impact of system parameters on the atmospheric dynamics. Such parameters tend to be known a priori for most targets. However, atmospheric parameters themselves, which are not often known a priori, are also expected to have first-order effects on the atmospheric dynamics. Similar to gravity, the metallicity also contributes in setting the photospheric pressure and therefore circulation regime, with metallicity and hotspot offsets expected to be inversely related \citep{roth2024}. Similarly, enhanced metallicities seek to reduce atmospheric flow and jet speeds, resulting in more inefficient day-night heat recirculation. Hence, knowledge of the atmospheric metallicity is imperative for interpreting the atmospheric dynamics, and may inhibit definitive constraints if not accounted for \citep{wasp17eclipsemap}.
    
    Ariel's chemical census will be particularly conducive to such an assessment. The use of phase curves to perform these measurements is also beneficial since the shape of the phase curve can itself be used to measure the metallicity, with metallicity expected to be proportional to the phase curve amplitude and inversely proportional to the phase curve offset 
    \citep{parmentier_crossfield}. With such a large dataset, these measurements can also be used in a data-model feedback loop to qualitatively benchmark the assumptions made in deriving these quantitative proportionalities. Atmospheric metallicity is expected to be somewhat correlated with mass \citep{welbanks2019}, and since our survey samples a wide mass range of 0.3--19.7 $\mathrm{M_J}$ (Figure \ref{fig:survey_total_param_space}), we expect to also survey a wide range of metallicities. Hence, this large-scale phase curve survey with Ariel would provide the most precise and comprehensive diagnostic assessment of the effect of metallicity on the atmospheric dynamics to date.
        
    \item \textbf{Atmospheric Chemistry and Clouds:} Beyond the generalised metallicity, individual chemical species can also carry first-order implications on the atmospheric dynamics. The presence of strong optical absorbers like TiO and VO in the photosphere can lead to thermal inversions, which may alter the thermal structure \citep{roth2024}. Condensates in the forms of clouds are already known to have a significant impact on {flux} maps through their modulation of the spatial thermal emission profile \citep{parmentier2016, beatty2019}{; such effects have already been observed for JWST eclipse maps \citep{bell2024, wasp43miri_eclipsemap}}. Ariel's photometric filters will allow the presence of these clouds to be identified, and their spatial distribution potentially mapped using our survey design. Synergistic observations with JWST MIRI/LRS may also allow the vibrational mode of the cloud to be measured for even better characterisation \citep{wakeford_sing2015, grant2023, inglis2024}. Our proposed survey targets span a large temperature range over which we expect to see significant transitions in the cloud species and properties \citep{gao2021}, enabling Ariel to comprehensively diagnose the impact of clouds on the atmospheric dynamics.
        
\end{enumerate}

In summary, a {survey such as this with Ariel would} well sample the parameter space of irradiated gas giants, which ensures the diversity needed in order to derive statistically significant population-level trends of their atmospheric dynamics. Crucially, it is evident from the above descriptions of these parameters that they often have degenerate impacts on the atmospheric dynamics. Inspection of the Rossby number equation, for example, shows that as far as atmospheric dynamics are concerned, decreasing the planetary radius and increasing the rotation rate have the same effect \citep{lewis_and_hammond}. Similarly, whilst higher irradiation temperature can result in smaller hotspot offsets, so too can higher metallicities, gravity, and even morning limb clouds \citep{showman2020}. These degeneracies are why the population-level trends of exoplanet atmospheric dynamics have to date been unable to be constrained using the data quantity available, and why quantitative target-by-target comparisons {often suffer from attribution degeneracies} since an ensemble of different parameter combinations can produce the same atmospheric dynamics \citep{wasp17eclipsemap}.

{This survey} with Ariel would constitute the largest assessment of exoplanet atmospheric dynamics to date, superseding the previous largest attempt using 29 Spitzer 4.5 $\upmu$m phase curves \citep{dang2025} by a factor of three. That survey found indications of population-level trends, but could not draw statistically significant conclusions due to both the small survey size, which inhibited their ability to control for all the aforementioned degenerate parameters, and problems with lack of uniformity between the datasets and analyses. Our three-times larger Ariel survey would allow for adequate sample sizes whilst still being able to control for all these parameters, whilst our uniform observing strategy would enable holistic interpretation. As such, this survey would revolutionise our understanding of exoplanet atmospheric dynamics and provide the most comprehensive data-driven benchmarking of model predictions ever performed.

\section{Conclusions}
\label{sec:conclusions}
Multidimensional characterisation of exoplanet atmospheres is rising in applicability and wide-spread application with the advent of new techniques and increase in data quality \citep[e.g.,][]{catwoman, theresa, harmonica, wasp18eclipsemap, wasp43miri_eclipsemap, murphy2024, murphy2025, espinoza2024, wasp43nirspec_eclipsemap, wasp17eclipsemap, lally2025, mukherjee2025}. Eclipse mapping is one such technique which can be used to map the dayside spatial emission profiles of transiting exoplanets, from which we can characterise their 3D (longitude-latitude-pressure) thermal patterns and constrain their atmospheric dynamics \citep{rauscher2007}.

JWST is currently spearheading the application of this technique, but as an oversubscribed general observatory, its ability to do this on a population-level is time-limited. Hence, we assessed the eclipse mapping potential of the upcoming ESA Ariel mission \citep{ariel_citation}. Ariel is a dedicated exoplanet mission set to observe $\sim$1000 transiting exoplanets over a four-year mission lifetime, making it a perfect candidate to facilitate population-level studies.

To assess Ariel's eclipse mapping potential, we began by taking existing JWST eclipse maps (WASP-18b, \citealt{wasp18eclipsemap}; WASP-43b, \citealt{wasp43miri_eclipsemap} and \citealt{wasp43nirspec_eclipsemap}; WASP-17b, \citealt{wasp17eclipsemap}; and HD 189733b, \citealt{lally2025}), along with a GCM output of HD 209458b for completeness of the canonical hot Jupiters \citep{hd209gcm}, and used them as test cases. By post-processing these maps into simulated Ariel light curves and retrieving on them, we benchmarked the data quantity needed for Ariel to derive qualitatively comparable maps to JWST, and compared the quantitative constraints achieved with each observatory. We outline the results of our test cases below.

\begin{itemize}
    \item HD 189733b and HD 209458b are two of the highest-ranking eclipse mapping targets \citep{boone2023}. We found that Ariel will be able to recover the same qualitative map as JWST using the same amount of data for both of these bright targets, while also expanding the wavelength coverage over which they are mappable. 

    \item WASP-18b ranks as 22nd best eclipse mapping target for JWST \citep{boone2023}. We found that Ariel will be able to map it using three eclipse observations, compared to one for JWST. Hence, we determined that Ariel will be capable of mapping all of JWST's highest-ranking targets using a feasible number of observations.
    
   \item WASP-17b is the lowest-ranking eclipse mapping target in our sample. We found that while Ariel will not be able to eclipse map it with adequate quantitative constraints using a reasonable number of repeated eclipse observations ($\leq$20), the use of single phase curve observations can map lower-ranking targets in a more time-efficient way. Hence, for Ariel to obtain a uniform dataset of all of these best mapping targets, phase curves are the most advantageous strategy in order to facilitate a population-level study of adequate size.
   
    \item WASP-43b is the only planet in our sample with eclipse maps constrained in latitude as well as longitude. We found that due to the small amplitude and short duration of latitudinal signals, smaller observatories like Ariel will not have the ability to robustly measure them in eclipse maps using a reasonable amount of data, requiring at minimum ten phase curves for this test case.
    
\end{itemize}

From these test cases, we determined that the most advantageous focus of a mapping survey with Ariel would be in constraining the longitudinal profiles of a large population of planets using a mixture of phase and eclipse mapping applied to full phase curves. From this, we would be able to characterise the first-order atmospheric dynamics of these targets and diagnose population-level trends.

In order to determine which targets {would be best suited to} such a survey, we used the methods of \citet{boone2023} to rank all the best mapping targets for Ariel. We determined that there are 138 unique targets that are mappable with Ariel using our proposed observing strategy. By applying selection criteria, we devised a survey of 87 targets with $P\leq4.2$ days that are mappable in at least 2/3 of the spectrographs, for which full phase curves would be observable using only a quarter of the Ariel mission lifetime \citep{ariel_target_list}.

{These survey targets span} a range of irradiation temperatures ($1370-4350$ K), planetary radii ($0.74-2.07 \ \mathrm{R_J}$), masses ($0.3-19.7 \ \mathrm{M_J}$){, and gravities (log$(g)=0.5-2.4$ ms$^{-2}$)}, {which would enable us to both sample and} control for the number of degenerate parameters that influence atmospheric dynamics at first-order. {These} 87 targets {would} supersede the previous largest survey of exoplanet atmospheric dynamics with Spitzer using 29 phase curves \citep{dang2025} by a factor of three. Hence, this survey would enable the most wide-ranging and comprehensive assessment of the population-level trends of transiting exoplanet atmospheric dynamics to date.

\section*{Acknowledgements}

We thank the anonymous referee for their helpful and constructive review. D.V acknowledges support from the ESA Archival Research Visitor Programme\footnote{\url{https://www.cosmos.esa.int/web/esdc/visitor-programme}}, which is designed to increase the scientific return from ESA space science missions by twice a year funding six archival research projects. D.V. also acknowledges funding STFC grant ST/X508263/1 and the University of Bristol School of Physics HH Potter PhD Scholarship Fund. H.R.W. was funded by UK Research and Innovation (UKRI) framework under the UK government’s Horizon Europe funding guarantee for an ERC Starter Grant [grant number EP/Y006313/1]. {D.V. also acknowledges the input and encouragement of the Ariel Science
Team and Ariel Consortium Phase Curve Working Group}.

%%%%%%%%%%%%%%%%%%%%%%%%%%%%%%%%%%%%%%%%%%%%%%%%%%
\section*{Data Availability}
The simulation code and data products are available at \href{https://doi.org/10.5281/zenodo.17245372}{10.5281/zenodo.17245372}, in addition to machine-readable versions of the full EMM rankings for JWST and Ariel, and the proposed survey targets.

%%%%%%%%%%%%%%%%%%%% REFERENCES %%%%%%%%%%%%%%%%%%

% The best way to enter references is to use BibTeX:

\bibliographystyle{mnras}
\bibliography{ARIEL_EMM} % if your bibtex file is called example.bib

% Alternatively you could enter them by hand, like this:
% This method is tedious and prone to error if you have lots of references
%\begin{thebibliography}{99}
%\bibitem[\protect\citeauthoryear{Author}{2012}]{Author2012}
%Author A.~N., 2013, Journal of Improbable Astronomy, 1, 1
%\bibitem[\protect\citeauthoryear{Others}{2013}]{Others2013}
%Others S., 2012, Journal of Interesting Stuff, 17, 198
%\end{thebibliography}

%%%%%%%%%%%%%%%%%%%%%%%%%%%%%%%%%%%%%%%%%%%%%%%%%%

%%%%%%%%%%%%%%%%% APPENDICES %%%%%%%%%%%%%%%%%%%%%

\appendix

\section{Best Eclipse Mapping Targets for Ariel}

Here, we tabulate the best eclipse mapping targets for the three Ariel spectrographs (NIRSpec, Table \ref{tab:emm_list_nirspec}; AIRS Ch0, Table \ref{tab:emm_list_airs_ch0}; and AIRS Ch1, Table \ref{tab:emm_list_airs_ch1}) as identified by their analytically-calculated eclipse mapping metrics \citep[EMMs,][]{boone2023}. We define a target to be ``eclipse mappable'' if it has an EMM$<90^{\circ}$ and $N_{\max} \geq 1$. We list the targets in order of their overall EMM.

\begin{table*}
	\centering
	\caption{Best eclipse mapping targets for Ariel NIRSpec. These have $\mathrm{EMM}<90^{\circ}$ and $N_{\max} \geq 1$ for this spectrograph. We list these targets in order of their overall EMM ranking, include relevant system parameters, along with our calculated EMM$_x$, EMM$_x$, and $N_{\max}$ values (see text for details).}
	\label{tab:emm_list_nirspec}
\begin{tabular}{lccccccccccc} % four columns, alignment for each
		\hline
        Planet Name & {Rank} & $T_{\mathrm{eq}}$ [K] & $R_{\mathrm{p}} \ [\mathrm{R}_{\mathrm{J}}]$ & $M_{\mathrm{p}} \ [\mathrm{M}_{\mathrm{J}}]$ & $P$ [days] & $b$ & $K$ (mag) & $N_{\mathrm{max}}$ & $\mathrm{EMM}_x$ [deg] & $\mathrm{EMM}_y$ [deg] & $\mathrm{EMM}$ [deg]\\
        \hline

    TOI-2109b & 1 & 3072 & 1.35 & 5.02 & 0.67 & 0.75 & 9.07 & 3.0 & 32.3 & 28.7 & 43.2\\
    
    WASP-189b & 2 & 2636 & 1.62 & 1.99 & 2.72 & 0.48 & 6.06 & 3.4 & 22.0 & 40.3 & 45.9\\
    
    KELT-20b & 3 & 2263 & 1.74 & 3.38 & 3.47 & 0.5 & 7.42 & 3.0 & 24.7 & 42.4 & 49.1\\
    
    KELT-7b & 4 & 2089 & 1.6 & 1.39 & 2.73 & 0.57 & 7.54 & 2.6 & 29.4 & 42.4 & 51.6\\
    
    TOI-1431b & 5 & 2395 & 1.49 & 3.12 & 2.65 & 0.88 & 7.44 & 2.9 & 46.7 & 24.8 & 52.8\\
    
    WASP-19b & 6 & 2117 & 1.42 & 1.15 & 0.79 & 0.67 & 10.48 & 2.2 & 36.3 & 40.7 & 54.5\\
    
    KELT-4Ab & 7 & 1822 & 1.7 & 0.9 & 2.99 & 0.69 & 8.69 & 2.1 & 39.5 & 41.3 & 57.1\\
\hline
	\end{tabular}
\end{table*}

\setcounter{table}{0} 
\begin{table*}
	\centering
	\caption{Best eclipse mapping targets for Ariel NIRSpec - Continued.}
	\label{tab:emm_list_nirspec}
	\begin{tabular}{lccccccccccc} % four columns, alignment for each
		\hline
        Planet Name & {Rank} & $T_{\mathrm{eq}}$ [K] & $R_{\mathrm{p}} \ [\mathrm{R}_{\mathrm{J}}]$ & $M_{\mathrm{p}} \ [\mathrm{M}_{\mathrm{J}}]$ & $P$ [days] & $b$ & $K$ (mag) & $N_{\mathrm{max}}$ & $\mathrm{EMM}_x$ [deg] & $\mathrm{EMM}_y$ [deg] & $\mathrm{EMM}$ [deg]\\
        \hline
        
        Qatar-1b & 64 & 1416 & 1.14 & 1.29 & 1.42 & 0.65 & 10.41 & 1.2 & 57.5 & 67.8 & 88.9\\
        
        WASP-20b & 65 & 1378 & 1.46 & 0.31 & 4.9 & 0.72 & 9.39 & 1.2 & 64.1 & 62.0 & 89.1\\
        
        WASP-50b & 66& 1394 & 1.17 & 1.47 & 1.96 & 0.67 & 9.97 & 1.2 & 60.1 & 66.6 & 89.7\\

    \hline
	\end{tabular}
\end{table*}

\begin{table*}
	\centering
	\caption{Best eclipse mapping targets for Ariel AIRS Ch0. These have $\mathrm{EMM}<90^{\circ}$ and $N_{\max} \geq 1$ for this spectrograph. We list these targets in order of their overall EMM ranking, include relevant system parameters, along with our calculated EMM$_x$, EMM$_x$, and $N_{\max}$ values (see text for details).}
	\label{tab:emm_list_airs_ch0}
	\begin{tabular}{lccccccccccc} % four columns, alignment for each
		\hline
        Planet Name & {Rank} & $T_{\mathrm{eq}}$ [K] & $R_{\mathrm{p}} \ [\mathrm{R}_{\mathrm{J}}]$ & $M_{\mathrm{p}} \ [\mathrm{M}_{\mathrm{J}}]$ & $P$ [days] & $b$ & $K$ (mag) & $N_{\mathrm{max}}$ & $\mathrm{EMM}_x$ [deg] & $\mathrm{EMM}_y$ [deg] & $\mathrm{EMM}$ [deg]\\
        \hline

        HD189733b & 1 & 1193 & 1.13 & 1.13 & 2.22 & 0.67 & 5.54 & 4.1 & 21.6 & 23.9 & 32.2\\

        WASP-189b & 2 & 2636 & 1.62 & 1.99 & 2.72 & 0.48 & 6.06 & 4.0 & 19.0 & 34.8 & 39.6\\
        
        KELT-20b & 3 & 2263 & 1.74 & 3.38 & 3.47 & 0.5 & 7.42 & 3.6 & 21.2 & 36.4 & 42.1\\
        
        HD209458b & 4 & 1477 & 1.39 & 0.73 & 3.52 & 0.49 & 6.31 & 3.6 & 20.6 & 36.9 & 42.2\\
        
        KELT-7b & 5 & 2089 & 1.6 & 1.39 & 2.73 & 0.57 & 7.54 & 3.0 & 25.6 & 36.8 & 44.8\\
        
        TOI-1431b & 6 & 2395 & 1.49 & 3.12 & 2.65 & 0.88 & 7.44 & 3.3 & 41.2 & 21.9 & 46.7\\
        
        TOI-2109b & 7 & 3072 & 1.35 & 5.02 & 0.67 & 0.75 & 9.07 & 2.5 & 36.8 & 32.7 & 49.3\\
\hline
	\end{tabular}
\end{table*}

\setcounter{table}{1} 
\begin{table*}
	\centering
	\caption{Best eclipse mapping targets for AIRS Ch0 - Continued.}
	\label{tab:emm_list}
	\begin{tabular}{lccccccccccc} % four columns, alignment for each
    	\hline
        Planet Name & {Rank} & $T_{\mathrm{eq}}$ [K] & $R_{\mathrm{p}} \ [\mathrm{R}_{\mathrm{J}}]$ & $M_{\mathrm{p}} \ [\mathrm{M}_{\mathrm{J}}]$ & $P$ [days] & $b$ & $K$ (mag) & $N_{\mathrm{max}}$ & $\mathrm{EMM}_x$ [deg] & $\mathrm{EMM}_y$ [deg] & $\mathrm{EMM}$ [deg]\\
        \hline

        HAT-P-7b & 64 & 2272 & 1.51 & 1.84 & 2.2 & 0.47 & 9.33 & 1.9 & 35.2 & 65.3 & 74.2\\

        HATS-56b & 65 & 1901 & 1.69 & 0.6 & 4.32 & 0.69 & 10.25 & 1.5 & 51.4 & 53.9 & 74.5\\
        
        WASP-3b & 66 & 2086 & 1.42 & 2.43 & 1.85 & 0.41 & 9.36 & 2.2 & 30.5 & 68.9 & 75.3\\
        
        WASP-90b & 67 & 1840 & 1.63 & 0.63 & 3.92 & 0.84 & 10.25 & 1.7 & 63.4 & 41.1 & 75.5\\
        
        WASP-119b & 68 & 1566 & 1.4 & 1.23 & 2.5 & 0.57 & 10.54 & 1.6 & 43.1 & 62.6 & 76.0\\
        
        TOI-3331Ab & 69 & 1487 & 1.16 & 2.27 & 2.02 & 0.57 & 9.91 & 1.6 & 43.8 & 63.0 & 76.7\\
        
        TOI-905b & 70 & 1192 & 1.17 & 0.67 & 3.74 & 0.82 & 9.45 & 1.6 & 63.3 & 43.9 & 77.1\\
\hline
	\end{tabular}
\end{table*}

\begin{table*}
	\centering
	\caption{Best eclipse mapping targets for Ariel AIRS Ch1. These have $\mathrm{EMM}<90^{\circ}$ and $N_{\max} \geq 1$ for this spectrograph. We list these targets in order of their overall EMM ranking, include relevant system parameters, along with our calculated EMM$_x$, EMM$_x$, and $N_{\max}$ values (see text for details).}
	\label{tab:emm_list_airs_ch1}
	\begin{tabular}{lccccccccccc} % four columns, alignment for each
		\hline
        Planet Name & {Rank} & $T_{\mathrm{eq}}$ [K] & $R_{\mathrm{p}} \ [\mathrm{R}_{\mathrm{J}}]$ & $M_{\mathrm{p}} \ [\mathrm{M}_{\mathrm{J}}]$ & $P$ [days] & $b$ & $K$ (mag) & $N_{\mathrm{max}}$ & $\mathrm{EMM}_x$ [deg] & $\mathrm{EMM}_y$ [deg] & $\mathrm{EMM}$ [deg]\\
		\hline

        HD189733b & 1 & 1193 & 1.13 & 1.13 & 2.22 & 0.67 & 5.54 & 5.2 & 17.7 & 19.6 & 26.4\\

        HD209458b & 2 & 1477 & 1.39 & 0.73 & 3.52 & 0.49 & 6.31 & 4.1 & 18.6 & 33.2 & 38.1\\
        
        WASP-189b & 3 & 2636 & 1.62 & 1.99 & 2.72 & 0.48 & 6.06 & 3.8 & 19.6 & 35.9 & 40.9\\
        
        KELT-20b & 4 & 2263 & 1.74 & 3.38 & 3.47 & 0.5 & 7.42 & 3.5 & 21.3 & 36.6 & 42.3\\
        
        KELT-7b & 5 & 2089 & 1.6 & 1.39 & 2.73 & 0.57 & 7.54 & 3.0 & 25.8 & 37.2 & 45.2\\
        
        WASP-69b & 6 & 988 & 1.11 & 0.29 & 3.87 & 0.65 & 7.46 & 2.7 & 30.5 & 35.7 & 46.9\\
        
        TOI-1431b & 7 & 2395 & 1.49 & 3.12 & 2.65 & 0.88 & 7.44 & 3.3 & 42.0 & 22.3 & 47.5\\
\hline
	\end{tabular}
\end{table*}

\setcounter{table}{2}
\begin{table*}
	\centering
	\caption{Best eclipse mapping targets for AIRS Ch1 - Continued.}
	\label{tab:emm_list}
	\begin{tabular}{lccccccccccc} % four columns, alignment for each
    	\hline
        Planet Name & {Rank} & $T_{\mathrm{eq}}$ [K] & $R_{\mathrm{p}} \ [\mathrm{R}_{\mathrm{J}}]$ & $M_{\mathrm{p}} \ [\mathrm{M}_{\mathrm{J}}]$ & $P$ [days] & $b$ & $K$ (mag) & $N_{\mathrm{max}}$ & $\mathrm{EMM}_x$ [deg] & $\mathrm{EMM}_y$ [deg] & $\mathrm{EMM}$ [deg]\\
        \hline

        HAT-P-6b & 64 & 1761 & 1.48 & 1.32 & 3.85 & 0.54 & 9.31 & 1.8 & 38.3 & 59.0 & 70.3\\
        
        HATS-1b & 65 & 1370 & 1.3 & 1.86 & 3.45 & 0.7 & 10.58 & 1.6 & 50.0 & 50.4 & 71.0\\
        
        HAT-P-13b & 66 & 1727 & 1.27 & 0.85 & 2.92 & 0.62 & 8.98 & 1.7 & 43.8 & 56.0 & 71.1\\
        
        TOI-1811b & 67 & 962 & 0.99 & 0.97 & 3.71 & 0.75 & 9.64 & 1.6 & 53.8 & 47.0 & 71.4\\
        
        TOI-622b & 68 & 1382 & 0.82 & 0.3 & 6.4 & 0.63 & 7.86 & 1.6 & 45.3 & 55.5 & 71.7\\
        
        HD221416b & 69 & 1198 & 0.84 & 0.19 & 14.28 & 0.67 & 6.04 & 1.6 & 48.0 & 53.8 & 72.1\\
        
        TOI-3331Ab & 70 & 1487 & 1.16 & 2.27 & 2.02 & 0.57 & 9.91 & 1.7 & 41.2 & 59.3 & 72.2\\
\hline
	\end{tabular}
\end{table*}

\setcounter{table}{2}
\begin{table*}
	\centering
	\caption{Best eclipse mapping targets for AIRS Ch1 - Continued.}
	\label{tab:emm_list}
	\begin{tabular}{lccccccccccc} % four columns, alignment for each
		\hline
        Planet Name & {Rank} & $T_{\mathrm{eq}}$ [K] & $R_{\mathrm{p}} \ [\mathrm{R}_{\mathrm{J}}]$ & $M_{\mathrm{p}} \ [\mathrm{M}_{\mathrm{J}}]$ & $P$ [days] & $b$ & $K$ (mag) & $N_{\mathrm{max}}$ & $\mathrm{EMM}_x$ [deg] & $\mathrm{EMM}_y$ [deg] & $\mathrm{EMM}$ [deg]\\
		\hline

        WASP-173Ab & 127 & 1871 & 1.2 & 3.69 & 1.39 & 0.4 & 10.0 & 1.8 & 35.0 & 79.7 & 87.0\\

        Qatar-2b & 128 & 1348 & 1.25 & 2.49 & 1.34 & 0.4 & 10.62 & 1.8 & 35.2 & 80.3 & 87.7\\
        
        TOI-3235b & 129 & 604 & 1.02 & 0.66 & 2.59 & 0.51 & 10.82 & 1.4 & 44.9 & 75.5 & 87.8\\
        
        WASP-12b & 130 & 2594 & 1.94 & 1.46 & 1.09 & 0.34 & 10.19 & 2.2 & 29.8 & 82.7 & 87.9\\
        
        TOI-3321b & 131 & 1620 & 1.39 & 0.55 & 3.65 & 0.42 & 9.67 & 1.7 & 37.1 & 80.8 & 88.9\\
        
        HAT-P-3b & 132 & 1186 & 0.94 & 0.65 & 2.9 & 0.46 & 9.45 & 1.5 & 41.0 & 79.0 & 89.0\\
        
        TOI-2587Ab & 133 & 1450 & 1.08 & 0.22 & 5.46 & 0.71 & 9.99 & 1.2 & 63.6 & 62.3 & 89.0\\
        
        NGTS-5b & 134 & 1058 & 1.14 & 0.23 & 3.36 & 0.66 & 11.61 & 1.2 & 58.8 & 67.2 & 89.3\\
        
        KELT-21b & 135 & 2053 & 1.59 & 3.91 & 3.61 & 0.42 & 10.09 & 1.7 & 37.9 & 81.1 & 89.5\\
        
        K2-34b & 136 & 1686 & 1.23 & 1.7 & 3.0 & 0.91 & 10.19 & 1.7 & 81.4 & 38.2 & 89.9\\

    \hline
	\end{tabular}
\end{table*}

\section{Best Eclipse Mapping Targets for JWST}

Here, we tabulate the best eclipse mapping targets for three JWST spectrographs (NIRISS/SOSS, Table \ref{tab:emm_list_soss}; NIRSpec/G395H, Table \ref{tab:emm_list_g395h}; and MIRI/LRS, Table \ref{tab:emm_list_miri}) as identified by their analytically-calculated eclipse mapping metrics \citep[EMMs,][]{boone2023}. We define a target to be ``eclipse mappable'' if it has an EMM$<90^{\circ}$ and $N_{\max} \geq 1$. We list the targets in order of their overall EMM. These tables are expanded versions of those presented in \citet{boone2023}, who present the top 15 targets for each of these spectrographs. We expand on them here to encourage future eclipse mapping work with JWST of a greater number of targets.

\begin{figure*}
    \centering
    \includegraphics[width=0.33\linewidth]{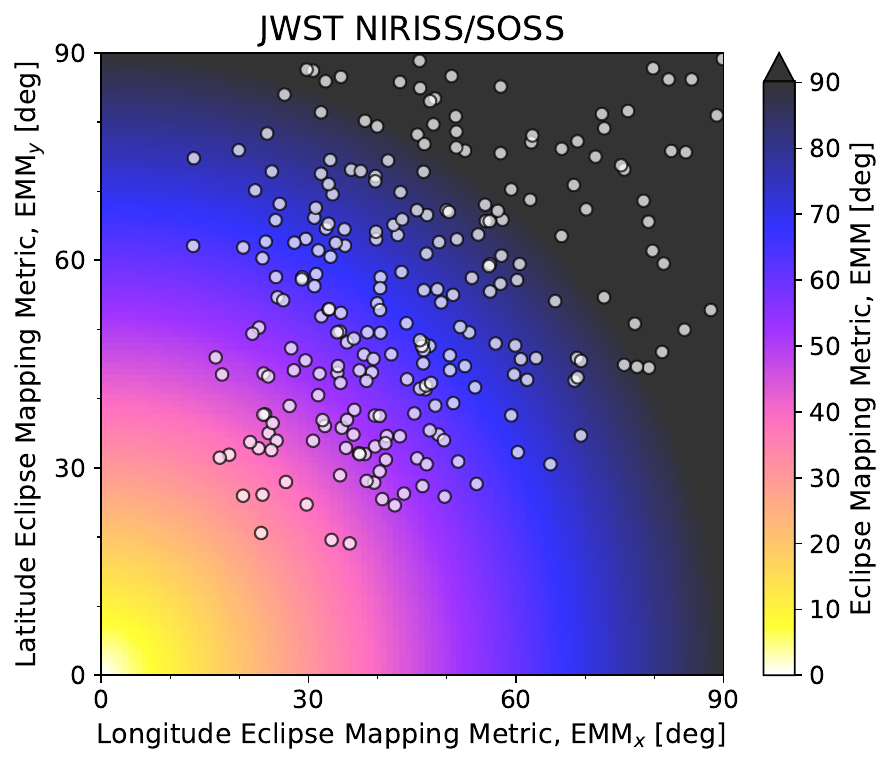}
    \includegraphics[width=0.33\linewidth]{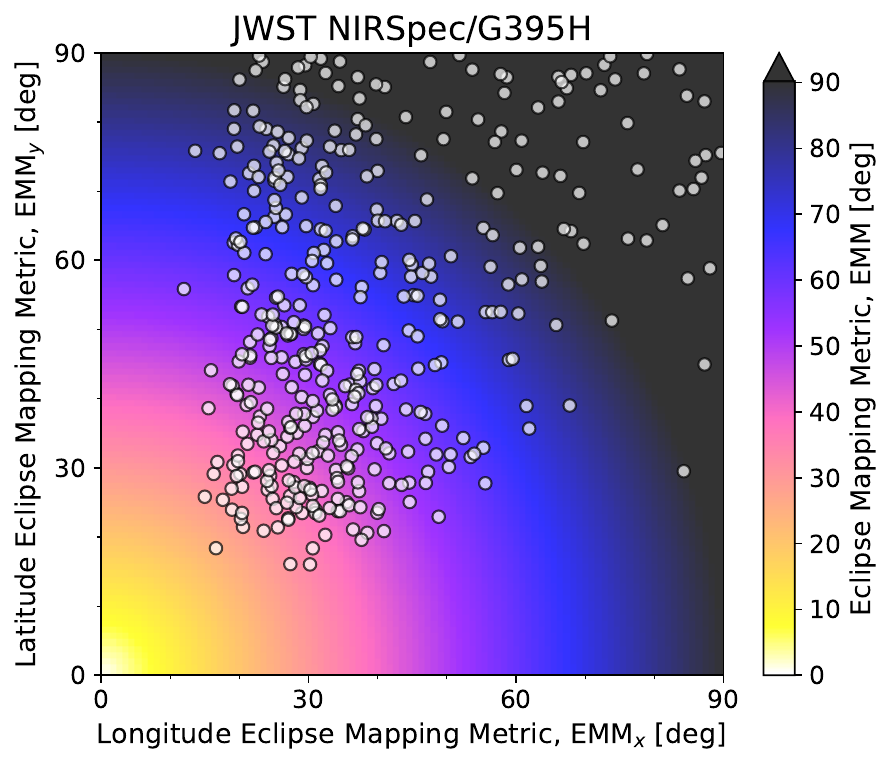}
    \includegraphics[width=0.33\linewidth]{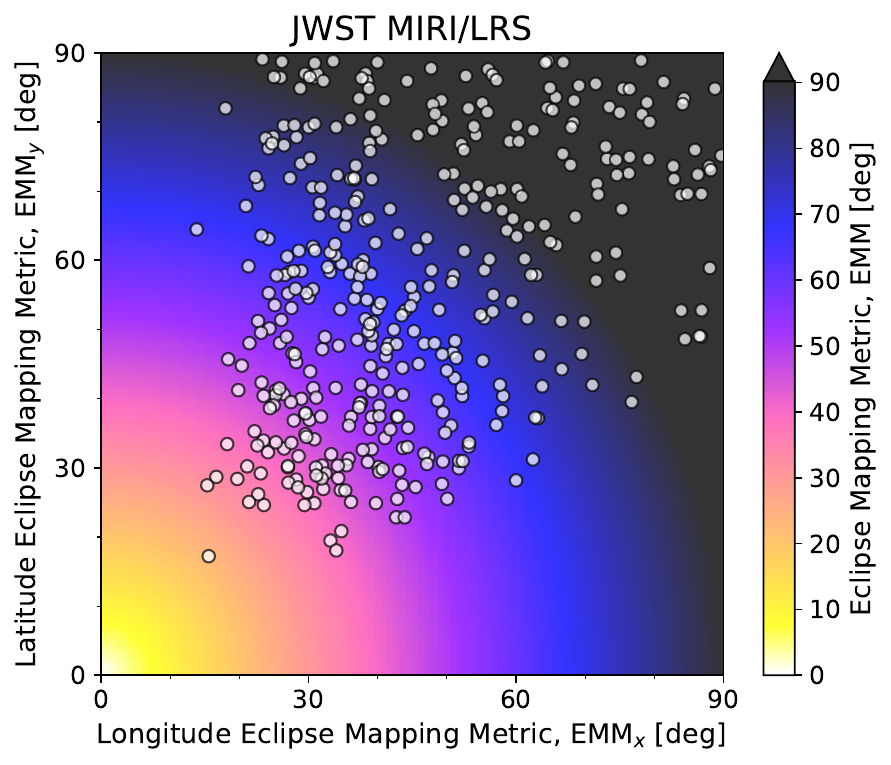}
    \caption{The eclipse mapping metric (EMM) parameter space for three JWST spectrographs: (left) NIRISS/SOSS, (middle) NIRSpec/G395H, and (right) MIRI/LRS. Targets increase in ranking going to the bottom left of the plots.}
    \label{fig:emm_plots_jwst}
\end{figure*}

\begin{table*}
	\centering
	\caption{100 best eclipse mapping targets for JWST NIRISS/SOSS. We list these targets in order of their overall EMM ranking, include relevant system parameters, along with our calculated EMM$_x$, EMM$_x$, and $N_{\max}$ values (see text for details). The full list of best eclipse mapping targets {for each instrument} can be found on Zenodo at \href{https://doi.org/10.5281/zenodo.17245372}{10.5281/zenodo.17245372}.}
	\label{tab:emm_list_soss}
\begin{tabular}{lccccccccccc} % four columns, alignment for each
		\hline
        Planet Name & {Rank} & $T_{\mathrm{eq}}$ [K] & $R_{\mathrm{p}} \ [\mathrm{R}_{\mathrm{J}}]$ & $M_{\mathrm{p}} \ [\mathrm{M}_{\mathrm{J}}]$ & $P$ [days] & $b$ & $K$ (mag) & $N_{\mathrm{max}}$ & $\mathrm{EMM}_x$ [deg] & $\mathrm{EMM}_y$ [deg] & $\mathrm{EMM}$ [deg]\\
		\hline

        TOI-2109 b & 1 & 3072 & 1.35 & 5.02 & 0.67 & 0.75 & 9.07 & 4.3 & 23.2 & 20.6 & 31.0\\

        HAT-P-70 b & 2 & 2552 & 1.87 & 6.78 & 2.74 & 0.62 & 8.96 & 4.1 & 20.5 & 26.0 & 33.1\\
        
        WASP-19 b & 3 & 2117 & 1.42 & 1.15 & 0.79 & 0.67 & 10.48 & 3.8 & 23.4 & 26.1 & 35.0\\
        
        WASP-189 b & 4 & 2636 & 1.62 & 1.99 & 2.72 & 0.48 & 6.06 & 4.4 & 17.2 & 31.5 & 35.8\\
        
        KELT-20 b & 5 & 2263 & 1.74 & 3.38 & 3.47 & 0.5 & 7.42 & 4.1 & 18.5 & 31.9 & 36.9\\
        
        WASP-167 b & 6 & 2363 & 1.58 & 8.0 & 2.02 & 0.77 & 9.76 & 3.4 & 29.7 & 24.7 & 38.6\\
        
        KELT-14 b & 7 & 1961 & 1.74 & 1.28 & 1.71 & 0.86 & 9.42 & 3.9 & 33.3 & 19.6 & 38.7\\
\hline
	\end{tabular}
\end{table*}

\setcounter{table}{0}
\begin{table*}
	\centering
	\caption{100 best eclipse mapping targets for JWST NIRISS/SOSS -- Continued.}
	% \label{tab:emm_list_soss}
	\begin{tabular}{lccccccccccc} % four columns, alignment for each
		\hline
        Planet Name & {Rank} & $T_{\mathrm{eq}}$ [K] & $R_{\mathrm{p}} \ [\mathrm{R}_{\mathrm{J}}]$ & $M_{\mathrm{p}} \ [\mathrm{M}_{\mathrm{J}}]$ & $P$ [days] & $b$ & $K$ (mag) & $N_{\mathrm{max}}$ & $\mathrm{EMM}_x$ [deg] & $\mathrm{EMM}_y$ [deg] & $\mathrm{EMM}$ [deg]\\
		\hline

        Qatar-1 b & 63 & 1416 & 1.14 & 1.29 & 1.42 & 0.65 & 10.41 & 2.1 & 37.4 & 44.1 & 57.8\\

        HATS-65 b & 64 & 1633 & 1.5 & 0.82 & 3.11 & 0.67 & 11.1 & 2.0 & 39.2 & 43.8 & 58.8\\
        
        TOI-2236 b & 65 & 1687 & 1.28 & 1.58 & 3.53 & 0.77 & 9.96 & 2.1 & 45.3 & 37.9 & 59.0\\
        
        TOI-157 b & 66 & 1591 & 1.29 & 1.18 & 2.08 & 0.8 & 10.89 & 2.1 & 47.5 & 35.4 & 59.3\\
        
        WASP-136 b & 67 & 1745 & 1.38 & 1.51 & 5.22 & 0.59 & 8.8 & 2.1 & 35.6 & 48.1 & 59.9\\
        
        CoRoT-11 b & 68 & 1741 & 1.43 & 2.33 & 2.99 & 0.81 & 11.25 & 2.1 & 48.8 & 34.8 & 59.9\\
        
        HAT-P-50 b & 69 & 1855 & 1.29 & 1.35 & 3.12 & 0.63 & 10.5 & 2.0 & 38.0 & 46.4 & 60.0\\
\hline
	\end{tabular}
\end{table*}

\begin{table*}
	\centering
	\caption{100 best eclipse mapping targets for JWST NIRSpec/G395H. We list these targets in order of their overall EMM ranking, include relevant system parameters, along with our calculated EMM$_x$, EMM$_x$, and $N_{\max}$ values (see text for details). The full list of best eclipse mapping targets {for each instrument} can be found on Zenodo at \href{https://doi.org/10.5281/zenodo.17245372}{10.5281/zenodo.17245372}.}
	\label{tab:emm_list_g395h}
	\begin{tabular}{lccccccccccc} % four columns, alignment for each
		\hline
        Planet Name & {Rank} & $T_{\mathrm{eq}}$ [K] & $R_{\mathrm{p}} \ [\mathrm{R}_{\mathrm{J}}]$ & $M_{\mathrm{p}} \ [\mathrm{M}_{\mathrm{J}}]$ & $P$ [days] & $b$ & $K$ (mag) & $N_{\mathrm{max}}$ & $\mathrm{EMM}_x$ [deg] & $\mathrm{EMM}_y$ [deg] & $\mathrm{EMM}$ [deg]\\
		\hline

        HD 189733 b & 1 & 1193 & 1.13 & 1.13 & 2.22 & 0.67 & 5.54 & 5.5 & 16.6 & 18.4 & 24.8\\

        KELT-4 A b & 2 & 1822 & 1.7 & 0.9 & 2.99 & 0.69 & 8.69 & 4.5 & 20.5 & 21.5 & 29.7\\
        
        KELT-20 b & 3 & 2263 & 1.74 & 3.38 & 3.47 & 0.5 & 7.42 & 5.2 & 15.0 & 25.8 & 29.9\\
        
        WASP-19 b & 4 & 2117 & 1.42 & 1.15 & 0.79 & 0.67 & 10.48 & 4.4 & 20.2 & 22.6 & 30.3\\
        
        HAT-P-70 b & 5 & 2552 & 1.87 & 6.78 & 2.74 & 0.62 & 8.96 & 4.5 & 18.9 & 24.0 & 30.5\\
        
        KELT-7 b & 6 & 2089 & 1.6 & 1.39 & 2.73 & 0.57 & 7.54 & 4.6 & 17.6 & 25.4 & 30.9\\
        
        WASP-43 b & 7 & 1379 & 0.93 & 1.78 & 0.81 & 0.66 & 9.27 & 4.3 & 20.4 & 23.5 & 31.1\\
\hline
	\end{tabular}
\end{table*}

\setcounter{table}{1}
\begin{table*}
	\centering
	\caption{100 best eclipse mapping targets for JWST NIRSpec/G395H -- Continued.}
	\begin{tabular}{lccccccccccc} % four columns, alignment for each
		\hline
        Planet Name & {Rank} & $T_{\mathrm{eq}}$ [K] & $R_{\mathrm{p}} \ [\mathrm{R}_{\mathrm{J}}]$ & $M_{\mathrm{p}} \ [\mathrm{M}_{\mathrm{J}}]$ & $P$ [days] & $b$ & $K$ (mag) & $N_{\mathrm{max}}$ & $\mathrm{EMM}_x$ [deg] & $\mathrm{EMM}_y$ [deg] & $\mathrm{EMM}$ [deg]\\
		\hline

        HAT-P-13 b & 63 & 1727 & 1.27 & 0.85 & 2.92 & 0.62 & 8.98 & 3.1 & 25.9 & 33.1 & 42.0\\

        TOI-1937 A b & 64 & 2096 & 1.25 & 2.01 & 0.95 & 0.87 & 11.23 & 3.6 & 36.4 & 21.1 & 42.1\\
        
        HATS-1 b & 65 & 1370 & 1.3 & 1.86 & 3.45 & 0.7 & 10.58 & 3.0 & 29.8 & 30.0 & 42.3\\
        
        TOI-640 b & 66 & 1747 & 1.77 & 0.88 & 5.0 & 0.89 & 9.24 & 3.8 & 37.7 & 19.6 & 42.5\\
        
        TOI-157 b & 67 & 1591 & 1.29 & 1.18 & 2.08 & 0.8 & 10.89 & 3.1 & 34.2 & 25.5 & 42.7\\
        
        WASP-119 b & 68 & 1566 & 1.4 & 1.23 & 2.5 & 0.57 & 10.54 & 3.2 & 24.3 & 35.3 & 42.8\\
        
        HATS-65 b & 69 & 1633 & 1.5 & 0.82 & 3.11 & 0.67 & 11.1 & 3.0 & 28.7 & 32.0 & 43.0\\
\hline
	\end{tabular}
\end{table*}

\begin{table*}
	\centering
	\caption{100 best eclipse mapping targets for JWST MIRI/LRS. We list these targets in order of their overall EMM ranking, include relevant system parameters, along with our calculated EMM$_x$, EMM$_x$, and $N_{\max}$ values (see text for details). The full list of best eclipse mapping targets {for each instrument} can be found on Zenodo at \href{https://doi.org/10.5281/zenodo.17245372}{10.5281/zenodo.17245372}.}
	\label{tab:emm_list_miri}
	\begin{tabular}{lccccccccccc} % four columns, alignment for each
		\hline
        Planet Name & {Rank} & $T_{\mathrm{eq}}$ [K] & $R_{\mathrm{p}} \ [\mathrm{R}_{\mathrm{J}}]$ & $M_{\mathrm{p}} \ [\mathrm{M}_{\mathrm{J}}]$ & $P$ [days] & $b$ & $K$ (mag) & $N_{\mathrm{max}}$ & $\mathrm{EMM}_x$ [deg] & $\mathrm{EMM}_y$ [deg] & $\mathrm{EMM}$ [deg]\\
		\hline

        HD 189733 b & 1 & 1193 & 1.13 & 1.13 & 2.22 & 0.67 & 5.54 & 5.9 & 15.6 & 17.2 & 23.2\\
        
        HD 209458 b & 2 & 1477 & 1.39 & 0.73 & 3.52 & 0.49 & 6.31 & 5.0 & 15.4 & 27.5 & 31.5\\
        
        WASP-69 b & 3 & 988 & 1.11 & 0.29 & 3.87 & 0.65 & 7.46 & 4.1 & 21.4 & 25.1 & 33.0\\
        
        KELT-20 b & 4 & 2263 & 1.74 & 3.38 & 3.47 & 0.5 & 7.42 & 4.7 & 16.7 & 28.7 & 33.2\\
        
        KELT-4 A b & 5 & 1822 & 1.7 & 0.9 & 2.99 & 0.69 & 8.69 & 3.9 & 23.5 & 24.6 & 34.1\\
        
        KELT-7 b & 6 & 2089 & 1.6 & 1.39 & 2.73 & 0.57 & 7.54 & 4.1 & 19.7 & 28.4 & 34.6\\
        
        WASP-43 b & 7 & 1379 & 0.93 & 1.78 & 0.81 & 0.66 & 9.27 & 3.8 & 22.7 & 26.2 & 34.7\\
\hline
	\end{tabular}
\end{table*}

\setcounter{table}{2}
\begin{table*}
	\centering
	\caption{100 best eclipse mapping targets for JWST MIRI/LRS - Continued.}
	\label{tab:emm_list_miri}
	\begin{tabular}{lccccccccccc} % four columns, alignment for each
		\hline
        Planet Name & {Rank} & $T_{\mathrm{eq}}$ [K] & $R_{\mathrm{p}} \ [\mathrm{R}_{\mathrm{J}}]$ & $M_{\mathrm{p}} \ [\mathrm{M}_{\mathrm{J}}]$ & $P$ [days] & $b$ & $K$ (mag) & $N_{\mathrm{max}}$ & $\mathrm{EMM}_x$ [deg] & $\mathrm{EMM}_y$ [deg] & $\mathrm{EMM}$ [deg]\\
		\hline

        WASP-84 b & 63 & 801 & 0.96 & 0.69 & 8.52 & 0.65 & 8.86 & 2.6 & 31.8 & 37.4 & 49.1\\

        HD 202772 A b & 64 & 2133 & 1.54 & 1.02 & 3.31 & 0.41 & 6.96 & 3.5 & 20.4 & 44.8 & 49.2\\

        WASP-18 b & 65 & 2504 & 1.24 & 10.2 & 0.94 & 0.37 & 8.13 & 3.9 & 18.4 & 45.7 & 49.3\\
        
        KELT-12 b & 66 & 1799 & 1.78 & 0.95 & 5.03 & 0.59 & 9.36 & 2.7 & 29.0 & 40.0 & 49.4\\
        
        WASP-31 b & 67 & 1574 & 1.55 & 0.48 & 3.41 & 0.78 & 10.65 & 2.6 & 38.6 & 30.8 & 49.4\\
        
        TOI-640 b & 68 & 1747 & 1.77 & 0.88 & 5.0 & 0.89 & 9.24 & 3.2 & 43.9 & 22.8 & 49.5\\
        
        WASP-100 b & 69 & 1949 & 1.33 & 1.26 & 2.85 & 0.81 & 9.67 & 2.6 & 40.2 & 29.4 & 49.8\\
\hline
	\end{tabular}
\end{table*}

% Don't change these lines
\bsp	% typesetting comment
\label{lastpage}
\end{document}